\documentclass{jfm}

\usepackage{graphicx}
\usepackage{natbib}
\usepackage{amsmath,epsfig}
\usepackage{color}
\usepackage{soul}

\ifCUPmtlplainloaded \else
  \checkfont{eurm10}
  \iffontfound
    \IfFileExists{upmath.sty}
      {\typeout{^^JFound AMS Euler Roman fonts on the system,
                   using the 'upmath' package.^^J}%
       \usepackage{upmath}}
      {\typeout{^^JFound AMS Euler Roman fonts on the system, but you
                   dont seem to have the}%
       \typeout{'upmath' package installed. JFM.cls can take advantage
                 of these fonts,^^Jif you use 'upmath' package.^^J}%
      }
  \else
  \fi
\fi

\ifCUPmtlplainloaded \else
  \checkfont{msam10}
  \iffontfound
    \IfFileExists{amssymb.sty}
      {\typeout{^^JFound AMS Symbol fonts on the system, using the
                'amssymb' package.^^J}%
       \usepackage{amssymb}%
       \let\le=\leqslant  
       \let\ge=\geqslant  
      }{}
  \fi
\fi

\ifCUPmtlplainloaded \else
  \IfFileExists{amsbsy.sty}
    {\typeout{^^JFound the 'amsbsy' package on the system, using it.^^J}%
     \usepackage{amsbsy}}
    {}
\fi

\newsavebox{\astrutbox}
\sbox{\astrutbox}{\rule[-5pt]{0pt}{20pt}}

\def\build#1_#2^#3{\mathrel{\mathop{\kern 0pt#1}\limits_{#2}^{#3}}}

\newtheorem{theorem}{Theorem}[section]

\newtheorem{proposition}[theorem]{Proposition}

\DeclareMathOperator\erf{erf}
\DeclareMathOperator\erfc{erfc}

\title[Modelling Lagrangian velocity and acceleration in turbulent flows]{Modelling Lagrangian velocity and acceleration in turbulent flows as infinitely differentiable stochastic processes}

\author[ ]%
{Bianca Viggiano$^{1,2}$, Jan Friedrich$^{1}$, Romain Volk$^1$, Mickael Bourgoin$^1$, Ra\'{u}l Bayo\'{a}n Cal$^{1,2}$, Laurent Chevillard$^1$\thanks{laurent.chevillard@ens-lyon.fr}}

\affiliation{$^1$Univ Lyon, Ens de Lyon, Univ Claude Bernard, CNRS, Laboratoire de Physique, 46 all\'ee d'Italie F-69342 Lyon, France\\
$^2$Department of Mechanical and Materials Engineering, Portland State University, Portland, Oregon, USA}

\pubyear{}
\volume{..}
\pagerange{..}
\date{..}

\begin{document}

\maketitle

\begin{abstract}

We develop a stochastic model for Lagrangian velocity as it is observed in experimental and numerical fully developed turbulent flows.  We define it as the unique statistically stationary solution of a causal dynamics, given by a stochastic differential equation. In comparison to previously proposed stochastic models, the obtained process is infinitely differentiable at a given finite Reynolds number, and its second-order statistical properties converge to those of an Ornstein-Uhlenbeck process in the infinite Reynolds number limit. In this limit, it exhibits furthermore intermittent scaling properties, as they can be quantified using higher-order statistics. To achieve this, we begin with generalizing the two-layered embedded stochastic process of \cite{Saw91} by considering an infinite number of layers. We then study, both theoretically and numerically, the convergence towards a smooth (i.e. infinitely differentiable) Gaussian process. To include intermittent corrections, we follow similar considerations as for the multifractal random walk of \cite{BacDel01}. We derive in an exact manner the statistical properties of this process, and compare them to those estimated from Lagrangian trajectories extracted from numerically simulated turbulent flows. Key predictions of the multifractal formalism regarding acceleration correlation function and high-order structure functions are also derived. Through these predictions, we understand phenomenologically peculiar behaviours of the fluctuations in the dissipative range, that are not reproduced by our stochastic process. The proposed theoretical method regarding the modelling of infinitely differentiability opens the route to the full stochastic modelling of velocity, including the peculiar action of viscosity on the very fine scales.
\end{abstract}

\normalsize



\section{Introduction}

Stochastic modelling of Lagrangian velocity and acceleration has a long history in the literature of turbulent flows (see  \cite{Pop90,PopChe90,Saw91,BorSaw94,WilSaw96,Pop02,MorDel03,SawYeu03,Bec03,Fri03,Rey03a,Rey05,LamPop07,MinChi14}, and references therein). Typical modelling approaches consist of proposing a random process in time for the velocity $v(t)$ of a tracer particle advected by a turbulent flow begins with reproducing the expected behaviour given by the standard phenomenology of turbulence.  At very large Reynolds number, in a sustained, statistically stationary, turbulent flow of characteristic large integral length scale $L$, (i) Lagrangian velocity itself is a statistically stationary process of finite variance $\langle v^2\rangle=\sigma^2$ and is correlated over a large time scale $T\propto L/\sigma$, (ii) it is non-differentiable (i.e. rough) such that the velocity increment variance $\langle (\delta_\tau v)^2\rangle$, where $\delta_\tau v(t)=v(t+\tau)-v(t)$, is proportional to $\tau$ as the scale $\tau$ becomes smaller. This is the standard dimensional picture of Lagrangian turbulence at infinite Reynolds numbers \citep{MonYag71,TenLum72}. Nonetheless, at a finite Reynolds number, let us stress that $v$ is regularized at small scales by viscosity, and an appropriate modelling must produce differentiable kinematic quantities.

From a stochastic point of view, we could wonder whether a random process $v(t)$ with $t\in\mathbb R$, and its respective dynamics ensuring causality could be built with the capability of reproducing these aforementioned statistical properties. More precisely, rephrased in terms inherited from the mathematics of stochastic differential equations, we would like to define such a process $v(t)$ as the solution of an evolution equation forced by a random force. {Henceforth, we will attribute the causality property to a given random process $v(t)$ if its infinitesimal increment $dv(t)\equiv v(t+dt)-v(t)$ over $dt$ is governed by the history of $v(t)$ (or any functionals of it) up to time $t$, and additional non anticipative filtering of the Wiener process (see for instance the textbook of \cite{Nua00}). In this context,  the simplest linear and Markovian stochastic evolution  is given by the so-called Ornstein-Uhlenbeck (OU) process} that reads
\begin{equation}\label{eq:OUIntro}
dv(t) = -\frac{1}{T}v(t) dt +\sqrt{\frac{2\sigma^2}{T}}W(dt),
\end{equation}
where $W(dt)$ is an instance of the increment over $dt$ of a Gaussian Wiener process.  It can be understood in an heuristic way as a collection of independent realizations of a zero-average Gaussian random variable of variance $dt$ (i.e., a white noise). The statistical properties of the unique solution $v(t)$ of this evolution (Eq. \ref{eq:OUIntro}) are precisely reviewed in Section \ref{sec:StandardOU}. We can nonetheless notice that since $v$ is defined as a linear operation on a Gaussian random force, it is necessarily Gaussian itself, and is indeed consistent with a finite variance process $\langle v^2\rangle=\sigma^2$ and the linear behaviour of its respective second-order structure function $\langle (\delta_\tau v)^2\rangle$ with $\tau$ representing the time delay (see the discussion in Section \ref{sec:StandardOU} and Eq. \ref{eq:StructFuncv1}).

Going beyond this simple phenomenology, and its respective stochastic modelling, we would like  to include finite Reynolds number effects, and  in particular acquire a stochastic description of the related acceleration process $a(t)=dv(t)/dt$. Notice that the stochastic evolution of $v(t)$ using a OU process (Eq. \ref{eq:OUIntro}) is typical of a non-differentiable process, and thus fails to reproduce proper statistical behaviours for $a$. To do so, we have to replace the white noise term $W(dt)$ entering in Eq. \ref{eq:OUIntro} by a finite-variance random force, correlated over a non-vanishing time scale $\tau_\eta$, that eventually depends on viscosity, known as the dissipative Kolmogorov time scale. If we furthermore assume that this random force is itself defined as the solution of an OU process of characteristic time scale $\tau_\eta$, we recover the two-layered embedded stochastic model of \cite{Saw91}. We review its statistical properties in Section \ref{sec:Sawford}. This model is appealing since it incorporates in a simple way the additional necessary time scale $\tau_\eta$ implied by the finite value of viscosity, or equivalently, the finite value of the Reynolds number. Both velocity and acceleration are statistically stationary and of finite variance in this framework, and the predicted acceleration correlation function reproduces in a consistent way the fact that it has to cross zero in the vicinity of $\tau_\eta$, before decaying towards 0 over $T$. Nonetheless, whereas the model gives an appropriate description of the velocity correlation function in both the inertial and dissipative ranges, further comparisons to numerical data (see respective discussions in \cite{Saw91,LamPop07}) underlined its limitations regarding the behaviour of the acceleration correlation function in the dissipative range, i.e. for time lags smaller than this zero-crossing time scale. 

Obviously, in the model of \cite{Saw91}, whereas velocity is differentiable, leading to a finite variance acceleration process, it is not twice differentiable: the obtained acceleration process is not a differentiable random function. This observation has strong implications on the shape of the acceleration correlation function.  In particular in the dissipative range: as observed in numerical data for both velocity and acceleration, and expected from the physical point of view when viscosity is finite, correlation functions of differentiable random functions are parabolic (or smoother) in the vicinity of the origin, whereas the predicted acceleration correlation function of  \cite{Saw91} behaves linearly. Modelling Lagrangian velocity by a two-layered embedded OU process, hence, appears to be too simplistic to reproduce the correlation structure of acceleration in the dissipative range.

For this reason, we found it relevant and original to develop and generalize the model of \cite{Saw91} in order to provide a meaning and answer to the following question: can we construct a causal stochastic process which is infinitely differentiable at a given finite Reynolds number, or equivalently at a given finite dissipative time scale $\tau_\eta$, consistent with the standard aforementioned phenomenology of turbulence in the inertial range (i.e. for scales $\tau_\eta\ll \tau\ll T$), and that converges towards an OU process (Eq. \ref{eq:OUIntro}) at infinite Reynolds numbers (or equivalently as $\tau_\eta\to 0$)?
We indeed develop in Sections \ref{Sec:GenNLayersSawford} and \ref{Sec:InfDiffProc} such a process. It is obtained as the generalization of the framework of  \cite{Saw91} to $n$ layers, the first layer corresponding to a Langevin process of characteristic time scale $T$, and then $n-1$ layers corresponding to the dynamics of the random forcing term given by Langevin processes of characteristic time scale $\tau_\eta$. Infinite differentiability is attained while iterating this procedure for an infinite number of layers $n\to \infty$, while properly normalizing the small time scale $\tau_\eta$ by a factor $\sqrt{n}$ to ensure a non-trivial convergence, as it is rigorously done in Section \ref{Sec:InfDiffProc}. We eventually end up with an infinitely differentiable causal random process, which is Gaussian, and derive in an exact fashion its statistical properties (listed in Proposition \ref{propcorrelationsInfinite}). We furthermore propose a first numerical illustration of this process in Section \ref{Sec:FirstNS}, through simulation of a time series of velocity and its respective acceleration, and comparison to theoretical expressions.  

As we quickly mentioned, since its dynamics is made of embedded linear operations on a Gaussian white noise, it is itself Gaussian. Such a Gaussian framework, in particular for acceleration, is at odds with experimental and numerical investigations of Lagrangian turbulence (see \cite{YeuPop89,VotSat98,PorVot01,MorMet01,MorDel02,MorDel03,CheRou03,Fri03,BifBof04,TosBod09,PinSaw12,BenLal19}, and references therein). As correctly predicted by \cite{Bor93}, the observed level of intermittency in the Lagrangian framework is found much higher than in the Eulerian framework \citep{Fri95}. 

To reproduce these highly non-Gaussian features of Lagrangian turbulence, we propose then to extend the construction of the current infinitely differentiable process  to include the intermittent, i.e. multifractal, nature of the fluctuations. To do so, we first revisit the construction of the so-called multifractal random walk (MRW) of \cite{BacDel01} that was shown in \cite{MorDel02} to reproduce several key aspects of Lagrangian intermittency. Compared to previously published investigations, we include, in an original way, the notion of causality in this non-Gaussian random walk. We design a stochastic evolution for the probabilistic model of the intermittency phenomenon (i.e. the multiplicative chaos) in \S \ref{Sec:CausalMRW}. We then proceed with deriving in a rigorous way its statistical properties, and list them in Propositions \ref{Prop:StatX_FOU} and {Section \ref{Sec:CausalMRW}}. Finite Reynolds number effects, and the implied infinitely differentiability, are then included in a similar fashion as in the first part of the article. Developments on this intermittent and infinitely differentiable process are proposed in Section \ref{Sec:MRWInfiniteDeriv}, and we highlight its statistical properties in Propositions \ref{propcorrelationsInfiniteX}, \ref{propcorrelationsInfiniteMRW} and \ref{propSFInfiniteMRW}. As we explain in Section \ref{Sec:CausalMRW}, including intermittency implies the introduction of a non-Markovian step, that is necessary to reproduce the high level of roughness (that we define precisely) implied by the multifractal structure of the trajectories: this asks for the design of a novel numerical algorithm able to simulate in an efficient way its time series. We propose then in Section \ref{Sec:AlgoSimulIDMRW} such an algorithm in which efficiency is based on its formulation in the Fourier space, allowing optimal consideration given its non-Markovian nature. Simulations of the time series of velocity and acceleration are then proposed in Section \ref{Sec:NumResInfMultiProcess}, where we compare the numerical estimation of their statistical properties to our theoretical predictions.

Section \ref{Sec:CompModDNSTotal} is devoted to the comparison of the statistical properties of the infinitely differentiable multifractal process to trajectories extracted from direct numerical simulations (DNSs) of the Navier-Stokes equations (see details on the database in Section \ref{Sec:CompModDNSDescDS}). To make this comparison transparent and reproducible, we explain in Section \ref{Sec:CalibModDNS} the chosen procedure to calibrate the model parameters $\tau_\eta$ and $T$, and their link to the physical parameters of the DNS data. Overall, we find  very good agreement between the statistical properties of the DNS data, and of those predicted by our theoretical approach. We nonetheless underline some discrepancies on the flatness of velocity increments in the dissipative range: As it is detailed in \S \ref{Sec:CompModDNSLimFlat}, the model does not reproduce the observed rapid increase of the flatness in the dissipative range, a behaviour which is known to be related to the very peculiar differential action of viscosity on the final damping of the singularities developed by the flow.

This motivates the final investigation that we propose in Section \ref{Sec:PredMultiForm} where we derive the corresponding predictions as they are obtained from the multifractal formalism \citep{Fri95}. As far as we know, this has never been done for the acceleration correlation function, and we take special care to quantify precisely the respective prediction for the Reynolds number dependence of acceleration variance (see Section \ref{Sec:PredMultiVarAcceRey}). Compared to the previous approach, aimed at building a stochastic process as the solution of a causal dynamical evolution, the multifractal formalism is not as complete from a probabilistic point of view: we do not obtain the time series of velocity and acceleration, but only model some of their statistical properties (i.e. their high-order structure functions). Once again, the calibration procedure is detailed (Section \ref{Sec:CalibCompMulti}), and proceed with the comparison to DNS data. We observe also an excellent agreement between predictions and estimations based on DNS data. In particular, which is our initial motivation, multifractal formalism, and its modelling of a fluctuating dissipative time scale, is able to reproduce this rapid increase of the flatness in the dissipative range. 

We gather conclusions and perspectives in \S \ref{Sec:Conclusion}.

\section{Ordinary and embedded Ornstein-Uhlenbeck processes as statistically stationary models for Lagrangian velocity and acceleration}

\subsection{Ordinary single-layered Ornstein-Uhlenbeck process}\label{sec:StandardOU}

Standard arguments developed in turbulence phenomenology \citep{TenLum72} lead to the consideration of, as a stochastic model for velocity of Lagrangian tracers, the Ornstein-Uhlenbeck (OU) process. In particular, such a process reaches a statistically stationary regime in which variance is finite and exponentially correlated. Let us denote such a process by $v_1(t)$, and define it as the unique stationary solution of the following stochastic differential equation, also called Langevin equation, 
\begin{equation}
 dv_1(t) =  -\frac{1}{T}v_1(t) dt + \sqrt{q}W(dt),
  \label{eq:OU}
\end{equation}
where $T$ is the turbulence (large) turnover time, $W(t)$ is a Wiener process, and $W(dt)$ its infinitesimal increment over $dt$ (i.e., independent instances of a Gaussian random variable, zero-average and of variance $dt$). It obeys the following rule of calculation (\textbf{cf.} \cite{Nua00}): for any appropriate deterministic functions $f$ and $g$, which follow in particular integrability conditions such that,
\begin{equation}
\left\langle \int_\mathcal A f(t)W(dt)\right\rangle=0,
  \label{eq:rule1}
\end{equation}
and
\begin{equation}
\left\langle \int_\mathcal A f(t)W(dt)\int_\mathcal B g(t)W(dt)\right\rangle=\int_{\mathcal A\cap \mathcal B} f(t)g(t)dt,
  \label{eq:rule2}
\end{equation}
where $\langle.\rangle$ stands for ensemble average, and $\mathcal A\cap \mathcal B$ is the intersection of the two ensembles $\mathcal A$ and $\mathcal B$.

The unique statistically stationary solution of the stochastic differential equation (SDE) provided in Eq. \ref{eq:OU} can be written conveniently as
\begin{equation}\label{eq:IntOU}
  v_1(t) = \sqrt{q}\int_{-\infty}^{t}e^{-(t-t')/T}W(dt')\;.
\end{equation}
Since $v_1$ is defined as a linear operation on the Gaussian white noise $W(dt)$, it is Gaussian itself. Following the rules given in Eqs. \ref{eq:rule1} and \ref{eq:rule2}, it is thus fully characterized by its average and correlation function. In particular, $v_1$ is a zero-average process, i.e. $\langle v_1 \rangle=0$, and is correlated as
\begin{equation}\label{eq:corrv1}
  \mathcal C_{v_1}(t_1-t_2)\equiv \langle v_1(t_1)v_1(t_2) \rangle =
  q\int_{-\infty}^{\textrm{min}(t_1,t_2)}  e^{-(t_1+t_2-2t)/T}dt
  = \frac{qT}{2} e^{-|t_1-t_2|/T}.\;
\end{equation}
Notice that $v_1$ is a finite variance process $\langle v_1^2\rangle =  qT/2$ (consider the value of the correlation function Eq. \ref{eq:corrv1} at equal times, $t_1=t_2$), and behaves at small scales as a Brownian motion, as it is required by dimensional arguments developed in the standard phenomenology of turbulence at infinite Reynolds number \citep{TenLum72}. To see this, define the velocity increment as
\begin{equation}\label{eq:Incrv1}
\delta_\tau v_1(t) \equiv v_1(t+\tau)-v_1(t),
\end{equation}
and notice that 
\begin{equation}\label{eq:StructFuncv1}
\left\langle \left(\delta_\tau v_1(t)\right)^2 \right\rangle =  2\left[\left\langle v_1^2\right\rangle-\mathcal C_{v_1}(\tau)\right]\build{\sim}_{\tau\rightarrow 0}^{}q|\tau|.
\end{equation}
The scaling behaviour given in Eq. \ref{eq:StructFuncv1} is typical of non-differentiable processes. Hence, the respective acceleration process $a_1(t) \equiv dv_1/dt$ is ill-defined (actually it is a random distribution). To circumvent this pathological behaviour, \cite{Saw91} has proposed to introduce the dissipative Kolmogorov time scale $\tau_\eta$, which will be discussed in the following Section.

\subsection{Embedded Ornstein-Uhlenbeck processes}

\subsubsection{Two layers: the Sawford model}\label{sec:Sawford}
Here, we follow the approach developed by  \cite{Saw91}. We consider the following embedded OU process $v_2(t)$:
\begin{equation}
   \label{eq:OU_embedded_v2}
  \frac{dv_2}{dt} = -\frac{1}{T}v_2(t)+ f_1(t),\;
\end{equation}
where $f_1(t)$ in an external random force that obeys itself an ordinary OU process, as it is discussed in the previous section \ref{sec:StandardOU}, but exponentially correlated over the small time scale $\tau_\eta$. It is thus defined as the unique solution of the following SDE:
\begin{equation}
 df_1(t) =  -\frac{1}{\tau_\eta}f_1(t) dt + \sqrt{q}W(dt).
\end{equation}
Hence, it is a zero-average Gaussian process, and its correlation function is given by
\begin{equation}\label{eq:corrf1}
  \mathcal C_{f_1}(\tau)\equiv \langle f_1(t)f_1(t+\tau) \rangle = \frac{q\tau_\eta}{2} e^{-|\tau|/\tau_\eta}.
\end{equation}
The unique statistically stationary solution of Eq. \ref{eq:OU_embedded_v2} is once again given by 
\begin{equation*}
  v_2(t) = \int_{-\infty}^{t}e^{-(t-t')/T}f_1(t')dt',\;
\end{equation*}
showing that $v_2$ is a zero-average Gaussian process, and correlated as
\begin{equation}\label{eq:corrv2def}
  \mathcal C_{v_2}(\tau)\equiv \langle v_2(t)v_2(t+\tau) \rangle = \int_{-\infty}^{t} \int_{-\infty}^{t+\tau} \; e^{-(2t+\tau-t_1-t_2)/T}\mathcal C_{f_1}(t_1-t_2)dt_1dt_2.
\end{equation}
Assuming without loss of generality $\tau\ge 0$ (recall that the correlation function of a statistically stationary process is an even function of its argument), splitting the integral entering in Eq. \ref{eq:corrv2def} over the dummy variable $t_2$ into the two sets $[-\infty,t]$ and $[t,t+\tau]$, and performing the remaining explicit double integral, we obtain the following expression: 
\begin{equation}  \label{eq:corr_v2_final}
  \mathcal C_{v_2}(\tau)= \frac{q \tau_{\eta}^2 T^2}{2(T^2-\tau_{\eta}^2)}\left[T e^{-|\tau|/T} - \tau_{\eta}e^{-|\tau|/\tau_{\eta}}\right],\;
\end{equation}
which is in agreement with the formula given by \cite{Saw91}.

The respective acceleration process $a_2(t)\equiv dv_2(t)/dt$, obtained from Eq. \ref{eq:OU_embedded_v2}, is accordingly a zero-average Gaussian process, and its correlation function is given by 
\begin{equation}\label{eq:corra2}
  \mathcal C_{a_2}(\tau)\equiv \langle a_2(t)a_2(t+\tau) \rangle=
  -\frac{\textrm{d}^2}{\textrm{d}\tau^2} \langle v_2(t)v_2(t+\tau) \rangle=
  \frac{q \tau_{\eta}^2 T^2}{2(T^2-\tau_{\eta}^2)}\left[ -\frac{1}{T}e^{-|\tau|/T} + \frac{1}{\tau_{\eta}}e^{-|\tau|/\tau_{\eta}}\right].\;
\end{equation}
Notice that the function $\mathcal C_{v_2}$ (Eq. \ref{eq:corr_v2_final}) is indeed twice differentiable at the origin, contrary to the function $\mathcal C_{v_1}$ (Eq. \ref{eq:corrv1}), such that $a_2$ has finite variance given by $\mathcal C_{a_2}(0)$ (Eq. \ref{eq:corra2}).

\subsubsection{Generalization to $n$ layers}\label{Sec:GenNLayersSawford}

By iterating the aforementioned procedure, we can consider similarly $n$ additional layers instead of  a single one, as it is proposed in the embedded Ornstein-Uhlenbeck process (Eq. \ref{eq:OU_embedded_v2}) by Sawford. Here, acceleration is not only a well defined random process, but also the velocity derivatives of order $n$. Once again, these additional layers will eventually be modeled as OU processes. A similar type of procedure has been adopted in \cite{ArrCab14} in a different context. The obtained embedded structure is defined using a set of $n$ coupled stochastic ODEs, with $n\ge 2$, that reads
\begin{align}
 \label{eq:OU_embedded_N_1}
   \frac{dv_n}{dt} &= -\frac{1}{T}v_n(t)+ f_{n-1}(t)\;\\
   \frac{df_{n-1}}{dt} &= -\frac{1}{\tau_{\eta}}f_{n-1}(t)+ f_{n-2}(t)\;\\
   &...&\\  
   \frac{df_{2}}{dt} &= -\frac{1}{\tau_{\eta}}f_{2}(t)+ f_{1}(t)\;\\
   df_1 &=  -\frac{1}{\tau_\eta}f_1(t) dt + \sqrt{q_{(n)}}W(dt)\;.
\label{eq:OU_embedded_N_N}
\end{align}
The remaining free parameter $q_{(n)}$ can be eventually chosen such that  \begin{equation}\label{eq:ConstVarVn}
\langle v_n^2\rangle = \sigma^2,
\end{equation}
independently of $\tau_\eta$ and/or the number of layers $n$, as it is required by the standard phenomenology of Lagrangian turbulence \citep{TenLum72}.

{We present in Proposition \ref{propcorrelationsN} the explicit computation of the correlation functions of  velocity $v_n$ and the respective acceleration $a_n$ in the statistically stationary regime, obtained from the set of equations \ref{eq:OU_embedded_N_1} to \ref{eq:OU_embedded_N_N} as $t\to \infty$. Their expressions are especially simple in the spectral domain, and read, considering $n\ge 2$ to ensure that acceleration is a well defined process,
\begin{equation}\label{CorrVelNFourText}
  \mathcal C_{v_n}(\tau) = q_{(n)}\int_{\mathbb R}e^{2i\pi \omega\tau}\frac{T^2}{1+4\pi^2T^2\omega^2}\left[\frac{\tau_\eta^{2}}{1+4\pi^2\tau_\eta^2\omega^2}\right]^{n-1}d\omega,
  \end{equation}
and
 \begin{equation}\label{CorrAccNFourText}
  \mathcal C_{a_n}(\tau)= q_{(n)}\int_{\mathbb R}4\pi^2\omega^2 e^{2i\pi \omega\tau}\frac{T^2}{1+4\pi^2T^2\omega^2}\left[\frac{\tau_\eta^{2}}{1+4\pi^2\tau_\eta^2\omega^2}\right]^{n-1}d\omega,
  \end{equation}
where the multiplicative factor $q_{(n)}$ (defined in Eq. \ref{ExplicitQN}) enforces the prescribed value of velocity variance (Eq. \ref{eq:ConstVarVn}). Let us notice that taking $n=2$ layers, the respective correlation of the process $v_2$ coincides with the one proposed in \cite{Saw91}, as it is recalled in Section~\ref{sec:Sawford}.
}

It is interesting to consider the limiting process $v$ or $a$ when the number of layers $n$ goes towards infinity from a physical point of view, which would give an example of a causal infinitely differentiable process, if such a process exists. It is indeed possible to show rigorously that the correlation function of $v_n$ (Eq. \ref{CorrVelNFourText}) looses its dependence on the time scale $\tau$. We then have $\mathcal C_{v_n}(\tau)\rightarrow \sigma^2$ for any $\tau\ge 0$ as $n\rightarrow \infty$. Thus, asymptotically, the limiting process does not decorrelate, which is at odds with the expected behaviour. We will see in the following section \ref{Sec:InfDiffProc} that by considering the re-scaled dissipative time scale $\tau_\eta/\sqrt{n-1}$ instead of $\tau_\eta$, the system of equations will converge towards a proper process with an appropriate correlation function as $n\rightarrow \infty$.

\subsection{Towards an infinitely differentiable causal process}\label{Sec:InfDiffProc}

Consider the following system of embedded differential equations:
\begin{align}
 \label{eq:OU_infinite_N_1}
   \frac{dv_n}{dt} &= -\frac{1}{T}v_n(t)+ f_{n-1}(t)\;\\
   \frac{df_{n-1}}{dt} &= -\frac{\sqrt{n-1}}{\tau_{\eta}}f_{n-1}(t)+ f_{n-2}(t)\;\\
   &...&\\  
   \frac{df_{2}}{dt} &= -\frac{\sqrt{n-1}}{\tau_{\eta}}f_{2}(t)+ f_{1}(t)\;\\
   df_1 &=  -\frac{\sqrt{n-1}}{\tau_\eta}f_1(t) dt + \sqrt{\alpha_{n}}W(dt)\;,
\label{eq:OU_infinite_N_N}
\end{align}
with 
\begin{equation}\label{eq:ChoiceAlphaN}
\alpha_n=\left( \frac{n-1}{\tau_\eta^2}\right)^{n-1}\frac{2\sigma^2e^{-\tau_\eta^2/T^2}}{T\erfc\left(\tau_\eta/T\right)},
\end{equation}
where we have introduced the error function $\erf(t) =(2/\sqrt{\pi})\int_0^te^{-s^2}ds$, and its respective complementary $\erfc(t)=1-\erf(t)$. The chosen white noise weight $\alpha_n$ (Eq. \ref{eq:ChoiceAlphaN}) ensures that the variance of the limiting process $v$ is finite with $\langle v^2\rangle=\sigma^2$.

{We summarize and derive in the Appendix (see Proposition~\ref{propcorrelationsInfinite}) the statistical properties of the unique statistically stationary solution of the set of embedded differential equations \ref{eq:OU_infinite_N_1} to \ref{eq:OU_infinite_N_N}. In particular, the velocity correlation function now reads
\begin{equation}\label{CorrVelNFour2}
  \mathcal C_{v_n}(\tau) = \frac{2\sigma^2e^{-\tau_\eta^2/T^2}}{T\erfc\left(\tau_\eta/T\right)}\int_{\mathbb R}e^{2i\pi \omega\tau}\frac{T^2}{1+4\pi^2T^2\omega^2}\left[\frac{1}{1+\frac{4\pi^2\tau_\eta^2\omega^2}{n-1}}\right]^{n-1}d\omega.
  \end{equation}
  }
{Whereas the function provided in Eq. \ref{CorrVelNFourText} does not converge towards a correlation function of a well-behaved stochastic process as the number of layers goes to infinity, Eq. \ref{CorrVelNFour2} does. In other words, through iteration of the set of embedded differential equations, \ref{eq:OU_infinite_N_1} to \ref{eq:OU_infinite_N_N}, over an infinite number of layers $n\rightarrow \infty$, we obtain an infinitely differentiable and causal Gaussian process, in which the velocity correlation function reads, in the stationary regime,
\begin{equation}\label{ExplicitCorrVLim}
\mathcal C_{v}(\tau)= \sigma^2\frac{e^{-|\tau|/T}}{2\erfc(\tau_\eta/T)}\left[1+\erf\left(\frac{|\tau|}{2\tau_\eta}-\frac{\tau_\eta}{T}\right)+e^{2|\tau|/T}\erfc\left(\frac{|\tau|}{2\tau_\eta}+\frac{\tau_\eta}{T}\right)\right].
  \end{equation}
Let us notice that indeed $\mathcal C_{v}(0)=\langle v^2\rangle=\sigma^2$. Furthermore, taking the second derivatives of Eq. \ref{ExplicitCorrVLim} and multiplying by the factor $-1/2$, we obtain the respective acceleration correlation function
\begin{align}\label{ExplicitCorrALim}
\mathcal C_{a}(\tau)= \frac{\sigma^2}{2T^2\erfc(\tau_\eta/T)}&\left[\frac{2T}{\tau_\eta\sqrt{\pi}}e^{-\left( \frac{\tau^2}{4\tau_\eta^2}+\frac{\tau_\eta^2}{T^2}\right)}-e^{-|\tau|/T}\left(1+\erf\left(\frac{|\tau|}{2\tau_\eta}-\frac{\tau_\eta}{T}\right)\right)\right. \notag\\
&\left.-e^{|\tau|/T}\erfc\left(\frac{|\tau|}{2\tau_\eta}+\frac{\tau_\eta}{T}\right)\right].
\end{align}
}

\subsection{A first numerical illustration}\label{Sec:FirstNS}

A first numerical illustration is proposed to observe numerically how the statistical characteristics of the Gaussian process $v_n$, typically its correlation function and the one of the associated acceleration for a given set of values of the parameters $\tau_\eta$ and $T$ go towards the limiting process $v$ (and given in Proposition \ref{propcorrelationsInfinite}) as the number of layers $n$ increases. {This limiting process $v$, being Gaussian and of zero average, is completely characterized by its correlation function (Eq. \ref{ExplicitCorrVLim}) in the statistically stationary regime, and could be obtained as a linear operation on the white Gaussian noise. Performing such a simulation is possible, although a causal kernel would need to be found such that the correlation function is consistent with Eq. \ref{ExplicitCorrVLim}. Although interesting, this is not a simple task and this perspective is kept for future investigations. Furthermore, in subsequent numerical simulations, the convergence towards the statistically steady state while solving the transient regime is observed. For these reasons, the set of stochastic differential equations  \ref{eq:OU_infinite_N_1} to \ref{eq:OU_infinite_N_N} for a given finite number of layers $n$ will be solved, and thus give a numerical estimation of the process $v_n$ and its statistical properties.}

\begin{figure}
\begin{center}
\epsfig{file=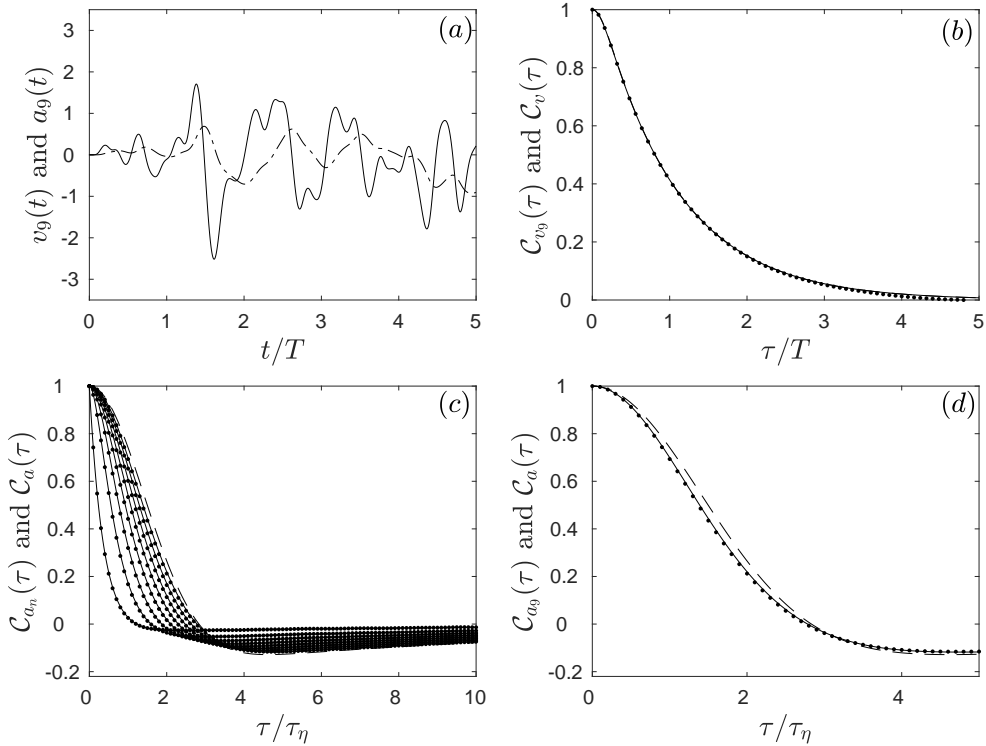,width=13cm}
\end{center}
\caption{\label{fig:NumOU} Numerical simulation of the set of equations \ref{eq:OU_infinite_N_1} to \ref{eq:OU_infinite_N_N} using $n=9$ layers, for $\tau_\eta=T/10$ and $\sigma^2=1$ (see text). (a) Typical time series of the obtained processes $v_9(t)$ (dashed line) and $a_9(t)$ (solid line), as a function of time $t$. (b) Respective velocity correlation functions $\mathcal C_{v_9}$, estimated from numerical simulations (dots), theoretically derived from Eq. \ref{CorrVelNFour2} (solid line), and the correlation function of the asymptotic process $\mathcal C_{v}$ which expression is provided in  Eq. \ref{ExplicitCorrVLim}. (c) Acceleration correlation functions $\mathcal C_{a_n}$ using $n$ layers, $n$ ranging from 2 to 9 (from left to right), using $\sigma^2=1$ and $\alpha_n=\alpha_9$ (Eq. $\ref{eq:ChoiceAlphaN}$). Numerical estimations from time series are displayed with dots, respective theoretical expressions starting from Eq. \ref{CorrVelNFour2} are represented with solid lines, and the asymptotic correlation function $\mathcal C_{a}$ (Eq.  \ref{ExplicitCorrALim}) is shown with a dashed line. For the sake of clarity, all curves are normalized by their values at the origin (i.e. the respective variances). (d) Similar plot as in (c), but only the layer $n=9$ is displayed, over a shorter range of time lag $\tau$.}
\end{figure}

We perform a numerical simulation of the set of equations \ref{eq:OU_infinite_N_1} to \ref{eq:OU_infinite_N_N} using $n=9$ layers, and for $\tau_\eta=T/10$. Choose for instance $T=1$, which is equivalent to dimensionalized time scales in units of $T$. Time integration is performed with a simple Euler discretization scheme. The choice for $dt$ is dictated by the smallest time scale of the system, here $\tau_\eta/\sqrt{n-1}$. Presently for $n=9$, we found the value $dt=\tau_\eta/100$ small enough to guarantee the appropriate behaviour. We take $\sigma^2=1$, and the respective weight $\alpha_9$ of the white noise is given in Eq. \ref{eq:ChoiceAlphaN}. Trajectories are then integrated over $10^4T$ and results are shown in Fig. \ref{fig:NumOU}.  We could have chosen to perform a simulation using more layers, although the simulation gets heavier, and as we will see, the statistical properties of the obtained process are observed very close to the asymptotic ones (as $n\to\infty$). Also, recall that the white noise weight $\alpha_{n+1}$ (Eq. \ref{eq:ChoiceAlphaN}) increases as $n^n$, so from a numerical point of view, if $n$ is chosen large, it may introduces additional rounding errors related to the double-precision floating-point format.

We display first in Fig. \ref{fig:NumOU}(a) an instance of the obtained processes $v_9(t)$ and its derivatives $a_9(t)$, over $5T$ after numerically integrating the equations  \ref{eq:OU_infinite_N_1} to \ref{eq:OU_infinite_N_N}. As claimed in Proposition \ref{propcorrelationsInfinite}, the process $v_9$ (which correlation function is given in Eq. \ref{CorrVelNFour2}) is $8$-times differentiable. Its first derivative $a_9(t)$ is consequently $7$-times differentiable; resulting in a smooth profile correlated over $\tau_\eta$. We could have performed a similar simulation using additional layers, although its estimated correlation functions of velocity and acceleration will eventually be very close to the asymptotic ones of $v$ (and provided in Proposition \ref{propcorrelationsInfinite}).

In Fig. \ref{fig:NumOU}(b), we present three curves corresponding to (i) the estimated correlation function $\mathcal C_{v_9}$ (dots), (ii) its theoretical expression (solid line), obtained when performing the integral entering in Eq. \ref{CorrVelNFour2} using a symbolic calculation software, and (iii) the asymptotic correlation function $\mathcal C_{v}$ given in Eq. \ref{ExplicitCorrVLim} (dashed line). The profiles collapse making it difficult to distinguish between these three curves. The velocity correlation functions  $\mathcal C_{v_n}$ depend weakly on $n$ (not shown). This can be understood easily since the dependence on $n$ is only really crucial in the dissipative scales; scales that are solely highlighted by a small scale quantity such as acceleration.

In this context, we present in Fig. \ref{fig:NumOU}(c) the corresponding estimated and theoretical curves $\mathcal C_{a_n}$ for $n$ ranging from $2$ to $9$ to observe and quantify the convergence of the acceleration correlation function towards its asymptotic regime. Recall that $\mathcal C_{a_2}$ corresponds to the prediction of \cite{Saw91} (see Eq. \ref{eq:corra2}), which is characteristic of the correlation function of a non-differentiable process ($\mathcal C_{a_2}$ is not twice differentiable at the origin). A perfect agreement between the numerical estimation based on random time series, and the theoretical expressions is observed and also derivable from Eq. \ref{CorrVelNFour2}. As the number of layers $n$ increases, the acceleration correlation functions become more and more curved at the origin, guaranteeing finite variance of higher order derivatives. We superpose on this figure the associated asymptotic correlation function $\mathcal C_{a}$ using a dashed line.  Its explicit expression is given in Eq. \ref{ExplicitCorrALim}. $\mathcal C_{a_9}$ is indeed very close to $\mathcal C_{a}$, as shown in Fig. \ref{fig:NumOU}(d). This shows that considering $n=9$ layers is enough to reproduce the statistical behaviours of the asymptotic process, at least for velocity and acceleration, which are our main concern.


\section{An infinitely differentiable causal process, asymptotically multifractal in the infinite Reynolds number limit}\label{Sec:InfDiffProcMulti}

We now elaborate on the system proposed in Eqs. \ref{eq:OU_infinite_N_1} to \ref{eq:OU_infinite_N_N}  in order to include intermittent, i.e. multifractal, corrections. We have to introduce more elaborate probabilistic objects to do so in the spirit of the multifractal random walk \citep{BacDel01}, applied to the Lagrangian context by \cite{MorDel02,MorDel03}. Recall that the zero-average process $v(t)$,  obtained as the limit when $n\rightarrow \infty$ of the causal system defining $v_n$ (Eqs. \ref{eq:OU_infinite_N_1} to \ref{eq:OU_infinite_N_N}), is Gaussian, thus fully characterized by its correlation function (given in Proposition \ref{propcorrelationsInfinite}). To go beyond this Gaussian framework, where linear operations on a Gaussian white noise $W(dt)$ are involved, we will consider in the sequel a non-linear operation while exponentiating a Gaussian field $X(t)$. Such logarithmic correlation structure guarantees multifractal behaviours (specified later). The so-obtained random field is ``$e^{\gamma X}$", where $\gamma$ is a free parameter of the theory that encodes the level of intermittency. This can be seen as a continuous and stationary version of the discrete cascade models developed in turbulence theory (see \cite{MenSre87,BenBif93,Fri95,ArnBac98} and references therein) and is known in the mathematical literature as a multiplicative chaos \citep{RhoVar14}. For recent applications of such a random distribution to the stochastic modelling of Eulerian velocity fields, see for instance \cite{PerGar16,CheGar19}. The purpose of this section is to generalize such a probabilistic approach to a causal context, and to include finite Reynolds number effects that guarantee differentiability below the Kolmogorov time scale $\tau_\eta$.

\subsection{A causal multifractal random walk}\label{Sec:CausalMRW}

Let us here review the stochastic modelling of the Lagrangian velocity proposed  by \cite{MorDel02,MorDel03}, which is based on the multifractal process of \cite{BacDel01}. This process can be considered as an OU process (Eq. \ref{eq:OU}) forced by a non-Gaussian uncorrelated random noise, and is called the multifractal random walk (MRW). Its dynamics reads
\begin{equation}  \label{eq:MRW}
 du_{1,\epsilon}(t) =  -\frac{1}{T}u_{1,\epsilon}(t)dt + \sqrt{q}e^{\gamma X_{1,\epsilon}(t)-\gamma^2\langle X_{1,\epsilon}^2\rangle}W(dt),
\end{equation}
where a new random field $X_{1,\epsilon}$ is introduced.  This random field is Gaussian, zero average, and taken independent of the white noise instance $W(dt)$, thus fully characterized by its correlation function. To reproduce intermittent corrections, as they have been observed in Lagrangian turbulence (see \citep{YeuPop89,VotSat98,PorVot01,MorMet01,MorDel02,MorDel03,CheRou03,BifBof04,TosBod09,PinSaw12,BenLal19}, and references therein), we demand the Gaussian field $X_{1,\epsilon}$ to be logarithmically correlated \citep{BacDel01}. Such a correlation structure implies in particular that the variance of $X_{1,\epsilon}$ diverges as $\epsilon\to 0$, making it difficult to give a proper mathematical meaning to such a field. This divergence is even amplified when considering its exponential, as it is proposed in Eq. \ref{eq:MRW}. Instead, we rely on an approximation procedure, at a given (small) parameter $\epsilon$, that will eventually play, loosely speaking, the role of the small time scale $\tau_\eta$ of turbulence. Such a logarithmic correlation structure has to be truncated over the large time scale $T$ in order to ensure a finite variance. These truncations are well understood from a mathematical perspective \citep{RhoVar14}, and a proper limit as $\epsilon\rightarrow 0$ leads to a well defined, canonical, random distribution.

Nonetheless, nothing is said in \cite{BacDel01} about causality. {Causal representations of multifractal random fields have been previously made by \cite{SchMar01} and \cite{BacMuz03}, yet these propositions are not defined as solutions of some stochastic evolutions.} In order to include this important physical constraint, we define the field $X_{1,\epsilon}$ as the unique statistically stationary solution of a stochastic differential equation, that will eventually be consistent with both truncations over the time scales $\epsilon$ and $T$, and a logarithmic behaviour in between. Being Gaussian, and independent of the white noise $W(dt)$ entering in Eq. \ref{eq:MRW}, such dynamics has to be defined as a linear operation on an independent instance of the Gaussian white noise, call it $\widetilde{W}(dt)$, such that $\langle W(dt)\widetilde{W}(dt')\rangle = 0$ at any time $t$ and $t'$. In this context, such a linear stochastic evolution has been proposed by \cite{Che17} and \cite{PerMor18}, and reads
\begin{equation}\label{eq:FracOUH}
dX_{1,\epsilon}(t)=-\frac{1}{T}X_{1,\epsilon}(t)dt -\frac{1}{2}\int_{-\infty}^t \left[ t-s+\epsilon\right]^{-3/2}\widetilde{W}(ds)dt + \epsilon^{-1/2}\widetilde{W}(dt).
\end{equation}
It can be seen as a fractional Ornstein-Uhlenbeck process of vanishing Hurst exponent \citep{Che17,PerMor18}. Remark also that the underlying integration over the past with a rapidly decreasing kernel that enters in the dynamics of $X_{1,\epsilon}$ (Eq. \ref{eq:FracOUH}) implies that we are dealing with non-Markovian processes. {A precise and comprehensive characterization of the statistical properties of the fields $X_{1,\epsilon}$ and its asymptotical log-correlated version $X_{1}\equiv \lim_{\epsilon\rightarrow 0}X_{1,\epsilon}$ can be found in Appendix~\ref{Prop:StatX_FOU}.}

Let us now focus on the statistical properties of the MRW that now includes a causal definition for the field $X_1$. We will work as much as possible, for the sake of presentation, in the asymptotic regime where we have taken the limit $\epsilon\rightarrow 0$. We keep in mind that the pointwise limit of such a process $u_1(t)=\lim_{\epsilon \to 0}u_{1,\epsilon}(t)$, where $u_{1,\epsilon}(t)$ is the unique statistically stationary solution of the SDE given in Eq. \ref{eq:MRW}, is not straightforward to acquire, since the random field $e^{\gamma X_{1,\epsilon}(t)-\gamma^2\langle X_{1,\epsilon}^2\rangle}$ becomes distributional in this limit \citep{RhoVar14}. We will thus be mainly concerned with statistical quantities of the asymptotic random process $u_1$, but will perform standard calculations using the classical field $u_{1,\epsilon}(t)$ if necessary and  convenient. Because we want to quantify the intermittent corrections implied by the this random distribution, we propose to compute the structure functions of the aforementioned stochastic model. Define thus the velocity increment as
\begin{equation}
\delta_\tau u_{1,\epsilon}(t) = u_{1,\epsilon}(t+\tau)-u_{1,\epsilon}(t).
\end{equation}
Accordingly, define the respective asymptotic structure functions as
\begin{equation}
\mathcal S_{u_1,m} (\tau)=\lim_{\epsilon\rightarrow 0}  \left\langle \left(u_{1,\epsilon}(t+\tau)-u_{1,\epsilon}(t)\right)^m\right\rangle.
\end{equation}
{In the following, we focus on the scaling properties of the structure functions of the causal MRW $u_{1}$. As a general remark, let us recall that the log-correlated field $X_1$ and the underlying white noise $W$ entering in the dynamics of $u_{1,\epsilon}$ are taken independently. This implies that all odd order structure functions vanish, namely $\mathcal S_{u_1,2m+1}=0$ with $m\in \mathbb N$.  Regarding the second-order structure function, it is the same as the one obtained from the OU process $v_1$ (Eq. \ref{eq:OU}), and given by,
\begin{equation}\label{eq:PropS2U1}
\mathcal S_{u_1,2} (\tau)=\mathcal S_{v_1,2} (\tau) = qT\left[1-e^{-\frac{|\tau|}{T}}\right]\build{\sim}_{\tau\rightarrow 0^+}^{}q\tau.
\end{equation}
On the contrary, the fourth-order structure function is impacted by intermittency, and we get, under the condition $4\gamma^2<1$,
\begin{equation}\label{eq:PropS4MRW}
\mathcal S_{u_1,4} (\tau)\build{\sim}_{\tau\rightarrow 0}^{}\frac{3}{1-6\gamma^2+8\gamma^4}q^2\tau^2\left(\frac{\tau}{T}\right)^{-4\gamma^2}e^{4\gamma^2c(0)},
\end{equation}
where the constant $c(0)$ is given in Eq. \ref{eq:EulerMarCst}. More generally, it is then possible to obtain an estimation of the $(2m)^{\text{\small{th}}}$ order structure functions that reads, for $2m(m-1)\gamma^2<1$,
\begin{equation}\label{eq:PropS2NMRW}
\mathcal S_{u_1,2m} (\tau)\build{\propto}_{\tau\rightarrow 0}^{}q^{m}\tau^m\left(\frac{\tau}{T}\right)^{-2m(m-1)\gamma^2},
\end{equation}
indicating that the causal MRW exhibits a lognormal spectrum. We gather all the proofs of these propositions in Appendix \ref{App:SQMRW}.}



\subsection{An infinitely differentiable causal Multifractal Random Walk}\label{Sec:MRWInfiniteDeriv}

Our proposition is herein made of a causal stochastic process representative of the statistical  behaviour of Lagrangian velocity in homogeneous and isotropic turbulent flows at a given finite Reynolds number (equivalently for a finite ratio $\tau_\eta/T$). We are demanding for a statistically stationary process, correlated over a large time scale $T$, infinitely differentiable (giving meaning to the respective acceleration process), acquiring  rough and intermittent behaviours as the small time scale $\tau_\eta$ goes to zero, i.e. in the infinite Reynolds number limit.

Assume $n\ge2$ and consider the following system of embedded differential equations
\begin{align}
 \label{eq:OU_MRW_infinite_N_1_Multi}
   \frac{du_{n,\epsilon}}{dt} &= -\frac{1}{T}u_{n,\epsilon}(t)+  e^{\gamma X_{n,\epsilon}(t)-\frac{\gamma^2}{2}\langle X_{n,\epsilon}^2\rangle} f_{n-1}(t)\;\\
   \frac{df_{n-1}}{dt} &= -\frac{\sqrt{n-1}}{\tau_{\eta}}f_{n-1}(t)+ f_{n-2}(t)\;\label{eq:OU_MRW_infinite_N_2_Multi} \\
   &...&\\  
   \frac{df_{2}}{dt} &= -\frac{\sqrt{n-1}}{\tau_{\eta}}f_{2}(t)+ f_{1}(t)\;\\
   df_1 &=  -\frac{\sqrt{n-1}}{\tau_\eta}f_1(t) dt + \sqrt{\beta_{n}}W(dt)\;,
\label{eq:OU_MRW_infinite_N_N_Multi}
\end{align}
with 
\begin{equation}\label{eq:ChoiceBetaNMulti}
\beta_n=\left( \frac{n-1}{\tau_\eta^2}\right)^{n-1}\frac{\sigma^2\sqrt{4\pi\tau_\eta^2}}{T\int_{0}^\infty e^{-\frac{h}{T}}e^{-h^2/(4\tau_\eta^2)}e^{\gamma^2\mathcal C_{X}(h)}dh}.
\end{equation}
In the system above, the causal process $X_{n,\epsilon}$ obeys the set of stochastic differential equations 
\begin{align}
 \label{eq:X_MRW_infinite_N_1_Multi}
   \frac{dX_{n,\epsilon}}{dt} &= -\frac{1}{T}X_{n,\epsilon}(t)+ \sqrt{\tilde{\beta}_n}\tilde{f}_{n-1,\epsilon}(t)\;\\
   \frac{d\tilde{f}_{n-1,\epsilon}}{dt} &= -\frac{\sqrt{n-1}}{\tau_{\eta}}\tilde{f}_{n-1,\epsilon}(t)+ \tilde{f}_{n-2,\epsilon}(t)\;\\
   &...&\\  
   \frac{d\tilde{f}_{2,\epsilon}}{dt} &= -\frac{\sqrt{n-1}}{\tau_{\eta}}\tilde{f}_{2,\epsilon}(t)+ \tilde{f}_{1,\epsilon}(t)\;\\
   d\tilde{f}_{1,\epsilon} &=  -\frac{\sqrt{n-1}}{\tau_\eta}\tilde{f}_{1,\epsilon}(t) dt  -\frac{1}{2}\int_{-\infty}^t \left[ t-s+\epsilon\right]^{-3/2}\widetilde{W}(ds)dt + \epsilon^{-1/2}\widetilde{W}(dt),\label{eq:X_MRW_infinite_N_N_Multi}
\end{align}
with 
\begin{equation}\label{eq:ChoiceBetaTildeNMulti}
\tilde{\beta}_n=\left( \frac{n-1}{\tau_\eta^2}\right)^{n-1}.
\end{equation}
where $W$ and $\widetilde{W}$ are two independent copies of the Wiener process.

Similar to the Gaussian infinitely differentiable process $v$ established in the first part, we show in the following Proposition \ref{propcorrelationsInfiniteMRW} that the process $u$, obtained once the procedure depicted in the set of embedded differential equations (Eqs. \ref{eq:OU_MRW_infinite_N_1_Multi} to \ref{eq:OU_MRW_infinite_N_N_Multi}) is iterated an infinite number of times $n\rightarrow \infty$, and when the small parameter $\epsilon$ goes to zero, converges to a well-defined limit.  Once again, the choice made for the white noise weight $\beta_n$ (Eq. \ref{eq:ChoiceBetaNMulti}) ensures that the variance of the limiting process $u$ is finite with $\langle u^2\rangle=\sigma^2$. Its precise value will become evident when we compute the correlation function $\mathcal C_f(\tau)=\langle f(t)f(t+\tau)\rangle$ of the force $f$ when $n\to \infty$  (see Eq. \ref{eq:CorrfInfinity}). 

Similarly, the precise choice for the coefficient  $\tilde{\beta}_n$ (Eq. \ref{eq:ChoiceBetaTildeNMulti}) entering in the dynamics of $X_{n,\epsilon}$ (Eq. \ref{eq:X_MRW_infinite_N_1_Multi}) is dictated by the necessity that in an asymptotic way, when both $\epsilon\to 0$ and $\tau_\eta\rightarrow 0$, and for any number of layers $n$,  $X_{n}$ gets logarithmically correlated in an appropriate manner. As far as the process $X_{n,\epsilon}$ is concerned, these limits can be taken in an arbitrary way since they commute. The small parameters $\epsilon$ and $\tau_\eta$ have a similar physical interpretation, they mimic finite Reynolds number effects. We define them a prior as separate entities and seek for limits independently for the sake of generality. More precisely $\epsilon$ is taken finite to make sense of the dynamics of $\tilde{f}_{1,\epsilon}$  as it is proposed in Eq. \ref{eq:X_MRW_infinite_N_N_Multi}. Remark finally that the multiplicative chaos entering in the dynamics of $u_{n,\epsilon}$ (Eq. \ref{eq:OU_MRW_infinite_N_1_Multi}) is renormalized by a smaller constant $\exp\left(\frac{\gamma^2}{2}\langle X_{n,\epsilon}^2\rangle\right)$ than in its non-differentiable version $u_{1,\epsilon}$ (Eq. \ref{eq:MRW}), where there typically exists a larger normalization constant $\exp\left(\gamma^2\langle X_{n,\epsilon}^2\rangle\right)$. It is related to the finite correlation of the of the term $f_{n-1}$ entering in Eq. \ref{eq:OU_MRW_infinite_N_1_Multi}, contrary to the dynamics proposed in Eq. \ref{eq:MRW} where a white noise $W(dt)$ enters.

As a general remark, notice that the dynamics depicted by  the set of embedded differential equations (Eqs. \ref{eq:OU_MRW_infinite_N_1_Multi} to \ref{eq:OU_MRW_infinite_N_N_Multi}) coincides with the dynamics of the Gaussian process $v_n$  (Eqs \ref{eq:OU_infinite_N_1} to \ref{eq:OU_infinite_N_N}) when we consider the particular value $\gamma=0$. In other words, the non intermittent limit of the process $u_{n,\epsilon}$ is Gaussian, and coincides with the process $v_n$ of Section \ref{Sec:InfDiffProc}.

Before establishing the statistical behaviour of the asymptotic process $u$, let us first focus on the statistical properties of $X_{n,\epsilon}$ {that we gather and derive in Proposition \ref{propcorrelationsInfiniteX}. Let us keep in mind that, whatever the ordering of the limits $n\rightarrow \infty$ and $\epsilon\rightarrow 0$, the correlation function of $X_{n,\epsilon}$ converges towards a well-defined function $\mathcal C_{X}(\tau) $ (Eq. \ref{CorrXLimFour}), of which its value at the origin diverges logarithmically with $\tau_\eta$ as $\tau_\eta\to 0$ (Eq. \ref{eq:VarIDXTauEtaDivLog}). Actually, in this limit of infinite Reynolds numbers, $\mathcal C_{X}(\tau) $ converges towards $\mathcal C_{X_1}(\tau)$ (Eq. \ref{eq:ConvXtoX_1asTauEta}), as expected.}

{We now proceed with the covariance structure of the limiting process $u$. We summarize and demonstrate in Proposition \ref{propcorrelationsInfiniteMRW} the main second-order statistical properties of velocity $u$ and acceleration $a$. We first derive the exact velocity correlation function $\mathcal C_u(\tau)$ in the joint commuting limit $\epsilon\to 0$ and $n\to \infty$ (Eq. \ref{eq:CorrelUInfinite}). This shows that, whereas $\mathcal C_u(\tau)$ depends weakly on intermittent corrections in the dissipative range, it loses this property as $\tau_\eta/T\to 0$ and coincides with the  correlation function of the OU process $\mathcal C_{v_1}(\tau)$ (Eq. \ref{eq:LimitBehaveCorrelU}). Similarly, the acceleration correlation function $\mathcal C_a(\tau)$ can be derived (Eq. \ref{eq:CorrelAUInfinite}). From there, we show that acceleration variance diverges as  $T/\tau_\eta$ as the Reynolds number increases (Eq. \ref{eq:PredVarAcceIDMRW}). }

Let us remark that the proposed stochastic model of velocity $u$, that we claim to be intermittent in a precise way and defined in the following Proposition \ref{propSFInfiniteMRW}, predicts that, as far as the covariance of $u$ is concerned, is similar to an Ornstein-Uhlenbeck process at infinite Reynolds number, independently of any intermittency corrections. This is consistent with the standard phenomenology of Lagrangian turbulence. The predicted acceleration variance (Eq. \ref{eq:PredVarAcceIDMRW}) does not exhibit either intermittent corrections: This precise behaviour of acceleration variance with respect to the Reynolds number is at odds with the extrapolations that can be made from numerical simulations (see \cite{IshKan07} and the discussion that we propose in Section \ref{Sec:CalibCompMulti}). We will see and develop in Section \ref{Sec:PredMultiForm} that the multifractal formalism allows  the understanding of how the velocity correlation does not get impacted by intermittency at infinite Reynolds numbers, whereas the acceleration variance does. 

{Let us now present the intermittent, i.e. multifractal, properties of the velocity process $u$, as they can be seen on higher-order structure functions (see Proposition \ref{propSFInfiniteMRW}). As it was shown previously, the correlations of $u$ and the OU process $v_1$ coincide as $\tau_\eta\to 0$. The same goes for the second-order structure function (Eq. \ref{eq:PredTheoS2InfiniteMulti}). Whereas showing that the fourth-order structure function of $u$ coincides with the one of the causal MRW process $u_1$ as first $\epsilon\to 0$ and then $\tau_\eta\to 0$ is obvious (Eq. \ref{eq:PropS4MRWInfinity}), the reversed order of limits is more involved. We nonetheless propose an approximation procedure that confirms that $u$ and $u_1$ possess the same intermittent properties (Eq. \ref{eq:PropS4MRWInfinityReverse}). All statements and proofs can be found in Proposition \ref{propSFInfiniteMRW} and Appendix \ref{App:SQMRWInfinity}.}

\subsection{A second numerical illustration}\label{Sec:NSU}

\subsubsection{An efficient algorithm under the periodic approximation}\label{Sec:AlgoSimulIDMRW}

In this Section we propose a numerical algorithm able to reproduce in a realistic and efficient fashion the statistical behaviour of the process $u$, which statistical properties are detailed in Propositions \ref{propcorrelationsInfiniteMRW}  and \ref{propSFInfiniteMRW}. As we have seen, the process $u$, contrary to the Gaussian process $v$ of Section \ref{Sec:InfDiffProc}, obeys a non Markovian dynamics. More precisely, the process $X(t)$ at a given time $t$, the limiting solution, as the number of layers $n$ goes to infinity and the small parameter $\epsilon$ goes to 0, of the system of embedded stochastic differential equations \ref{eq:X_MRW_infinite_N_1_Multi} to \ref{eq:X_MRW_infinite_N_N_Multi}, requires the knowledge of its entire past. It is thus tempting to use the discrete Fourier transform to solve its dynamics. We will incidentally generate periodical solutions of this non Markovian dynamics. Since we will consider in the sequel very long trajectories, of order $10^5$ times the largest time scale $T$ of the process, all aliasing effects will be negligible.  This periodic approximation is well justified. {As argued in Section \ref{Sec:FirstNS}, simulations of the limiting process with $n\to\infty$ require the causal factorization of covariance functions of underlying Gaussian components, a procedure which is not simple. Furthermore, the limit $\epsilon\to 0$ is also complicated to obtain from a numerical point of view, and therefore, we will perform simulations for a finite $n$ number of layers, and for a finite $\epsilon>0$.}

Consider first an estimator for the discrete process $\widehat{X}_{n,\epsilon}[t]$ of the continuous solution $X_{n,\epsilon}(t)$ of the coupled system Eqs. \ref{eq:X_MRW_infinite_N_1_Multi} to \ref{eq:X_MRW_infinite_N_N_Multi}. {Let us introduce the convolution product $\ast$, which is defined as, for any two functions $g_1$ and $g_2$,
$$ \left(g_1\ast g_2\right)(\tau) = \int_{\mathbb R}g_1(t)g_2(\tau-t)dt,$$
with the corresponding short-hand notation,
$$g^{\ast n}=\underbrace{g \ast g \ast \cdots \ast g}_{n}.$$
In the statistically stationary regime, the continuous expression of the Gaussian process $X_{n,\epsilon}(t)$ reads
\begin{equation}\label{eq:ContinuousXn}
X_{n,\epsilon}(t)=\sqrt{\widetilde{\beta}_n} \left(g_T\ast g_{\frac{\tau_\eta}{\sqrt{n-1}}}^{\ast (n-1)}\ast \left( h_\epsilon+\epsilon^{-1/2}\delta\right) \ast  \widetilde{W} \right)(t), 
\end{equation}
where the multiplicative factor $\tilde{\beta}_n$ is given in Eq. \ref{eq:ChoiceBetaTildeNMulti}, and recall that $g_\tau(t) = e^{-t/\tau}1_{t\ge 0}$. We also include $h_\epsilon(t) = -\frac{1}{2}(t+\epsilon)^{-3/2}1_{t\ge 0}$ and $\delta(t)$ stands for the Dirac delta function.}

{Now in the discrete setting, call} $N$ the number of collocation points, $T_{\mbox{\tiny{tot}}}$ the total length of the simulation, and $\Delta t$ the timestep. As already mentioned, make sure that $T_{\mbox{\tiny{tot}}}=N\Delta t \gg T$ to prevent {from} aliasing errors. {In the aforementioned periodic framework, the discrete estimator $\widehat{X}_{n,\epsilon}[t]$ of the continuous solution $X_{n,\epsilon}(t)$ (Eq. \ref{eq:ContinuousXn})  reads}
\begin{equation}\label{eq:DiscreteXn}
\widehat{X}_{n,\epsilon}[t]=\sqrt{\widetilde{\beta}_n} \mbox{DFT}^{-1}\left(  \mbox{DFT}\left(g_T \right)\mbox{DFT}^{n-1}\left(g_{\frac{\tau_\eta}{\sqrt{n-1}}} \right) \mbox{DFT}_c\left(h_\epsilon \right)  \mbox{DFT}\left(\widetilde{W} \right) \right)[t]\times (\Delta t)^{n}, 
\end{equation}
{where we have introduced the discrete Fourier transform (DFT). It also enters in the expression given in Eq. \ref{eq:DiscreteXn}, properly discretizing and periodizing forms of the continuous functions $g_\tau(t)$ at various time scales $\tau$ and $h_\epsilon(t)$. Notice that in the continuous framework, $\int_\mathbb R h_\epsilon(t)dt=-\epsilon^{-1/2}$ is the value at the origin of frequencies of the Fourier transform (FT) of $h_\epsilon$, such that $\mbox{FT}(h_\epsilon+\epsilon^{-1/2}\delta)(\omega)=\mbox{FT}(h_\epsilon)(\omega)-\mbox{FT}(h_\epsilon)(0)$. This justifies the short-hand notation  $\mbox{DFT}_c(h_\epsilon)[\omega] = \mbox{DFT}(h_\epsilon)[\omega] -\mbox{DFT}(h_\epsilon)[0]$ in Eq. \ref{eq:DiscreteXn}.} Finally, we have noted  $\widetilde{W}[t]$ an instance of the white noise field,  comprised of $N$ independent Gaussian random variables of zero average and variance $\Delta t$. The $(\Delta t)^{n}$ factor originates from the convolution by the kernel $g_T(t)$ and $(n-1)$ convolutions by the kernel $g_{\frac{\tau_\eta}{\sqrt{n-1}}}$.

In a similar manner, the numerical, discretized and periodized, estimator $\widehat{u}_{n,\epsilon}$ of the continuous solution $u_{n,\epsilon}$ of the coupled system Eqs. \ref{eq:OU_MRW_infinite_N_1_Multi} to \ref{eq:OU_MRW_infinite_N_N_Multi} {in the statistically stationary regime, which reads
\begin{equation}\label{eq:ContinuousUn}
u_{n,\epsilon}(t)=\sqrt{\beta_n} \left(g_T\ast g_{\frac{\tau_\eta}{\sqrt{n-1}}}^{\ast (n-1)}\ast \left( \frac{e^{\gamma \widehat{X}_{n,\epsilon}}}{e^{\frac{\gamma^2}{2}\langle \widehat{X}_{n,\epsilon}^2 \rangle}} W\right) \right)(t), 
\end{equation}
can be written as}
\begin{equation}\label{eq:DiscreteUn}
\widehat{u}_{n,\epsilon}[t]=\sqrt{\beta_n} \mbox{DFT}^{-1}\left(  \mbox{DFT}\left(g_T \right)\mbox{DFT}^{n-1}\left(g_{\frac{\tau_\eta}{\sqrt{n-1}}} \right) \mbox{DFT}\left(\frac{e^{\gamma \widehat{X}_{n,\epsilon}}}{e^{\frac{\gamma^2}{2}\langle \widehat{X}_{n,\epsilon}^2 \rangle}} W \right) \right)[t]\times (\Delta t)^{n-1}, 
\end{equation}
where $\beta_n$ is provided in Eq. \ref{eq:ChoiceBetaNMulti}, and recall that the white noise $W$ is independent of $\widetilde{W}$ that enters in Eq. \ref{eq:DiscreteXn}. The fact that we multiply by $(\Delta t)^{n-1}$ the overall expression \ref{eq:DiscreteUn}, instead of $(\Delta t)^{n}$ (as in Eq. \ref{eq:DiscreteXn}), originates from the white (i.e. distributional) nature of $W$, whereas $\widetilde{W}$ is already smoothed out by the kernel $h_\epsilon$.

The timestep $\Delta t$ has to be chosen smaller than the smallest scale of motion, that is $\frac{\tau_\eta}{\sqrt{n-1}}$. Furthermore, we are interested in performing a realistic simulation of the limiting process $u$, obtained in the limit $\epsilon \to 0$, at a given finite $\tau_\eta$. A convenient choice for $\epsilon$ is to take it proportional to $\Delta t$, such that both of them go to 0 in the continuous limit. In subsequent simulations, we find appropriate to choose 
\begin{equation}\label{eq:ChoiceDeltatEpsilon}
\Delta t=\frac{\tau_\eta}{200\sqrt{n-1}} \mbox{ and } \epsilon = 5\Delta t.
\end{equation}
This choice gives numerical stability and a proper illustration of the exact statistical quantities provided in Propositions \ref{propcorrelationsInfiniteMRW}  and \ref{propSFInfiniteMRW}  for the range of investigated values of $\tau_\eta$ (see the following Section \ref{Sec:NumResInfMultiProcess}). To prevent  aliasing errors, we work with a large number of collocation points $N=2^{32}$, such that $T_{\mbox{\tiny{tot}}}=N\Delta t$ is always much larger than $T$. 

\begin{figure}
\begin{center}
\epsfig{file=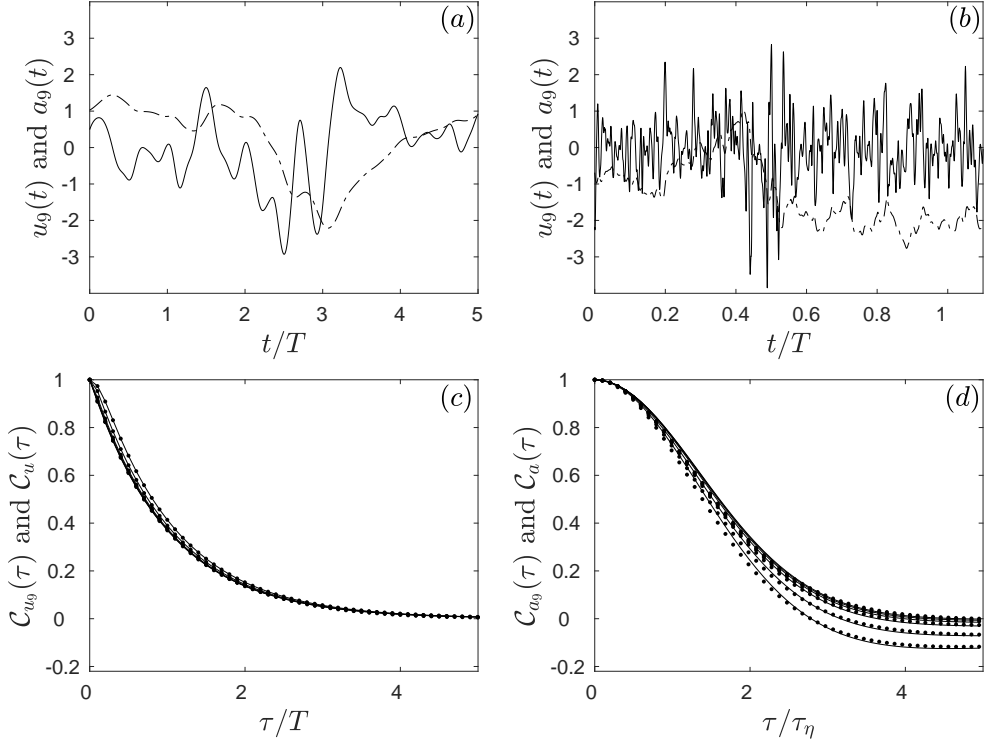,width=13cm}
\end{center}
\caption{\label{fig:NumOUMulti} Numerical simulation, in a periodical fashion, of the set of equations \ref{eq:OU_MRW_infinite_N_1_Multi} to \ref{eq:OU_MRW_infinite_N_N_Multi} using $n=9$ layers, for 6 values of $\tau_\eta$, that is $T/\tau_\eta=$ 10, 20, 50, 100, 200, 500, and $\sigma^2=1$. See the description of the algorithm in Section \ref{Sec:AlgoSimulIDMRW}, and the choice made for other parameters in Section \ref{Sec:NumResInfMultiProcess}. (a) Typical time series of the obtained processes $u_9(t)$ (dashed line) and $a_9(t)$ (solid line), as a function of time $t$, for $T/\tau_\eta=10$. For the sake of comparison, all time series are normalized by their standard deviation. (b) Similar time series as in (a), but for $T/\tau_\eta=500$. (c) Respective velocity correlation functions $\mathcal C_{u_9}$ for the six different values of $\tau_\eta$, estimated from numerical simulations (dots), and compared to their asymptotic theoretical prediction $\mathcal C_{u}$ (Eq. \ref{eq:CorrelUInfinite}) (solid line). (d) Respective acceleration correlation functions $\mathcal C_{a_9}$ and compared to the asymptotic correlation function $\mathcal C_{a}$ (Eq. \ref{eq:CorrelAUInfinite}). For the sake of clarity, all curves are normalized by their values at the origin (i.e. the respective variances).}
\end{figure}

\subsubsection{Numerical results and comparisons to theoretical predictions}
\label{Sec:NumResInfMultiProcess}

Without loss of generality, take $T=1$. We numerically perform the (discrete) Fourier transforms as they are detailed in Eqs. \ref{eq:DiscreteXn} and  \ref{eq:DiscreteUn}, using 6 values for $T/\tau_\eta$, that is 10, 20, 50, 100, 200 and  500. Keeping in mind that $\tau_\eta$ is a fairly good representation of the Kolmogorov time scale, these values correspond to an extended range of Reynolds numbers.
Choosing for $\Delta t$ and $\epsilon$ the values depicted in Eq. \ref{eq:ChoiceDeltatEpsilon}, working with $N=2^{32}$ collocation points and $n=9$ layers, we find in the worst scenario corresponding to the smallest $\tau_\eta$ a total time of simulation $T_{\mbox{\tiny{tot}}}=N\Delta t \approx 10^4T$, preventing any aliasing effects. As it will be precisely quantified when we will discuss intermittent corrections, we find the particular value
\begin{equation}\label{eq:ChoiceGamma}
\gamma^2=0.085,
\end{equation}
representative of the level of intermittency as it is seen in numerical simulations of the Navier-Stokes equations, consistent with previous estimations (see \cite{MorDel02,CheRou03,BifBof04,CheCas12} and references therein). Forthcoming statistical quantities are averaged over three independent instances of these trajectories.

\begin{figure}
\begin{center}
\epsfig{file=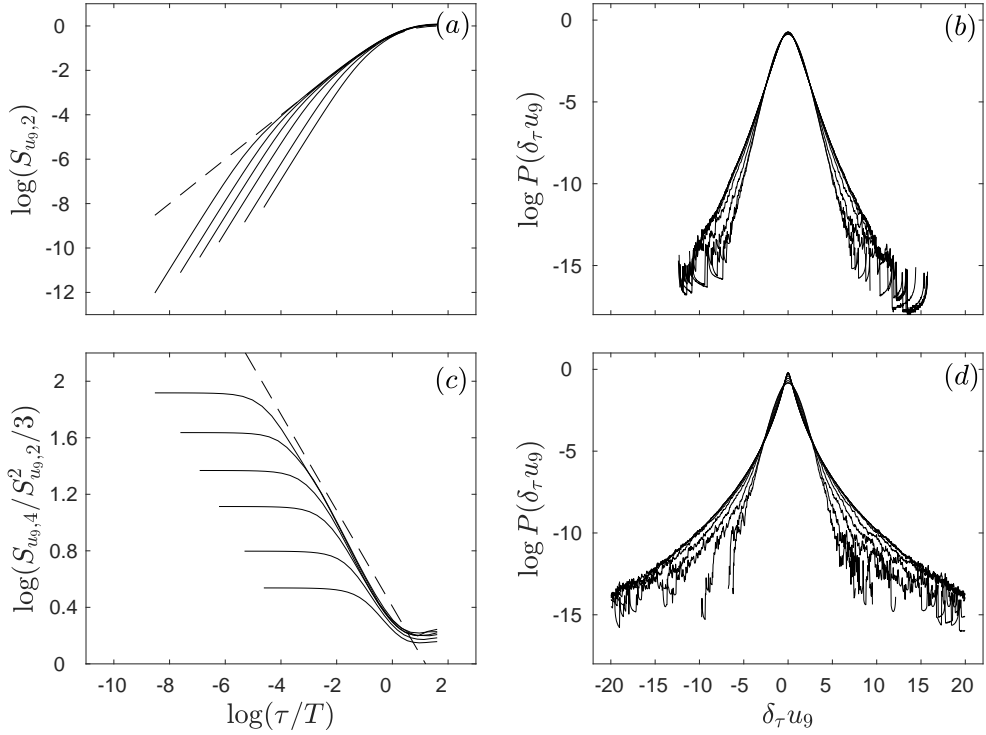,width=13cm}
\end{center}
\caption{\label{fig:NumOUMultiStructPdfs} Illustration of the behaviour of higher order statistics of the processes studied in Fig. \ref{fig:NumOUMulti}. (a) Logarithmic representation of the second-order structure function, estimated from the times series of the six different values of $\tau_\eta$ (solid lines), and compared to their asymptotical prediction provided in Eq. \ref{eq:PredTheoS2InfiniteMulti} (dashed line). (c) Similar logarithmic process as in (a), but for the flatness of velocity increments. We superimpose the theoretical prediction based on Eq. \ref{eq:PropS4MRW} (see the devoted discussion in Section \ref{Sec:NumResInfMultiProcess}). (b) Estimation of the Probability density Functions of velocity increments for scales logarithmically spanned across the accessible range of scales displayed in (a) and (c), and for $\tau_\eta/T=1/10$. (d) Similar plot as in (b), but for $\tau_\eta/T=1/500$. }
\end{figure}

For the sake of clarity, we omit the hat on the simulated discrete version of $u_{9,\epsilon}$, and display in Fig. \ref{fig:NumOUMulti}(a) and (b) two instances of this stochastic process for the largest $\tau_\eta=T/10$ (lowest Reynolds number) and the smallest $\tau_\eta=T/500$ (highest Reynolds number) ratios of the small over the large time scales. Velocity is represented using a dot-dashed line, whereas the respective acceleration with a solid line. All time series are divided by their respective standard deviation for the sake of comparison. In the low Reynolds number case (Fig. \ref{fig:NumOUMulti}(a)), we observe that indeed velocity is correlated over $t$, whereas acceleration is correlated over a shorter time scale $\tau_\eta$. In the highest Reynolds number case (Fig. \ref{fig:NumOUMulti}(b)), we can definitely observe the scale decoupling between the large $T$ and the small $\tau_\eta$ time scales. Also, notice that the statistics of acceleration are evidently non Gaussian. This is a manifestation of the intermittency phenomenon, which is modeled by the multiplicative chaos that enters into the construction. These non Gaussian fluctuations would be enhanced by a higher value for $\gamma$ (data not shown) than the one chosen presently (Eq. \ref{eq:ChoiceGamma}). We will come back to this point while discussing Fig. \ref{fig:NumOUMultiStructPdfs}.

We present in Fig. \ref{fig:NumOUMulti} the velocity (c) and acceleration (d) correlation functions. Results from the numerical simulation of Eqs. \ref{eq:DiscreteXn} and  \ref{eq:DiscreteUn} for the six  values of $\tau_\eta$ are displayed using dots, we superimpose the theoretical expressions provided in Eqs. \ref{eq:CorrelUInfinite} and \ref{eq:CorrelAUInfinite}. Concerning the velocity correlations (Fig. \ref{fig:NumOUMulti}(c)), we can notice the striking agreement between the numerical estimation based on time series of $u_{9,\epsilon}$ and the limiting theoretical expression (Eq. \ref{eq:CorrelUInfinite}), as it was already observed in the Gaussian case (Fig. \ref{fig:NumOU}). Furthermore, as expected, the dependence on $\tau_\eta$ is very weak. This can be easily understood once realizing that velocity is a large scale quantity, mostly governed by the physics taking place at $T$. To this regard, acceleration correlation functions will highlight the physics ruling phenomena  which occur at $\tau_\eta$ and are displayed in Fig. \ref{fig:NumOUMulti}(d). All curves are normalized by the respective value at the origin (i.e. the acceleration variance). The low Reynolds number case (largest $\tau_\eta$) is easily recognizable; this is the curve going the most negative after the zero-crossing. As $\tau_\eta$ decreases, $\mathcal C_a(\tau)$ is closer to 0. This is consistent with the constraint that the integral of this curve has to vanish, as a consequence of statistical stationarity. Once again, the collapse of the numerically estimated $\mathcal C_{a_9}(\tau)$ (dots) on the limiting theoretical expression given in Eq. \ref{eq:CorrelAUInfinite} (solid line) is excellent.

Let us now focus on the precise quantification on the intermittency phenomenon. We display in Figs. \ref{fig:NumOUMultiStructPdfs}(a) and (c) the behaviour across scales $\tau$ of the structure functions $\mathcal S_{u_{n,\epsilon},m}=\langle (\delta_\tau u_{n,\epsilon})^m \rangle$ of the simulated process $u_{n,\epsilon}$. We then compare them to our theoretical predictions (Proposition \ref{propSFInfiniteMRW}) obtained in the asymptotic regime $n\to \infty$, $\epsilon \to 0$, $\tau_\eta\to 0$ and finally $\tau\to 0$ (limits are taken in this very order).

We present in Fig. \ref{fig:NumOUMultiStructPdfs}(a)  the scaling behaviour of the second order structure function $\mathcal S_{u_{9,\epsilon},2}(\tau)=\langle (\delta_\tau u_{9,\epsilon})^2 \rangle$ (solid lines) for the 6 values of $\tau_\eta$ that we formerly detailed. Notice that  in this representation, $\mathcal S_{u_{9,\epsilon},2}(\tau)$ is normalized by $2\langle u^2_{9,\epsilon}\rangle$, such that it goes to unity at large arguments $\tau\gg T$. We recover at small scales $\tau\ll \tau_\eta$ the dissipative behaviour $\mathcal S_{u_{9,\epsilon},2}(\tau)\propto \tau^2$, which is a consequence of the differentiable nature of the process. In the inertial range $\tau_\eta\ll \tau\ll T$, as expected by our theoretical prediction (Eq. \ref{eq:PredTheoS2InfiniteMulti}), we get a behaviour similar to an OU process, that is $\mathcal S_{u_{9,\epsilon},2}(\tau)\propto \tau$. We superimpose using a dashed line the expected behaviour from an OU process, namely $S_{u_{1},2}(\tau)=2\langle u^2_{1}\rangle \left(1-e^{-|\tau|/T}\right)$. We indeed observe that it describes with great accuracy the scaling behaviour of $\mathcal S_{u_{9,\epsilon},2}(\tau)$ in the inertial range and at larger scales.  The second order statistics of $u_{9,\epsilon}$ are well described by our asymptotic predictions in this range of scales. Similar conclusions were obtained while describing velocity correlation function in Fig. \ref{fig:NumOUMulti}(c). 

As mentioned in Proposition \ref{propSFInfiniteMRW}, only fourth-order statistics and higher are impacted by intermittency. To check this, we represent in Fig. \ref{fig:NumOUMultiStructPdfs}(c)  the scaling behaviour of the flatness of velocity increments, that is $\mathcal S_{u_{9,\epsilon},4}/\mathcal S_{u_{9,\epsilon},2}^2$ (solid lines), and for the 6 different values of $\tau_\eta$, in a logarithmic fashion. As shown, flatnesses are normalized by 3, i.e. the value obtained for Gaussian processes. As we can observe, flatnesses are close to 3 at large scales $\tau\ge T$, and then increase in the inertial range as a power-law, before saturating in the dissipative range $\tau\le \tau_\eta$. This saturation is typical of differentiable processes: a Taylor series of increments makes the dependence on $\tau$  disappear. We superimpose on this plot, using a dashed line, the theoretical prediction that we made for MRW (Eq. \ref{eq:PropS4MRW}) without the unjustified additional free parameter. We indeed see that the power-law exponent is given by $-4\gamma^2$, and that the multiplicative constant is close to the one derived for the non-differentiable MRW (Eq. \ref{eq:PropS4MRW}). This theoretical prediction seems to be more and more representative of the intermittent properties of $u_{9,\epsilon}$ as $\tau_\eta$ gets smaller and smaller. This indicates that the constant $c_{\gamma,4}$ which is tedious to compute in an exact fashion (but easily accessible in the approximative framework developed in Appendix \ref{App:SQMRWInfinity}) for the infinitely differentiable MRW (Eq. \ref{eq:PropS4MRWInfinityReverse}) is the same as in the non differentiable case (Eq. \ref{eq:PropS4MRW}).  This shows that the limits $\epsilon\to 0$ and $\tau_\eta\to 0$ commute at the fourth-order too (Eqs. \ref{eq:PropS4MRWInfinity} and \ref{eq:PropS4MRWInfinityReverse}). This remains to be done on a rigorous ground.

Finally, to illustrate the intermittent behaviour of the process $u_{9,\epsilon}$, we display in Figs. \ref{fig:NumOUMultiStructPdfs}(b) and (d) the probability density functions (PDFs) of velocity increments at various scales, from large to small: (b) $\tau_\eta/T=1/10$ and (d) $\tau_\eta/T=1/500$. We indeed observe the continuous shape deformation of these PDFS as the scales $\tau$ decreases in length, being Gaussian at large scales $\tau\ge T$, and strongly non-Gaussian in the dissipative range. In a consistent manner with the behaviour of flatnesses (Fig. \ref{fig:NumOUMultiStructPdfs}(c)), the acceleration PDF, obtained when $\tau\ll \tau_\eta$, is less and less Gaussian as $\tau_\eta$ diminishes in size.

\section{Comparison to Direct Numerical Simulations}\label{Sec:CompModDNSTotal}

\subsection{Description of the datasets}\label{Sec:CompModDNSDescDS}

\begin{table}
  \begin{center}
\def~{\hphantom{0}}
  \begin{tabular}{lccccccc}
      Origin & Resolution &  $\mathcal R_\lambda$  & $\tau_K$  & $T_L$ & number of trajectories   &   $dt$ & Duration \\[3pt]
       Turbase & $512^3$ & 185   & 0.0470 & 0.7736 & 126720 & $4.10^{-3}$ & $17.063\, T_L$\\
       JHTDB  & $1024^3$ & 418 & 0.0424 & 1.3003 & 32768 & $2.10^{-3}$ & $7.692\, T_L$\\
  \end{tabular}
  \caption{Summary of relevant physical parameters of the two sets of DNS data. Resolution of the Eulerian fields, Taylor based Reynolds number $\mathcal R_\lambda$ and Kolmogorov dissipative time scale $\tau_K$ (Eq. \ref{eq:deftauKForAll}) are provided in relevant publications (see text). The Lagrangian integral time scale $T_L$  is defined in Eq. \ref{eq:defTLForAll} and is computed from our statistical estimation of the velocity correlation function.}
  \label{tab:DNSParam}
  \end{center}
\end{table}

We consider in this article two sets of data that have been made freely accessible to the public. We focus our attention to statistically homogeneous and isotropic numerical flows obtained by solving the Navier-Stokes equations in a periodic box. Lagrangian trajectories are then extracted from the time evolution of the Eulerian fields while integrating the positions of tracer particles, initially distributed homogeneously in space. The first set concerns a direct numerical simulation (DNS) at a moderate Taylor based Reynolds number $\mathcal R_\lambda=185$, referenced in \cite{BecBif06,BecBif11}, which can be downloaded from \textsc{https://turbase.cineca.it/}. The second dataset concerns a higher Taylor based Reynolds number $\mathcal R_\lambda=418$, hosted at JHTDB (see \textsc{http://turbulence.pha.jhu.edu}). Details on this DNS and how to proceed to extract the Lagrangian trajectories can be found in \cite{LiPer08,YuKan12}. Relevant parameters and specificities of these datasets and of the Lagrangian trajectories are given in Table \ref{tab:DNSParam}.

\subsection{Definition and estimation of the Lagrangian integral time scale}\label{Sec:DiscussEstTL}

Let us now make a connection between the present modelling approach, and its parameters, and numerical investigations. To do so, we have to consider quantities that can be extracted from DNS data, and show how to relate them to the free parameters entering in the definition of the stochastic process $u$, which are at a given Reynolds number $\tau_\eta$, $T$ and $\gamma$. 

Call  $T_L$ the Lagrangian integral time scale, defined as the integral of the velocity correlation function, i.e.
\begin{equation}\label{eq:defTLForAll}
T_L = \int_{0}^\infty \frac{\mathcal C_u(\tau)}{\mathcal C_u(0)}d\tau,
\end{equation}
where $u$ stands for any Lagrangian velocity components extracted from DNS data, or the present stochastic model.

On the one hand, the definition of $T_L$ (Eq. \ref{eq:defTLForAll}) is appealing because it can be applied to and estimated from velocity time series coming indifferently  from DNS or the model. On the other hand, it requires proper statistical convergence of the velocity correlation $\mathcal C_u(\tau)$ that is especially difficult to get from DNS at large time scales $\tau$ close to the velocity decorrelation time scale. This is even more true when considering experimental data (see a recent discussion on this by \cite{HucMac19}) in which the duration of trajectories are usually shorter. Moreover, on the entire accessible statistical sample, made of tens (even one hundred in the moderate Reynolds number case) thousands trajectories for each three velocity components, we have observed a non negligible level of anisotropy for both sets of data, the standard deviation of the variance of the three velocity components is of order of $20\%$ of the average variance. We found this level of anisotropy surprising given the isotropic and periodic boundary conditions of the advecting flow. We are forced to reach the conclusion that in both cases, trajectories are not long enough to guarantee statistical isotropy. This has consequences on the estimation of $T_L$. Nonetheless, and because we expect ultimately that the flow, and incidentally its Lagrangian trajectories, is isotropic, we average the velocity correlation function over the three components, keeping in mind that the lack of statistical convergence can imply a non negligible error on the estimation of this large time scale. We gather our findings in Table \ref{tab:DNSParam}. Notice that this observed anisotropy on the velocity variance  has weak impact on the acceleration correlation function once normalized by its value at the origin (data not shown). This can be understood by realizing that acceleration is governed by the small scales of the flow, whereas velocity by the large ones.

\begin{figure}
\begin{center}
\epsfig{file=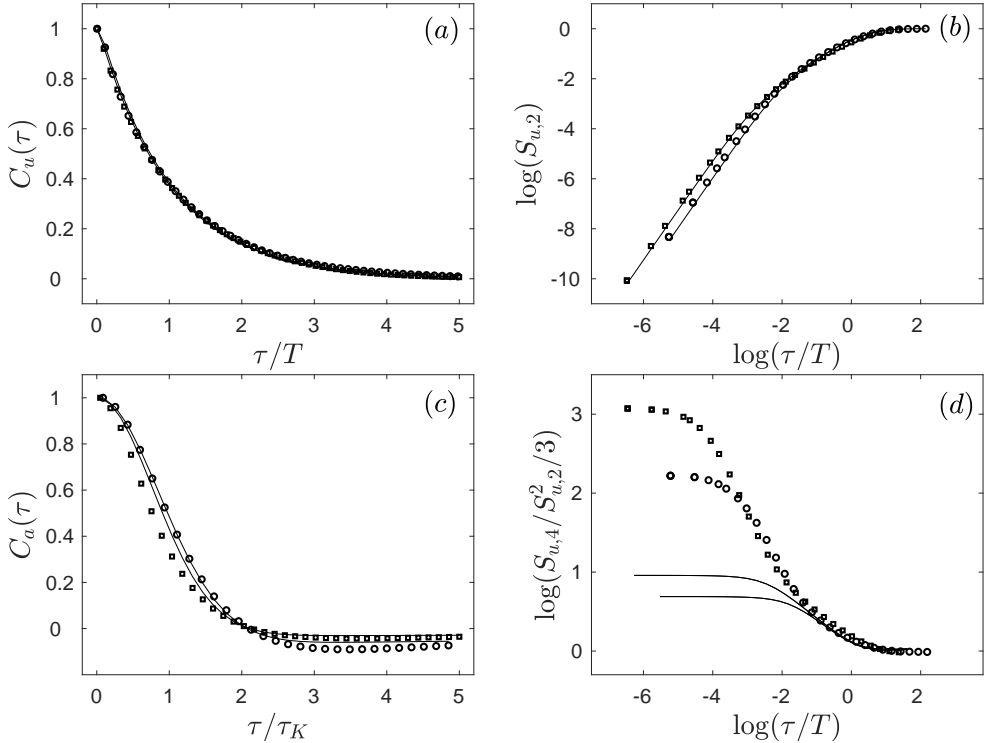,width=13cm}
\end{center}
\caption{\label{fig:CompModDns180_480} Comparison of DNS data to model predictions. (a) Estimation of the velocity correlation function from DNS data (using $\circ$ for $\mathcal R_\lambda = 185$ and $\Box$ for $\mathcal R_\lambda = 418$. We superimpose theoretical predictions using Eq.  \ref{eq:CorrelUInfinite}, for the set of values of the parameters $\tau_\eta$ and $T$ given by our calibration procedure presented in Section \ref{Sec:CalibModDNS}, and for a prescribed value for $\gamma$ (Eq. \ref{eq:ChoiceGamma}). Time lags are normalized by the calibrated time scale $T$. (b) Same plot as in (a) but for the second-order structure function. (c) Similar plot as in (a) and (b) but for the acceleration correlation function, normalized by its value at the origin. Superimposed theoretical predictions are based on the exact expression Eq. \ref{eq:CorrelAUInfinite}. (d) Similar plot as in (a) and (b) but for the flatnesses of velocity increments. Theoretical prediction are obtained thanks to a numerical estimation of velocity time series of the model, in the spirit of Section \ref{Sec:NumResInfMultiProcess}, with the values of the free parameters obtained from  our calibration procedure presented in Section \ref{Sec:CalibModDNS} and for a prescribed value for $\gamma$ (Eq. \ref{eq:ChoiceGamma}). }
\end{figure}

\subsection{Statistical analysis of the DNS datasets}\label{Sec:StatAnalysisDNS}

We display in Figs. \ref{fig:CompModDns180_480}(a) and (c) the numerical estimation of velocity and acceleration correlation functions based on the Lagrangian trajectories extracted from DNS, at moderate Reynolds number $\mathcal R_\lambda=185$ (using open circles $\circ$) and at high Reynolds number $\mathcal R_\lambda=418$ (using open squares $\Box$). As $\mathcal C_u(\tau)$ is concerned (Fig. \ref{fig:CompModDns180_480}(a)), we normalize time lags $\tau$ by a large time scale $T$ coming from the adopted calibration procedure of our model, and that we properly define in Section \ref{Sec:CalibModDNS}. At this level of discussion, keep in mind that $T$ is very close to $T_L$ (Eq. \ref{eq:defTLForAll}). Concerning $\mathcal C_a(\tau)$ (Fig. \ref{fig:CompModDns180_480}(c)), we normalize time lags $\tau$ by the Kolmogorov time scale $\tau_K$ that reads
\begin{equation}\label{eq:deftauKForAll}
\tau_K = \sqrt{\frac{\nu}{\langle \varepsilon \rangle}},
\end{equation}
where $\nu$ is the kinematic viscosity and  $\langle \varepsilon \rangle$ the average viscous dissipation per unit of mass. Interestingly, we observe that, in this representation where scales are normalized by $\tau_K$, $\mathcal C_a(\tau)$ crosses zero at a Reynolds number independent time scale. Call such a scale $\tau_0$, thus defined by $\mathcal C_a(\tau_0)=0$. Indeed, this was already observed in numerical and laboratory flows \citep{YeuPop07,HucMac19}: the zero-crossing time scale of acceleration has a universal (i.e. Reynolds number independent) behaviour with respect to the Kolmogorov time scale $\tau_K$ (Eq. \ref{eq:deftauKForAll}), such that 
\begin{equation}\label{eq:EmpirConstApprox}
\tau_0 \approx 2.2\,\tau_K,
\end{equation}
in the range of investigated Kolmogorov time scales. In our case and to be more precise, we find  $\tau_0 = 2.11\,\tau_K$ at $\mathcal R_\lambda=185$, and $\tau_0 = 2.14\,\tau_K$ at $\mathcal R_\lambda=418$, indeed very close to previous findings of \cite{YeuPop07} (Eq. \ref{eq:EmpirConstApprox}). In the sequel, we will use this fact to fully calibrate our model, in particular while relating its free parameter $\tau_\eta$ to the characteristics of the numerical flows. We will revisit this point in Section \ref{Sec:CalibModDNS}.

Similarly, we display the scaling behaviour of the second-order structure function $\mathcal S_{u,2}$ (Fig. \ref{fig:CompModDns180_480}(b)) and of the flatness of velocity increments (Fig. \ref{fig:CompModDns180_480}(d)). We can easily observe the three expected ranges of scales: the dissipative one with $\mathcal S_{u,2}(\tau)\propto \tau^2$, the inertial one with $\mathcal S_{u,2}(\tau)\propto \tau$, and the saturation towards $2\langle u^2\rangle$ at larger scales. Concerning the flatness, similar behaviour  is observed, saturation at the Gaussian value 3 at large scales, and a power-law behaviour in the inertial range, reminiscent of the intermittency phenomenon. We furthermore observe a more rapid increase in the intermediate dissipative range, and then a Reynolds number dependent saturation towards the flatness of acceleration. This is a known effect of the fine structure of turbulence, linked to subtle differential action of viscosity that depends on the local regularity of the velocity field \citep{CheRou03,CheCas05,CheCas06,ArnBen08,BenBif10,CheCas12}.
This phenomenon is well reproduced by the phenomenology of the intermittency phenomenon developed in the framework of the multifractal formalism \citep{PalVul87,Fri95}. We will develop these ideas in Section \ref{Sec:PredMultiForm}.

\subsection{Discussions on the Reynolds number dependence of the zero-crossing time scale of the acceleration correlation function}

\subsubsection{Model predictions of the zero-crossing time scales}

It is clear from previous developments that the present model, both for its Gaussian version $v$ (Proposition \ref{propcorrelationsInfinite} and Fig. \ref{fig:NumOU}(d)) and for its intermittent generalization $u$ (Proposition \ref{propcorrelationsInfiniteMRW} and Fig. \ref{fig:NumOUMulti}(d)), predicts this aforementioned zero-crossing time scale $\tau_0$ of the acceleration correlation function, as a function of its parameters $\tau_\eta$ and $T$. At this level of discussion, we neglect the influence of the intermittency parameter $\gamma$ in this picture. Indeed, even if in the intermittent framework the parameter enters explicitly in the form of the correlation function (Eq. \ref{eq:CorrelAUInfinite}), it has only a very weak influence on its overall shape, even in the dissipative range (data not shown). Thus, given the low value of $\gamma$ (Eq. \ref{eq:ChoiceGamma}) that makes the predicted intermittent acceleration correlation function (Eq. \ref{eq:CorrelAUInfinite}) indiscernible from its Gaussian approximation (Eq. \ref{ExplicitCorrALim}), we pursue further theoretical discussions neglecting these non-Gaussian effects. It is moreover convenient since in this case, $\mathcal C_a(\tau)$ has an explicit form (Eq. \ref{ExplicitCorrALim}), that makes its dependence present on $\tau_\eta$ and $T$.

Further inspection of the numerical results presented in  Fig. \ref{fig:NumOUMulti}(d) when $\tau_\eta$ is varying shows that this predicted zero-crossing time scale depends in a non trivial way on $\tau_\eta$. Actually, keeping only the leading terms entering in Eq. \ref{ExplicitCorrALim} as $\tau_\eta$, we can observe that asymptotically, this time scale behaves as 
\begin{equation}\label{eq:PredZerOUinfGauss}
\tau_0\build{\sim}_{\tau_\eta\to 0}^{}2\tau_\eta\sqrt{\log\left(\frac{T}{\sqrt{\pi}\tau_\eta}\right)}.
\end{equation}
Taking into account the empirical fact that the zero-crossing time scale is proportional to the Kolmogorov time scale $\tau_K$ in a universal way (Eq. \ref{eq:EmpirConstApprox}), this shows that $\tau_\eta$, up to logarithmic corrections, has the same Reynolds number dependence as $\tau_K$, and thus can be considered as a dissipative time scale. Interestingly, for the process proposed by Sawford (Section \ref{sec:Sawford}), named here $v_2$, such a zero-crossing time scale can be exactly derived from Eq. \ref{eq:corra2}. In this case, we obtain $\tau_0 = \tau_{\eta}\frac{\log(T/\tau_{\eta})}{1-\frac{\tau_{\eta}}{T}}$. The present prediction for $\tau_0$ (Eq. \ref{eq:PredZerOUinfGauss}) made with an infinitely differentiable process can be seen as an improvement of the model by Sawford, since the parameter $\tau_\eta$ is closer to $\tau_K$.

\subsubsection{The proposed calibration procedure of models parameters}\label{Sec:CalibModDNS}

As explained in the preceding Section, we can neglect in this discussion all possible intermittent effects, and work in a convenient way with the explicit second-order statistical properties of the Gaussian process $v$ (Proposition \ref{propcorrelationsInfinite}). To determine the free parameters of the model $\tau_\eta$, given the characteristic scales of the DNS $\tau_K$ and $T_L$, we solve the nonlinear system of coupled equations
\begin{align}
T_L &= T\frac{e^{-\tau_\eta^2/T^2}}{\erfc\left(\tau_\eta/T\right)}\label{eq:CalibSyst1}\\
\mathcal C_a(\alpha \tau_K) &= 0\label{eq:CalibSyst2},
\end{align}
where the exact expression of $T_L$ in Eq. \ref{eq:CalibSyst1} can be easily obtained from Eq. \ref{CorrVelLimFour2}, the explicit expression of $\mathcal C_a$ is provided in Eq. \ref{ExplicitCorrALim}, and $\alpha$ being equal to 2.11 at $\mathcal R_\lambda=185$, and 2.14 at $\mathcal R_\lambda=418$. This is our calibration procedure. Using a standard numerical solver of nonlinear equations and the values of $(\tau_K,T_L)$ provided in Table \ref{tab:DNSParam}, we look for the solution of the system of equations \ref{eq:CalibSyst1}  and \ref{eq:CalibSyst2}, and get $(\tau_\eta/\tau_K,T/T_L)=(0.6335,0.9562)$ for $\mathcal R_\lambda=185$, and $(0.5759,0.9791)$ for $\mathcal R_\lambda=418$.

\subsection{Comparison of model predictions to DNS data}\label{Sec:CompModDNSLimFlat}

Having performed the calibration procedure depicted in Section \ref{Sec:CalibModDNS}, and obtained the respective values for the free parameters $\tau_\eta$ and $T$, we compare the predictions of the present model to data.

We represent theoretical second-order statistics in Figs. \ref{fig:CompModDns180_480}(a) and (b) using solid lines. We indeed observe an almost perfect collapse with the statistical estimations based on DNS data.

Let us focus now on the acceleration correlation function (Fig. \ref{fig:CompModDns180_480}(c)). At a moderate Reynolds number $\mathcal R_\lambda=185$, we can see that the agreement is excellent in the dissipative range, i.e. for scales smaller that the zero-crossing time scale $\tau_0$. We can also observe a slight disagreement above  $\tau_0$. This can be due to the lack of statistical convergence at large scales that induces an overestimation of the integral time scale $T_L$, as we discussed in Section \ref{Sec:DiscussEstTL}. Only a specially devoted DNS simulation, that would be run over several tens of large turnover time scale could show us whether model predictions can be improved. At the current level of precision, we can consider that overall agreement with second-order statistics is satisfactory at this Reynolds number. At a higher Reynolds number $\mathcal R_\lambda=418$, further discrepancies can be seen in the dissipative range. This is very probably due to intermittency effects, that are negligible in the model, but not in DNS. To see this more clearly, let us focus on the flatness of velocity increments.

We superimpose in Fig.  \ref{fig:CompModDns180_480}(d) using solid lines the theoretical predictions that can be made from the model for flatnesses using the prescribed value $\gamma^2$ (Eq. \ref{eq:ChoiceGamma}). To get these theoretical predictions, that are tedious to obtain in an analytical fashion, we perform additional numerical simulations of time series of the model, as it is done in Section \ref{Sec:NumResInfMultiProcess}, for the calibrated values of the parameters $\tau_\eta$ and $T$ obtained in Section \ref{Sec:CalibModDNS}. We observe a very good agreement in the inertial range, showing that the chosen value for the intermittency coefficient $\gamma$ (Eq. \ref{eq:ChoiceGamma}) is realistic of DNS. Unfortunately, as we already noticed in Section \ref{Sec:StatAnalysisDNS}, the model is unable to reproduce the rapid increase of intermittency in the dissipative range. To go further in this direction, we propose to derive the predictions of the multifractal formalism in the following Section \ref{Sec:PredMultiForm} concerning the behaviour of the flatnesses in this range of scales.

\section{Predictions of the multifractal formalism regarding the acceleration correlation function} \label{Sec:PredMultiForm}

An alternative method of modelling the velocity and acceleration correlation functions consists in directly proposing their functional forms. We will thus construct models of the statistical behaviours of velocity, that will take into account the various range of scales pointed out by the phenomenology of turbulence, namely the inertial and dissipative ranges (with additional intermittent corrections). Doing so, we will end up with an explicit form of the velocity correlation function, or equivalently the second order structure function, without building up the underlying stochastic process. Compared to the previous construction of a stochastic process, from which we were deducing its statistical behaviour, this approach appears only partial from a probabilistic point of view: we model the velocity correlation function (from which we deduce the acceleration correlation function) and higher-order moments of velocity increments, but we do not characterize completely the velocity process itself. To this regard, the following probabilistic description is not complete, but will allow us, in particular, to understand in a fine way the rapid increase of the velocity increment flatness across the dissipative range, which is depicted in Fig.  \ref{fig:CompModDns180_480}(d). 

\subsection{The Batchelor parametrization of the second order structure function}\label{sec:BatchelorS2}

Let us begin with proposing a simple model for the velocity correlation function, or equivalently a model of the second moment of velocity increments. Concerning the Eulerian framework, \cite{Bat51} proposed a simple form for the second order structure function that includes the inertial behaviour $\langle (\delta_\ell u)^2\rangle \sim \ell^{2/3}$ and the dissipative one $\langle (\delta_\ell u)^2\rangle \sim \ell^{2}$, with an additional polynomial interpolation relating these two behaviours across the Kolmogorov dissipative length scale (see for instance \cite{Men96,CheCas06,CheCas12} for developments on this matter and references therein). A similar procedure can be adapted to the Lagrangian framework, that would include the respective inertial behaviour $\langle (\delta_\tau v)^2\rangle \sim \tau$ and the dissipative one $\langle (\delta_\tau v)^2\rangle \sim \tau^{2}$, as it was considered by \cite{CheRou03,ArnBen08,BenBif10,CheCas12}. Such a form reads, assuming $\tau\ll T$,
\begin{equation*}
\mathcal S_2(\tau)=\langle (\delta_\tau v)^2\rangle = 2\sigma^2\frac{\frac{\tau}{T}}{\left[1+\left(\frac{\tau}{\tau_{\eta}}\right)^{-\delta}\right]^{\frac{1}{\delta}}},
\end{equation*}
where $\tau_\eta$ is the typical dissipative (Kolmogorov) time scale, and $\sigma^2=\langle v^2\rangle$. The additional free parameter $\delta$ governs the transition between the inertial and dissipative ranges of scales. For instance, as far as the Eulerian framework is concerned, the value $\delta=2$ was chosen by \cite{Bat51}. We will see that the value $\delta=4$ will eventually reproduce in a appropriate manner the behaviour of the statistical quantities in the Lagrangian framework, as it was chosen in \cite{ArnBen08}. At large scales, $\tau$ of the order of $T$ and larger, we could think about multiplying the proposed form Eq. \ref{eq:BatchelorS2} by a cut-off function of characteristic time scale $T$, as it was proposed in \cite{BosChe12}. Such a procedure is necessary to ensure a smooth transition towards decorrelation. It is indeed required that $\mathcal S_2(\tau)$ goes to $2\sigma^2=2\langle v^2\rangle$ as $\tau\rightarrow \infty$. Incidentally, it will also make the integral of the velocity correlation function $C_v(\tau)\equiv \sigma^2-\mathcal S_2(\tau)/2$ converge, as it is required when assuming stationary statistics. Recall furthermore that we will be interested in looking at the second derivatives of $\mathcal S_2$ in order to describe the acceleration correlation, for which statistical stationarity implies that its integral over time lags $\tau$ vanishes. To this regard, multiplying by a cut-off function of characteristic time scale $T$ turns out to be too schematic.  Instead, we will be using the following ad-hoc form, for any time lags $\tau\ge 0$,
\begin{equation}\label{eq:BatchelorS2}
\mathcal S_2(\tau)=\langle (\delta_\tau v)^2\rangle = 2\sigma^2\frac{1-e^{-\frac{\tau}{T}}}{\left[1+\left(\frac{\tau}{\tau_{\eta}}\right)^{-\delta}\right]^{\frac{1}{\delta}}}.
\end{equation}

Correspondingly, the acceleration correlation function is given by (half) the second derivatives of Eq. \ref{eq:BatchelorS2}, and we get, written in a convenient form,
\begin{equation}\label{eq:AcceBatchelorS2}
\mathcal C_a(\tau)\equiv  \frac{1}{2} \frac{d^2\mathcal S_2(\tau)}{d\tau^2}.
\end{equation}

\subsection{Including intermittency corrections using the multifractal formalism}

The multifractal formalism \citep{Fri95} provides a convenient theoretical framework to generalize the approach of Batchelor (Eq. \ref{eq:BatchelorS2}) such that  inclusion of intermittent corrections are possible, and consistent with high-order structure functions. Mostly developed for the Eulerian framework, it has been then adapted to the Lagrangian framework by several authors and compared with great success to experimental and numerical data (see \cite{Bor93,CheRou03,BifBof04} and references therein). Here, we follow mainly the approach reviewed in \cite{CheCas12}, where we furthermore include the smooth behaviour at large scales that we motivated in Section \ref{sec:BatchelorS2}.

\subsubsection{Second-order structure function and implied acceleration correlation using the language of the multifractal formalism}

In few words, arguments developed in this context concern the probabilistic modelling of the Lagrangian velocity increment, defined by  $\delta_{\tau}v(t)=v(t+\tau)-v(t)$. In a similar spirit as the Batchelor parametrization of the second-order structure function (Eq. \ref{eq:BatchelorS2}), taking into account expected behaviours in the inertial and dissipative ranges, we get the following explicit expression for $\tau\ge 0$
\begin{equation}\label{eq:MultiS2}
\mathcal S_2(\tau)=\langle (\delta_\tau v)^2\rangle = 2\sigma^2\int_{h_{\text{min}}}^{h_{\text{max}}}\frac{\left(1-e^{-\frac{\tau}{T}}\right)^{2h}}{\left[1+\left(\frac{\tau}{\tau_{\eta}(h)}\right)^{-\delta}\right]^{\frac{2(1-h)}{\delta}}}\mathcal P_h^{(\tau)}(h)dh,
\end{equation}
which can be regarded as a generalization of the parametrization used in Eq. \ref{eq:BatchelorS2} to a non-unique exponent $h$ that eventually fluctuates according to its probability density $\mathcal P_h^{(\tau)}$ at a given scale $\tau$. Actually, we can recover exactly Eq. \ref{eq:BatchelorS2} while assuming a unique (non-fluctuating, i.e. deterministic) exponent $h=1/2$, that corresponds to a distributional density $\mathcal P_h^{(\tau)}$ equals to the Dirac delta function centered on this unique value $1/2$. Remark also that we included in such a generalization (Eq. \ref{eq:MultiS2}) a possible dependence of the dissipative scale $\tau_\eta(h)$ on this fluctuating exponent $h$, that remains to be determined.

The dissipative time scale entering in this formulation (Eq. \ref{eq:MultiS2}) has a natural dependence on the exponent $h$. Following the arguments developed for the Eulerian framework by \cite{PalVul87,Nel90}, and adapted to the Lagrangian one in \cite{Bor93} (and reviewed in \cite{CheCas12} with corresponding notations), we assume that 
\begin{equation}\label{eq:TauEtaHMulti}
    \tau_{\eta}(h)=T\left(\frac{\tau_\eta}{T}\right)^{\frac{2}{2h+1}},
\end{equation}
where, to simplify notations, we call $\tau_\eta\equiv \tau_{\eta}(1/2)$ the value of the fluctuating dissipative time scale $ \tau_{\eta}(h)$ (Eq. \ref{eq:TauEtaHMulti}) at the very particular value $h=1/2$. Finally, the fluctuating exponent $h$ is characterized by its probability density function at a given scale $\tau$, namely
\begin{equation}\label{eq:dist1}
    P_{h}^{(\tau)}(h)=\frac{1}{\mathcal{Z}(\tau)}\frac{\left(1-e^{-\frac{\tau}{T}}\right)^{1-\mathcal D^L(h)}}{[1+(\frac{\tau}{\tau_{\eta}(h)})^{-\delta}]^{(D^L(h)-1)/\delta}}
\end{equation}
normalized in a appropriate manner using
\begin{equation}\label{eq:norm1}
   \mathcal{Z}(\tau)=\int_{h_{\text{min}}}^{h_{\text{max}}}\frac{\left(1-e^{-\frac{\tau}{T}}\right)^{1-\mathcal D^L(h)}}{[1+(\frac{\tau}{\tau_{\eta}(h)})^{-\delta}]^{(\mathcal D^L(h)-1)/\delta}}\text{d}h.
\end{equation}

Besides the two obvious free parameters $T$ and $\tau_\eta$ of this model of the second-order structure function (Eq. \ref{eq:MultiS2}) that will be calibrated in units of $T_L$ and $\tau_K$  in a similar fashion as it is presented in Section \ref{Sec:CalibModDNS}, the multifractal formalism \citep{Fri95} requires the introduction of a parameter function $\mathcal D^L(h)$. It acquires the status of a singularity spectrum asymptotically at infinite Reynolds number (i.e. when  $\tau_\eta$ goes to 0) and then at vanishing scales $\tau \rightarrow 0$. It eventually governs the level of fluctuations of the exponent $h$ around its average value, that we expect to be $\langle h\rangle=1/2$. Several forms have been proposed in the literature (see \cite{Fri95}). We make a simple quadratic choice for $\mathcal D^L(h)$, which is known as a \textit{lognormal} approximation, parametrized by the intermittency coefficient $\gamma^2$ (Eq. \ref{eq:ChoiceGamma}), that reads 
\begin{equation}\label{eq:DHLN}
\mathcal D^L(h) = 1-\frac{(h-1/2-\gamma^2)^2}{2\gamma^2},
\end{equation} 
such that we enforce a linear behaviour of $ \mathcal{S}_2(\tau)$ with $\tau$ in the inertial range (in the appropriate infinite Reynolds number limit). To make a connection with the notations chosen in \citep{CheRou03,CheCas12}, this corresponds to $c_1^L=1/2+c_2^L$ for $c_2^L=\gamma^2$.

Correspondingly, the correlation function of acceleration $\mathcal C_a(\tau)$ can be defined as (half) the derivatives of the second order structure function (Eq. \ref{eq:sf2}). Using the following notations,
\begin{equation}\label{eq:sf2}
   \mathcal{S}_2(\tau) = \frac{1}{\mathcal{Z}(\tau)}\int_{h_{\text{min}}}^{h_{\text{max}}} \mathcal{Q}(h,\tau) dh, \hspace{5mm}  \text{where  } \hspace{5mm}  \mathcal{Q}(\tau,h) = \frac{\left(1-e^{-\frac{\tau}{T}}\right)^{2h+1-\mathcal D^L(h)}}{[1+(\frac{\tau}{\tau_{\eta}(h)})^{-\delta}]^{(2(1-h)+\mathcal D^L(h)-1)/\delta}},
\end{equation}
we get
\begin{align}\label{eq:corraMulti}
    \mathcal{C}_{a}(\tau)&= \left(\frac{\mathcal{Z}^{\prime}(\tau)^2}{\mathcal{Z}(\tau)^3}-\frac{1}{2}\frac{\mathcal{Z}^{\prime\prime}(\tau)}{\mathcal{Z}(\tau)^2} \right)\int_{h_{\text{min}}}^{h_{\text{max}}} \mathcal{Q}(h,\tau)dh - \frac{\mathcal{Z}^{\prime}(\tau)}{\mathcal{Z}(\tau)^2}\int_{h_{\text{min}}}^{h_{\text{max}}} \frac{\partial \mathcal{Q}(h,\tau)}{\partial \tau}dh\nonumber \\
    &+\frac{1}{2\mathcal{Z}(\tau)}\int_{h_{\text{min}}}^{h_{\text{max}}} \frac{\partial^2 \mathcal{Q}(h,\tau)}{\partial \tau^2}dh.
\end{align}
The form given in Eq. \ref{eq:corraMulti} can be then considered as a model for the correlation function of acceleration, at a given Reynolds number (which can be estimated as the value of $(T/\tau_\eta)^2$), and that includes intermittent corrections (using a non vanishing value for $\gamma^2$). Remaining integrals entering in Eq. \ref{eq:corraMulti} are evaluated numerically using standard numerical integration algorithms.

\subsubsection{Higher-order structure functions and their scaling behaviour}

Let us give the corresponding prediction for the structure function $\mathcal S_{2m}(\tau)$ of order $2m$, that will eventually enter in the expression of the velocity increment flatness. Note that statistics of increment are expected and observed symmetrical, making odd-order moments vanish. It reads
\begin{equation}\label{eq:MultiSm}
\mathcal S_{2m}(\tau)=\langle (\delta_\tau v)^{2m}\rangle = (\sqrt{2}\sigma)^{2m}\frac{(2m)!}{m!2^m}\int_{h_{\text{min}}}^{h_{\text{max}}}\frac{\left(1-e^{-\frac{\tau}{T}}\right)^{2mh}}{\left[1+\left(\frac{\tau}{\tau_{\eta}(h)}\right)^{-\delta}\right]^{\frac{2m(1-h)}{\delta}}}\mathcal P_h^{(\tau)}(h)dh,
\end{equation}
where the additional combinatorial factor originates from the moment of order $2m$ of a zero-average unit-variance Gaussian random variable that enters in the more complete probabilistic description detailed in \cite{CheCas12}.

In the dissipative range, such that $\tau\ll \tau_\eta$, $\mathcal S_{2m}(\tau)$ (Eq. \ref{eq:MultiSm}) behaves in a consistent manner with its Taylor's development, that is $\mathcal S_{2m}(\tau)=\langle a^{2m}\rangle \tau^{2m}+o(\tau^{2m})$. In the inertial range, i.e. for $\tau_\eta\ll \tau\ll T$, we recover the standard prediction of the multifractal formalism, that relates the power-law behaviour of the structure functions to the functional shape of the parameter function $\mathcal D^L(h)$ through a Legendre transform \citep{Fri95}. We have, in the proper ordering of limits,
\begin{equation}\label{eq:MultiSmLegendre}
\lim_{\tau_\eta\to 0}\mathcal S_{2m}(\tau)\build{\sim}_{\tau \to 0}^{} c_{\gamma,2m} (\sqrt{2}\sigma)^{2m}\frac{(2m)!}{m!2^m}\left( \frac{\tau}{T}\right)^{\min_h \left[ 2mh+1-\mathcal D^L(h)\right]},
\end{equation}
where the remaining multiplicative constant could be computed while pushing forward the underlying steepest-descent calculation, techniques that we develop in the Section \ref{Sec:PredMultiVarAcceRey}. Assuming then a quadratic form for the parameter function $\mathcal D^L(h)$ (Eq. \ref{eq:DHLN}), once again this could be done for other choices \citep{Fri95}, we obtain the following intermittent behaviour
\begin{equation}\label{eq:MultiSmLegendreLN}
\lim_{\tau_\eta\to 0}\mathcal S_{2m}(\tau)\build{\sim}_{\tau \to 0}^{} c_{\gamma,2m} (\sqrt{2}\sigma)^{2m}\frac{(2m)!}{m!2^m}\left( \frac{\tau}{T}\right)^{\left( 1+2\gamma^2\right)m-2\gamma^2m^2},
\end{equation}
which power-law exponent $\zeta_{2m}\equiv\left( 1+2\gamma^2\right)m-2\gamma^2m^2$ corresponds exactly to the one obtained for the infinitely differentiable multifractal random walk of Section \ref{Sec:MRWInfiniteDeriv} (where the scaling behaviour of its structure functions at infinite Reynolds number can be found in Proposition \ref{propSFInfiniteMRW}).

\subsubsection{Derivation of the Reynolds number dependence of the acceleration variance}\label{Sec:PredMultiVarAcceRey}

Let us now give the Reynolds number dependence, or equivalently the dependence on the free parameters $\tau_\eta$ and $T$, of the acceleration variance, and the scaling behaviour of $\mathcal S_{2m}(\tau)$ with $\tau$ at infinite Reynolds number (i.e. for $\tau_\eta\to 0$). As it is detailed in \cite{CheCas12}, or simply deduced from Eq. \ref{eq:MultiSm} using $\mathcal S_{2}(\tau)=\langle a^2\rangle \tau^2 +o(\tau^2)$, we have
\begin{equation}\label{eq:PredVarAcceMulti}
\langle a^2\rangle = \frac{2\sigma^2}{T^2}\frac{1}{\mathcal Z(0)}\int_{h_{\text{min}}}^{h_{\text{max}}} \left(\frac{\tau_\eta}{T}\right)^{2\frac{2(h-1)+1-\mathcal D^L(h)}{2h+1}}dh,
\end{equation}
with
\begin{equation}\label{eq:PredVarAcceMultiZ0}
\mathcal Z(0)=\int_{h_{\text{min}}}^{h_{\text{max}}} \left(\frac{\tau_\eta}{T}\right)^{2\frac{1-\mathcal D^L(h)}{2h+1}}dh.
\end{equation}
Follow then a steepest-descent procedure. Compute first the minimum and the minimizer of the exponents entering in Eqs. \ref{eq:PredVarAcceMulti} and \ref{eq:PredVarAcceMultiZ0}, using for $\mathcal D^L$ the expression provided in Eq. \ref{eq:DHLN}. Notice that $\min_h \frac{1-\mathcal D^L(h)}{2h+1}=0$ and assume $\gamma^2<2-\sqrt{3}$ to guarantee the positivity of these real-valued minimizers, a condition which is fulfilled by the empirical value of the intermittency coefficient (Eq. \ref{eq:ChoiceGamma}). To get an estimation of the remaining multiplicative constant following this steepest-descent calculation, perform a Taylor series of the exponents entering in Eqs. \ref{eq:PredVarAcceMulti} and \ref{eq:PredVarAcceMultiZ0} around their respective minimizer up to second order, and finally approximate the remaining Gaussian integrals extending the integration range over $h\in \mathbb R$. We eventually obtain the following exact equivalent as the Reynolds number goes to infinity:
\begin{equation}\label{eq:PredVarAcceMultiSD}
\langle a^2\rangle \build{\sim}_{\tau_\eta\to 0}^{} \frac{2\sigma^2}{T^2}
\frac{\left[1-4\gamma^2+\gamma^4\right]^{\frac{1}{4}}}{\sqrt{1+\gamma^2}}
\left(\frac{\tau_\eta}{T}\right)^{\frac{\gamma^2-1+\sqrt{1-4\gamma^2+\gamma^4}}{\gamma^2}}.
\end{equation}
We can see that the multifractal prediction of acceleration variance (Eq. \ref{eq:PredVarAcceMultiSD}) does exhibit an intermittent correction, as it was already derived in a very similar way by \cite{Bor93,SawYeu03}. For a more detailed  comparison to DNS data, we invite the reader to the following Section \ref{Sec:CalibCompMulti}. At this stage, let us notice that, whereas structure functions at infinite Reynolds number obtained from the multifractal formalism (Eq. \ref{eq:MultiSmLegendreLN}) and from the infinitely differentiable MRW (Proposition \ref{propSFInfiniteMRW}) behave in a very similar way, predicted acceleration variances differ by intermittent corrections (compare Eq. \ref{eq:PredVarAcceMultiSD} and Eq. \ref{eq:PredVarAcceIDMRW}).

\subsection{Calibration of the free parameters and comparisons to DNS data}\label{Sec:CalibCompMulti}

\begin{figure}
\begin{center}
\epsfig{file=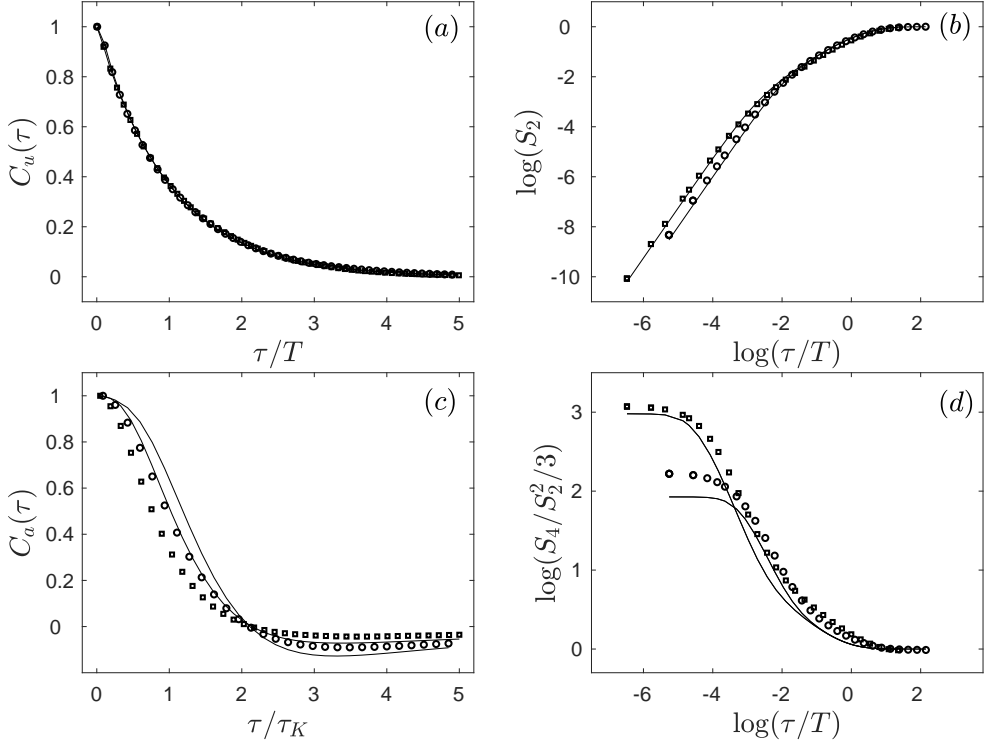,width=13cm}
\end{center}
\caption{\label{fig:ExactMultiRe} Comparison of DNS data to model predictions, similar to Fig. \ref{fig:CompModDns180_480}, but for multifractal predictions. (a) Estimation of the velocity correlation function from DNS data ($\circ$ and $\Box$ as in Fig. \ref{fig:CompModDns180_480}). We superimpose theoretical predictions based on the multifractal  parametrization of the second order structure function (Eq. \ref{eq:MultiS2}),  for the set of values of the parameters $\tau_\eta$ and $T$ given by our calibration procedure presented in Section \ref{Sec:CalibCompMulti}, and for a prescribed value for $\gamma$ (Eq. \ref{eq:ChoiceGamma}) and $\delta=4$. Time lags are normalized by the calibrated time scale $T$. (b) Same plot as in (a) but for the second-order structure function. (c) Similar plot as in (a) and (b) but for the acceleration correlation function, normalized by its value at the origin. Superimposed theoretical predictions are based on the exact expression given in Eq. \ref{eq:corraMulti}. (d) Similar plot as in (a) and (b) but for the corresponding flatnesses of velocity increments. Theoretical predictions are obtained from the expression given in Eq. \ref{eq:MultiSm}.}
\end{figure}

We adopt the same calibration of the free parameters $\tau_\eta$ and $T$ as it is depicted in Section \ref{Sec:CalibModDNS}. We numerically solve the nonlinear problem consisting in obtaining  $\tau_\eta$ and $T$ from the empirical value of $T_L$ and the appropriate zero-crossing of acceleration time scale given in unit of $\tau_K$. It is thus very similar to solving the system of Eqs. \ref{eq:CalibSyst1} and \ref{eq:CalibSyst2}, but notice there that moreover  the integral time scale $T_L$ predicted from the model has to be computed numerically using a standard integration scheme of the expression provided in Eq. \ref{eq:MultiS2}. To give a hint to the numerical algorithm that looks for zeros of functions, as it is required while solving this nonlinear problem, we can make a simple prediction for the zero-crossing of acceleration time scale $\tau_0$. Using the Batchelor's parametrization of the second-order structure function (Eq. \ref{eq:BatchelorS2}), and the corresponding prediction of the acceleration correlation function (Eq. \ref{eq:AcceBatchelorS2}), we expect that a good approximation of $\tau_0$ would be given by 
\begin{equation}\label{eq:PredZerBat}
\tau_0 \build{\approx}_{\tau_\eta\to 0}^{}\tau_\eta\left(\frac{\delta-1}{2}\right)^{-\frac{1}{\delta}},
\end{equation}
showing that indeed the free parameter $\tau_\eta$ is expected to be proportional to the Kolmogorov dissipative time scale $\tau_K$.

Using the physical parameters of the DNS data provided in Table \ref{tab:DNSParam}, assuming furthermore $\gamma^2=0.085$ (Eq. \ref{eq:ChoiceGamma}) and $\delta=4$, we look for the solution of this aforementioned nonlinear  system of equations (similar to Eqs. \ref{eq:CalibSyst1}  and \ref{eq:CalibSyst2}). We finally retrieve $(\tau_\eta/\tau_K,T/T_L)=(2.7596,0.9927)$ for $\mathcal R_\lambda=185$, and $(2.6106,0.9983)$ for $\mathcal R_\lambda=418$.

Having in hand the calibrated values for the parameters $\tau_\eta$ and $T$, we now compare to DNS data. Similar to Figs.  \ref{fig:CompModDns180_480}(a), (b), and (c) we represent in Figs. \ref{fig:ExactMultiRe}(a), (b) and (c) the predictions of the velocity  correlation function $\mathcal{C}_{v}(\tau)$, the second-order structure function $\mathcal S_2(\tau)$ and the acceleration correlation function $\mathcal{C}_{a}(\tau)$, all based on the multifractal  parametrization of the second-order structure function (Eq. \ref{eq:MultiS2}), and its second derivative (Eq. \ref{eq:corraMulti}). As far as velocity is concerned, we observe a perfect agreement between predictions and DNS data, for both correlation  (Fig.  \ref{fig:ExactMultiRe}(a)) and second-order structure function (Fig.  \ref{fig:ExactMultiRe}(b)).

Concerning the acceleration correlation function $\mathcal{C}_{a}(\tau)$ (Fig.  \ref{fig:ExactMultiRe}(c)), we observe that predictions overestimate slightly the observed negative values after the zero-crossing. Interestingly, we observed an opposite behaviour with the former depicted infinitely differentiable process (Fig. \ref{fig:CompModDns180_480}(c)). Below the zero-crossing time scale, predictions overestimate the decrease of correlation, although the dependence on the Reynolds number goes in the good direction. Compared to the performance of the stochastic process depicted in Section \ref{Sec:MRWInfiniteDeriv}, and displayed in Fig. \ref{fig:CompModDns180_480}(c), we can see that predictions based on the multifractal formalism do not perform as well. As we will see, the strength of the multifractal formalism lies in the possibility to understand and model the rapid increase of the flatness in the intermediate dissipative range. We are thus led to the conclusion that this rapid increase, coming from the differential action of viscosity, does not explain the discrepancies that we can observe between DNS and models.

Let us now focus on the intermittency corrections, as it is well quantified by the flatness of velocity increments. We compare in Fig.  \ref{fig:ExactMultiRe}(d) the flatness of increments, based on DNS and on the current multifractal model using the expression given in Eq. \ref{eq:MultiSm}. We can see that multifractal predictions reproduce accurately the overall shape of the flatness, including the rapid increase in the intermediate dissipative range, for both Reynolds numbers. Recall that this very dissipative behaviour is not reproduced by the stochastic approach of Section \ref{Sec:MRWInfiniteDeriv}, and displayed in Fig. \ref{fig:CompModDns180_480}(d), We can notice furthermore a slight shift between numerical and theoretical curves: this indicates that the large time scale associated with intermittent corrections is slightly larger that the one associated to the velocity correlation time scale. This could be included in the expressions of structure functions (Eqs. \ref{eq:MultiS2} and \ref{eq:MultiSm}) at the price of introducing another ad-hoc free parameter of order unity, without further justifications (data not shown). Nonetheless, we can see that, overall, the present multifractal model reproduces in good agreement DNS data, both in the inertial and dissipative ranges. 

Let us go back to the predicted variance of acceleration (Eq. \ref{eq:PredVarAcceMultiSD}) and its comparison to data. To this purpose, we will articulate this discussion around the compilation of DNS data at various Reynolds numbers performed by \cite{IshKan07}, and their comparison to an empirical form proposed by \cite{Hil02}. To make the discussion short and simple, use the prescribed value for $\gamma$ (Eq. \ref{eq:ChoiceGamma}), and write the predicted variance (Eq. \ref{eq:PredVarAcceMultiSD}) as $\langle a^2\rangle\propto (\sigma/T)^2(\tau_\eta/T)^{-1-0.155}$, which is the standard non-intermittent phenomenological prediction, enhanced by an intermittent correction of order $(\tau_\eta/T)^{-0.155}$. The calibration procedure used here confirms that $\tau_\eta$ has, in a good approximation, the same Reynolds number dependence as $\tau_K$. Furthermore, $T$ is very close to $T_L$, such that $T\propto L/\sigma$, where $L$ is the large length scale of the flow, and recall that $\sigma$ is the velocity standard deviation. Using $\langle \varepsilon\rangle \propto \sigma^3/L$, we can rewrite the empirical form for $\langle a^2\rangle$ proposed by  \cite{IshKan07} (see their equation 5.10 and the respective discussion) in units of $(\sigma/T)^2$. This empirical form of \cite{IshKan07} consists in the sum of two power-laws, a dominant one at large Reynolds numbers of order  $(\sigma/T)^2(\tau_\eta/T)^{-1.25}$, and a subdominant one of order  $(\sigma/T)^2(\tau_\eta/T)^{-1.11}$. We can see that the present theoretical prediction, i.e. $(\sigma/T)^2(\tau_\eta/T)^{-1.155}$, using Eq. \ref{eq:PredVarAcceMultiSD} with the prescribed value for $\gamma$ (Eq. \ref{eq:ChoiceGamma}) lies in between these two power-laws. As we noticed in the former Section \ref{Sec:PredMultiVarAcceRey}, such a prediction of the multifractal formalism has already been derived by \cite{Bor93,SawYeu03}, and compared to a compilation of DNS data in \cite{SawYeu03,YeuPop06}: derived in a very similar way as we do, although based on a different choice for the parameter function $\mathcal D^L(h)$ (Eq. \ref{eq:DHLN}),  the acceleration variance was predicted to behave as $(\sigma/T)^2(\tau_\eta/T)^{-1.135}$, which is very close to the present prediction, and was shown to reproduce accurately the trends observed in DNS. We are led to the conclusion that, given the available range of Reynolds numbers accessible in DNS, corrections to standard phenomenological arguments for the acceleration variance as they are observed in DNS data are consistent with implied corrections by the intermittency phenomenon.

\subsection{Further considerations regarding the prediction of the multifractal formalism}

Let us develop here  the modelling of the differential action of viscosity and the implied dependence of the dissipative length and time scales on the local exponent $h$, as it is proposed in particular in \cite{PalVul87,Nel90,Bor93}. Rephrased in terms of time scales, similar arguments could be developed for length scales, we can estimate the extension of the range on which the dissipative time scale $\tau_\eta(h)$ varies. Actually, it will turn out to be more appropriate to estimate this range in a logarithmic fashion. This is due to the fact that the probability density function of $\log \left(\tau_{\eta}(h)/T\right)$ is eventually very close to a Gaussian function as $\tau_{\eta}/T\rightarrow 0$, and is thus well characterized by its average and standard deviation.  Using Eq. \ref{eq:TauEtaHMulti}, we get
\begin{equation}\label{eq:LogTauEtaHMulti}
   \log \left(\frac{\tau_{\eta}(h)}{T}\right)=\frac{2}{2h+1}\log\left(\frac{\tau_\eta}{T}\right),
\end{equation}
such that the respective moments of order $q\in \mathbb N$ are given by,
\begin{equation}
  \left\langle \left(\log \left(\frac{\tau_{\eta}(h)}{T}\right)\right)^q\right\rangle=\frac{1}{\mathcal Z(0)}\int_{h_{\text{min}}}^{h_{\text{max}}} \left(\frac{2}{2h+1}\right)^q\left(\frac{\tau_\eta}{T}\right)^{\frac{2\left[ 1-\mathcal D^L(h)\right]}{2h+1}}dh\,\log^q\left(\frac{\tau_\eta}{T}\right),
\end{equation}
where the normalization constant $\mathcal Z(0)$ is defined as the limit when $\tau\rightarrow 0$ of the expression given in Eq. \ref{eq:norm1}. To simplify expressions, and work with explicit functions instead of integrals, assume for this discussion 
$h_{\text{min}}=-1/2$ and $h_{\text{max}}=+\infty$. Make the change of variable $x=(2h+1)/2$   to obtain
\begin{equation}
  \left\langle \left(\log \left(\frac{\tau_{\eta}(h)}{T}\right)\right)^q\right\rangle=\frac{1}{\mathcal Z(0)}\int_{0}^{\infty} \frac{1}{x^q}\left(\frac{\tau_\eta}{T}\right)^{\frac{1-\mathcal D^L(x-1/2)}{x}}dx\,\log^q\left(\frac{\tau_\eta}{T}\right).
\end{equation}
Assuming then for $\mathcal D^L$ a quadratic approximation (Eq. \ref{eq:DHLN}) with given parameter $\gamma^2$, using a symbolic calculation software, we obtain as $\tau_{\eta}/T\rightarrow 0$
\begin{equation}\label{eq:meanLog}
  \left\langle \log \left(\frac{\tau_{\eta}(h)}{T}\right)\right\rangle=\frac{1}{1+\gamma^2}\log\left(\frac{\tau_\eta}{T}\right) + O(1),
\end{equation}
and
\begin{equation}\label{eq:varLog}
 {  \left\langle \left(\log \left(\frac{\tau_{\eta}(h)}{T}\right)\right)^2\right\rangle- \left\langle \log \left(\frac{\tau_{\eta}(h)}{T}\right)\right\rangle^2}={\frac{\gamma^2}{(1+\gamma^2) ^3}\log\left(\frac{T}{\tau_\eta}\right)} +O(1).
\end{equation}
Keeping in mind that $\gamma^2=0.085$ (Eq. \ref{eq:ChoiceGamma}) remains small compared to unity, these former considerations show that in a logarithmic representation, the dissipative time scale fluctuates over an extended range, centered on a time scale close to $\log \tau_\eta$ (Eq. \ref{eq:meanLog}), and of width proportional to $\sqrt{\log(T/\tau_\eta)}$ (Eq. \ref{eq:varLog}), or equivalently proportional to $\sqrt{\log \mathcal R_e}$. The extension of such an intermediate dissipative range and its respective Reynolds number dependence has been already predicted by similar, although different, arguments in \cite{CheCas05}. It is here re-derived based on the multifractal modelling using Eq.  \ref{eq:TauEtaHMulti}. Although such a predicted extension of the intermediate dissipative range (a width that behaves as $\sqrt{\log \mathcal R_e}$ in this logarithmic representation) can be considered as large, it differs in nature with, and is narrower than, other predictions. For example, \cite{YakSre05} attributes a dynamical significance to length scales that behave as $\mathcal R_e^{-1}$. Such small length scales, once reformulated in a Lagrangian context, have no significance as far as variance of the logarithm of $\tau_\eta(h)$ is concerned, or equivalently at this level of description, as it is given by the flatness of velocity increments. Similarly, in \cite{Dub19}, much emphasis is attributed to the scale obtained while taking $h\rightarrow -1/2$ in Eq. \ref{eq:TauEtaHMulti} corresponding to a vanishing time scale (or correspondingly in a Eulerian framework, taking $h\rightarrow -1$ in the multifractal parametrization of the Kolmogorov dissipative length scale). Once again, the present derivation of the intermediate dissipative range gives no significance to such a small time scale, i.e. its probability of appearance is vanishingly small as the Reynolds number gets large. Finally, it is claimed in \cite{BuaPum19}, based on the behaviour of the tails of the probability density functions of velocity gradients, that much smaller time of length scales are involved in the dynamics. Once again, the implication of the existence of these very fine length or time scales cannot be quantified using only the flatness of velocity increments. Actually, extreme events of gradients (or acceleration), as they are observed in the tails of their probability density, can be modeled using the probabilistic approach of \cite{CasGag90}, as it is reviewed, and related to the language of the multifractal formalism, by \cite{CheCas12}.

Let us conclude this digression by justifying our estimation of the width of the intermediate dissipative range based on logarithmic scales (and incidentally the moments of the logarithm of the dissipative time scales as given in Eqs. \ref{eq:meanLog} and  \ref{eq:varLog}). Further calculations, similar to the ones performed in  Eqs. \ref{eq:meanLog} and  \ref{eq:varLog} based on a quadratic approximation for $\mathcal D^L(h)$ (Eq. \ref{eq:DHLN}), show that the respective flatness of $\log (\tau_{\eta}(h)/T)$ (once centered in an appropriate way) behaves as $3+O(\log^{-1} (\tau_{\eta}/T))$, showing that the logarithm of the fluctuating dissipative time scale behaves in an asymptotic way as a Gaussian random variable, thus properly characterized by its mean and variance.

\section{Conclusions and perspectives}\label{Sec:Conclusion}

Let us summarize our original findings in the context of the stochastic modelling of Lagrangian velocity and acceleration.

First, we have proposed, for the first time as far we know, a stochastic dynamics, which is causal, infinitely differentiable at a given Reynolds number, or equivalently in a good approximation, for a given finite ratio of a dissipative time scale $\tau_\eta$ over a large one $T$. This process, that we called $u$, is defined as the limit $n\to \infty$ of the $n$-layered embedded process $u_n$ (Eqs. \ref{eq:OU_MRW_infinite_N_1_Multi} to \ref{eq:OU_MRW_infinite_N_N_Multi}). Its second-order statistical properties are derived analytically and results are gathered in Proposition \ref{propcorrelationsInfiniteMRW}. We furthermore included in a causal and exact way some intermittent properties, given an intermittent coefficient $\gamma$ (Eq. \ref{eq:ChoiceGamma}). As intermittency disappears, i.e. if we take $\gamma=0$,  we recover a Gaussian process that we noted by $v$, of which causal dynamics is discussed in Section \ref{Sec:InfDiffProc}, and of which second-order statistical properties are listed in Proposition \ref{propcorrelationsInfinite}. At infinite Reynolds number, i.e. when $\tau_\eta\to 0$, both processes converge towards a statistically stationary and  finite variance causal process, which is a (Gaussian) Ornstein-Uhlenbeck process concerning $v$ and a multifractal random walk concerning $u$. As far as the multifractal version $u$ is concerned, we have computed in an exact fashion the intermittent behaviour of its structure functions, and results are gathered in Proposition \ref{propSFInfiniteMRW}. Using an efficient algorithm designed in Section \ref{Sec:AlgoSimulIDMRW}, we have shown that such processes are easily to simulate, and we have been able to compare with great success our theoretical predictions to numerical simulations of the underlying dynamics.

We have then analyzed Lagrangian trajectories extracted from a set of DNS of the Navier-Stokes equations (see Table \ref{tab:DNSParam} where important physical parameters of the simulations are gathered) and compared their statistical properties to those of $u$ in Fig. \ref{fig:CompModDns180_480}. Following a calibration procedure (Section \ref{Sec:CalibModDNS}) that relates in a transparent and reproducible way the free parameters of the model $\tau_\eta$ and $T$ to the empirical values of the Kolmogorov time scale $\tau_K$ and of the integral one $T_L$, we are then able to reproduce with great accuracy the statistical properties of the DNS trajectories. We nonetheless observed some discrepancies below the zero-crossing time scale of the acceleration correlation function (Fig. \ref{fig:CompModDns180_480}(c)), and the flatness of velocity increments at similar dissipative time scales (Fig. \ref{fig:CompModDns180_480}(d)).

To push forward our understanding of the observed rapid increase of the flatness in the intermediate dissipative range, and on the way explore some new types of prediction for the acceleration correlation function, we have recalled and developed a phenomenological procedure mostly based on the multifractal formalism (see Section \ref{Sec:PredMultiForm}). This alternative approach differs from building up a stochastic process, as it was done for $u$. Instead, it proposed the modelling of the some chosen statistical properties such as structure functions. Nonetheless, it allows the derivation of new predictions for the acceleration correlation function and flatness of velocity increments, that reproduce in a very accurate way DNS data (see the proposed discussions on the results displayed in Figs. \ref{fig:ExactMultiRe}(c) and (d)). In particular, the theoretically predicted flatness reproduces its rapid increase in the intermediate dissipative range, a phenomenon that is related to the differential action of viscosity depending on the local singular strength of velocity, as it is modeled by the parametrization of \cite{PalVul87,Nel90,Bor93}.

From a perspective point of view, it would be useful to analyze a specifically designed DNS, and its Lagrangian trajectories, where special care has been taken to resolve in an appropriate and fine way the range of dissipative scales. Also, at the price of being limited in terms of Reynolds numbers, it would be much appreciated to work with numerous trajectories, each of them lasting far longer that the Lagrangian integral time scale $T_L$. Only then would we be able to discriminate between schematic modelling aspects and lack of numerical resolution. Also, both current theoretical approaches shed new light on the interpretation of experimental data in this range of time scales where viscosity dominates, and open the route to an original characterization of the influence of  possible large scale anisotropic situations. Finally, it would be much welcome, from the theoretical side, to include this differential action of viscosity as it is modeled by \cite{PalVul87,Nel90,Bor93} into the stochastic approach that ends up with $u$ and developed in Section \ref{Sec:MRWInfiniteDeriv}. Up to today, we do not know how to model in a stochastic viewpoint (and to provide the respective causal dynamics) this tricky action of viscosity, {we can nonetheless conclude that a simple linear filtering at small scales fails at reproducing such a behaviour. A natural idea would be to weight the filtering at the scale of order $\tau_\eta$ by a function of the multifractal random field. This remains to be explored and we leave these aspects for future investigations.}

\section*{Acknowledgements}
The authors thank the PSMN (P\^ole Scientifique de Mod\'elisation Num\'erique) computing centre of ENS de Lyon for numerical resources. We thank J. Bec, L. Biferale and F. Toschi for providing and making available DNS data. Similarly, we thank people involved in maintaining and making available the JHTDB.

B.V. and R.B.C are supported by U.S. National Science Foundation grant (GEO-1756259). R.B.C is also thankful for the support provided through the Fulbright Scholar Program. J.F. acknowledges funding from the Humboldt Foundation within a Feodor-Lynen fellowship. B.V., J.F., R.V., M.B., L.C.  benefit from the financial support of the Project IDEXLYON of the University of Lyon in the framework of the French program “Programme Investissements d'Avenir” (ANR-16-IDEX-0005).  L.C. is supported by ANR grants \textsc{Liouville} ANR-15-CE40-0013 and by the Simons Foundation Award ID: 651475. 

\section*{Declaration of interests}

The  authors  report  no  conflict  of  interest.

\appendix

{\section{Propositions concerning infinitely differentiable causal stochastic processes}
\label{App:PROPS}}

\begin{proposition}\label{propcorrelationsN}
 {Assume $n\ge 2$. Then the correlation functions of velocity and acceleration are given by
 \begin{equation}\label{CorrVelN}
  \mathcal C_{v_n}(\tau) = q_{(n)}\left(G_T\star G_{\tau_\eta}^{\star (n-1)}\right)(\tau),
  \end{equation}
and
 \begin{equation}\label{CorrAccN}
  \mathcal C_{a_n}(\tau)= -\frac{d^2\mathcal C_{v_n}(\tau)}{d\tau^2},
  \end{equation}
where we have introduced the correlation product $\star$, which is defined as, for any two functions $g_1$ and $g_2$,
$$ \left(g_1\star g_2\right)(\tau) = \int_{\mathbb R}g_1(t)g_2(t+\tau)dt,$$
with the corresponding short-hand notation,
$$g^{\star n}=\underbrace{g \star g \star \cdots \star g}_{n},$$
and the response function of the OU process at a given time scale $\tau$ (here $\tau=T$ or $\tau=\tau_\eta$)
\begin{equation}\label{eq:TransFoncOU}
t\in \mathbb R\mapsto G_\tau (t) = \frac{\tau}{2}e^{-|t|/\tau}.
\end{equation} 
For the sake of completeness, we also provide the spectral view of the correlation functions of velocity and acceleration (Eqs. \ref{CorrVelN} and \ref{CorrAccN}), which is especially useful when seeking their explicit expression for a given layer $n$, once injected into a symbolic calculation software. We have
\begin{equation}\label{CorrVelNFourApp}
  \mathcal C_{v_n}(\tau) = q_{(n)}\int_{\mathbb R}e^{2i\pi \omega\tau}\frac{T^2}{1+4\pi^2T^2\omega^2}\left[\frac{\tau_\eta^{2}}{1+4\pi^2\tau_\eta^2\omega^2}\right]^{n-1}d\omega,
  \end{equation}
and
 \begin{equation}\label{CorrAccNFourApp}
  \mathcal C_{a_n}(\tau)= q_{(n)}\int_{\mathbb R}4\pi^2\omega^2 e^{2i\pi \omega\tau}\frac{T^2}{1+4\pi^2T^2\omega^2}\left[\frac{\tau_\eta^{2}}{1+4\pi^2\tau_\eta^2\omega^2}\right]^{n-1}d\omega.
  \end{equation}
To finish with this proposition, we state the implied expression for the constant $q_{(n)}$ to ensure the physical constraint on velocity variance (Eq. \ref{eq:ConstVarVn}) by Parseval's identity,
\begin{equation}\label{ExplicitQN}
  \frac{\sigma^2}{q_{(n)}}= \int_{\mathbb R}\frac{T^2}{1+4\pi^2T^2\omega^2}\left[\frac{\tau_\eta^{2}}{1+4\pi^2\tau_\eta^2\omega^2}\right]^{n-1}d\omega.
  \end{equation}}
\end{proposition}

 \textit{Proof.} 
Rephrased in the language of linear systems theory (see for instance \cite{Pap91}), the system of equations Eqs. \ref{eq:OU_embedded_N_1} to \ref{eq:OU_embedded_N_N} defines a series of linear filters with a stochastic input. This explains the expression given for the velocity correlation of $v_n$ (Eq. \ref{CorrVelN}). 

We compute the correlation function of $v_n$, as it was done in Eq. \ref{eq:corrv2def} in a more straightforward manner, and drawing a connection with the approach adopted to present the model of Sawford (Section \ref{sec:Sawford}).  We obtain 
  \begin{equation*}
  \mathcal C_{v_n}(\tau)= \int_{-\infty}^{0} \int_{-\infty}^{\tau} \; e^{-(\tau-t_1-t_2)/T}\mathcal C_{f_{n-1}}(t_1-t_2)dt_1dt_2,
\end{equation*}
which can be formally rewritten as 
  \begin{align*}
  \mathcal C_{v_n}(\tau) &= \int_{\mathbb R^2} g_T(\tau+t_2)g_T(t_1)\mathcal C_{f_{n-1}}(t_1-t_2)dt_1dt_2\\
  &=\int_{\mathbb R^2} g_T(\tau+t_1+t_2)g_T(t_1)\mathcal C_{f_{n-1}}(t_2)dt_1dt_2\\
  &=\int_{\mathbb R} \left(g_T\star g_T\right)(\tau+t_2)\mathcal C_{f_{n-1}}(t_2)dt_2\\
  &=\left(g_T\star g_T\star \mathcal C_{f_{n-1}}\right)(\tau),
\end{align*}
where $g_T(t)=e^{-t/T}1_{t\ge 0}$. Noticing that $G_T(t)= \left(g_T\star g_T\right)(t)$, we arrive at the proposition made in Eq. \ref{CorrVelN} after iterating the procedure for the $n-1$ remaining layers. The equivalent form of the velocity correlation in the spectral space (Eq. \ref{CorrVelNFourApp}) is a consequence of the convolution theorem, and that the Fourier transform of $G_T$ is a Lorentzian function.  \textit{End of proof.}

\begin{proposition}\label{propcorrelationsInfinite}
{Take $n\ge 2$. Using the results of Proposition \ref{propcorrelationsN}, we have
\begin{equation}\label{CorrVelNFour2App}
  \mathcal C_{v_n}(\tau) = \frac{2\sigma^2e^{-\tau_\eta^2/T^2}}{T\erfc\left(\tau_\eta/T\right)}\int_{\mathbb R}e^{2i\pi \omega\tau}\frac{T^2}{1+4\pi^2T^2\omega^2}\left[\frac{1}{1+\frac{4\pi^2\tau_\eta^2\omega^2}{n-1}}\right]^{n-1}d\omega,
  \end{equation}
such that
\begin{equation}\label{CorrVelLimFour2}
\mathcal C_{v}(\tau)\equiv  \lim_{n\rightarrow \infty}\mathcal C_{v_n}(\tau) = \frac{2\sigma^2e^{-\tau_\eta^2/T^2}}{T\erfc\left(\tau_\eta/T\right)}\int_{\mathbb R}e^{2i\pi \omega\tau}\frac{T^2}{1+4\pi^2T^2\omega^2}e^{-4\pi^2\tau_\eta^2\omega^2}d\omega.
  \end{equation}
We get
\begin{equation}\label{ExplicitCorrVLimApp}
\mathcal C_{v}(\tau)= \sigma^2\frac{e^{-|\tau|/T}}{2\erfc(\tau_\eta/T)}\left[1+\erf\left(\frac{|\tau|}{2\tau_\eta}-\frac{\tau_\eta}{T}\right)+e^{2|\tau|/T}\erfc\left(\frac{|\tau|}{2\tau_\eta}+\frac{\tau_\eta}{T}\right)\right],
  \end{equation}
with the particular value $\mathcal C_{v}(0)=\langle v^2\rangle=\sigma^2$. Concerning the acceleration correlation function, take (minus) the second derivative of $\mathcal C_{v}$ (Eq. \ref{ExplicitCorrVLimApp}) and obtain
\begin{align}\label{ExplicitCorrALimApp}
\mathcal C_{a}(\tau)= \frac{\sigma^2}{2T^2\erfc(\tau_\eta/T)}&\left[\frac{2T}{\tau_\eta\sqrt{\pi}}e^{-\left( \frac{\tau^2}{4\tau_\eta^2}+\frac{\tau_\eta^2}{T^2}\right)}-e^{-|\tau|/T}\left(1+\erf\left(\frac{|\tau|}{2\tau_\eta}-\frac{\tau_\eta}{T}\right)\right)\right. \notag\\
&\left.-e^{|\tau|/T}\erfc\left(\frac{|\tau|}{2\tau_\eta}+\frac{\tau_\eta}{T}\right)\right].
\end{align}}
\end{proposition}

  \textit{Proof.} 
 By Lebesgue's dominated convergence, we can safely commute $\lim_{n\rightarrow \infty}$ and the indefinite integral that enter in the expression given in Eq. \ref{CorrVelNFour2App}. Recall that $(1+x/n)^n$ tends to $e^x$ as $n\rightarrow \infty$, and get to Eq. \ref{CorrVelLimFour2}. Express then  Eq. \ref{CorrVelLimFour2} in the physical space as a convolution, and perform the remaining integral to arrive at Eq. \ref{ExplicitCorrVLimApp}. The expression in Eq. \ref{ExplicitCorrALimApp}, the acceleration correlation function, also follows.   \textit{End of proof.}

\begin{proposition}\label{Prop:StatX_FOU}
{(On the statistical properties of the fields $X_{1,\epsilon}$ and its asymptotical log-correlated version $X_{1}\equiv \lim_{\epsilon\rightarrow 0}X_{1,\epsilon}$)}

{Recall first the definition of the OU-kernel $g_\tau(t) = e^{-t/\tau}1_{t\ge 0}$, where $1_{t\ge 0}$ stands for the indicator function of positive reals, and the associated response function $G_\tau(t) = (g_\tau\star g_\tau)(t) = \frac{\tau}{2}e^{-|t|/\tau}$ (Eq. \ref{eq:TransFoncOU}). We will also need its derivative, which reads as $G'_\tau(t)  = -\frac{t}{2|t|}e^{-|t|/\tau}$.}

{The unique solution $X_{1,\epsilon}$ of the dynamics given in Eq. \ref{eq:FracOUH} is a zero-average Gaussian process, that reaches a statistically stationary regime at large time $t$, independently of the initial condition. In this statistically steady state, $X_{1,\epsilon}$ is thus fully characterized by its correlation function that reads
\begin{align}
\mathcal C_{X_{1,\epsilon}}(\tau) &= -\int_{0}^{\infty}\left[ G_T'(\tau+h)-G_T'(\tau-h)\right]\frac{dh}{h+\epsilon+\sqrt{\epsilon(h+\epsilon)}}\label{eq:CXEps1}\\
&=-e^{-|\tau|/T}\int_0^{|\tau|}\frac{\sinh\left(h/T\right)dh}{h+\epsilon+\sqrt{\epsilon(h+\epsilon)}}+\cosh(|\tau|/T)\int_{|\tau|}^{\infty}\frac{e^{-h/T}dh}{h+\epsilon+\sqrt{\epsilon(h+\epsilon)}}.\label{eq:CXEps2}
\end{align}
In particular, we have 
\begin{align}
\mathcal C_{X_{1,\epsilon}}(0) = \langle X_{1,\epsilon}^2 \rangle&= \int_{0}^{\infty}\frac{e^{-h/T}dh}{h+\epsilon+\sqrt{\epsilon(h+\epsilon)}}\label{eq:VarXEps}\\
&\build{=}_{\epsilon\to 0}^{} \log\left(\frac{1}{\epsilon}\right)+O(1).\label{eq:VarXEpsDivLog}
\end{align}
In the asymptotic regime $\epsilon\rightarrow 0$, whereas the variance of $X_{1,\epsilon}$ diverges, its correlation function at a given time lag $|\tau|>0$ remains a bounded function of $\epsilon$. This defines an asymptotic zero-average Gaussian process $X_{1}$ of infinite variance, but with a bounded covariance for $|\tau|>0$. We obtain
\begin{align}
\mathcal C_{X_{1}}(\tau) =\lim_{\epsilon\rightarrow 0} \mathcal C_{X_{1,\epsilon}}(\tau) &= -\int_{0}^{\infty}\left[ G_T'(\tau+h)-G_T'(\tau-h)\right]\frac{dh}{h}\label{eq:CX1}\\
&=-e^{-|\tau|/T}\int_0^{|\tau|}\sinh\left(h/T\right)\frac{dh}{h}+\cosh(|\tau|/T)\int_{|\tau|}^{\infty}e^{-h/T}\frac{dh}{h}\label{eq:CX2}\\
&= \log^+\left( \frac{T}{|\tau|}\right)+c(|\tau|),\label{eq:CXDivLog}
\end{align}
where $\log^+(x)=\log \left(\max (x,1)\right)$ and $c(|\tau|)$ is a bounded function of its argument such that it goes to 0 as $|\tau|\rightarrow \infty$. Of special interest is the value of $c$ at the origin. We obtain
\begin{equation}\label{eq:EulerMarCst}
c(0) = \int_0^{\infty}e^{-y}\log(y)dy\approx -0.577216,
\end{equation}
and is known as (minus) the Euler-Mascheroni constant.}

{The corresponding spectral representation of the correlation function of the limiting process $X_{1}$ is given by
\begin{align}
\mathcal C_{X_{1}}(\tau) &= \int_{\mathbb R}e^{2i\pi\omega \tau}2\pi^2|\omega|\frac{T^2}{1+4\pi^2T^2\omega^2}d\omega.\label{eq:CXFour}
\end{align}}
\end{proposition}

  \textit{Proof.} 

Arguments developed in \cite{Che17} can be easily adapted to show the expression of the correlation function of $X_{1,\epsilon}$ at a given finite $\epsilon$ (Eqs. \ref{eq:CXEps1} and \ref{eq:CXEps2}) (see \cite{PerMor18} for full derivation). The expression of its variance (Eq. \ref{eq:VarXEps}) is a consequence of Eq. \ref{eq:CXEps2}. To see the logarithmic divergence with respect to $\epsilon$ (Eq. \ref{eq:VarXEpsDivLog}), split the integral entering in Eq. \ref{eq:VarXEps} in two over $[0,\epsilon]$ and $[\epsilon,\infty]$ and observe that the first term tends to a bounded constant as $\epsilon\rightarrow 0$. Subtract then from the second term the quantity $\int_\epsilon^{\infty}e^{-h/T}dh/h$ and observe that the overall quantity remains bounded as $\epsilon\rightarrow 0$. This shows the logarithmic divergence since this is the case for this subtracted quantity (performing an integration by parts over the dummy variable $h$).

Similarly, expressions for the correlation function of the limiting process $X_1$ (Eqs. \ref{eq:CX1} and \ref{eq:CX2}) are shown in \cite{Che17} and \cite{PerMor18}. Remark that the first integral on the RHS of Eq. \ref{eq:CX2} vanishes as $\tau\rightarrow 0$, and observe (again by integration by parts) that the second integral diverges logarithmically with $\tau$, showing the small scale diverging behaviour depicted in Eq. \ref{eq:CXDivLog}. To prove the overall shape of $\mathcal C_{X_{1}}$ as it is given in Eq. \ref{eq:CXDivLog}, we have to show that the function $c$ is indeed bounded and goes to 0 at large arguments. It is easy to see that once the logarithmic diverging behaviour is subtracted to the full expression, only bounded terms remain, which makes $c$ bounded too. At large arguments, re-organize the terms in a proper way to see the convergence towards 0.

{To show} the spectral representation of the correlation function (Eq. \ref{eq:CXFour}), use $G_T'(t)=\int e^{2i\pi \omega t}2i\pi\omega T^2/(1+4\pi^2\omega^2T^2)d\omega$ and inject into Eq. \ref{eq:CX1}. Perform then the remaining integral over the dummy variable $h$ using the known result $\int_0^{\infty}\sin(u)/u\,du=\pi/2$, and get Eq. \ref{eq:CXFour}. {As a final remark, whereas the regularization procedure over $\epsilon$ used in Eq. \ref{eq:FracOUH} may appear somehow arbitrary, and has some impact on the functional form of the correlation function $\mathcal C_{X_{1,\epsilon}}(\tau)$ (Eqs. \ref{eq:CXEps1} and \ref{eq:CXEps2}), this dependence disappears in the limit $\epsilon\to 0$. In other words, the same correlation function  $\mathcal C_{X_{1}}(\tau)$  (Eqs. \ref{eq:CX1} and \ref{eq:CX2}) would have been obtained using another regularization procedure as long as the divergent behaviours of variance (Eq. \ref{eq:VarXEpsDivLog}) and covariance  (Eq. \ref{eq:CXDivLog}) are ensured. This canonical behaviour of the limiting process $X_1$ is consistent with the conclusions of \cite{RobVar10} and \cite{RhoVar14}.}

  \textit{End of proof.}

  \begin{proposition}\label{propcorrelationsInfiniteX}{(On the statistical properties of the fields $X_{n,\epsilon}$ and its asymptotical behaviour)}

{The unique solution $X_{n,\epsilon}$ of the dynamics given in Eq. \ref{eq:X_MRW_infinite_N_1_Multi} is a zero-average Gaussian process, and reaches a statistically stationary regime at large time $T$, independent of the initial condition. In this statistically steady state, $X_{n,\epsilon}$ is thus fully characterized by its correlation function,  conveniently expressed in spectral space. We have 
\begin{align}\label{eq:CXNEpsilonFour}
\mathcal C_{X_{n,\epsilon}}(\tau) &= \int_{\mathbb R}e^{2i\pi\omega \tau}4\pi \omega \frac{T^2}{1+4\pi^2T^2\omega^2}\left[\frac{1}{1+\frac{4\pi^2\tau_\eta^2\omega^2}{n-1}}\right]^{n-1}\left(\int_0^\infty \frac{\sin(2\pi\omega h)dh}{h+\epsilon+\sqrt{\epsilon(h+\epsilon)}}\right)d\omega,
\end{align}
such that
\begin{align}\label{CorrXLimFour}
\mathcal C_{X}(\tau) &\equiv  \lim_{n\rightarrow \infty} \lim_{\epsilon\rightarrow 0}\mathcal C_{X_{n,\epsilon}}(\tau) =  \lim_{\epsilon\rightarrow 0} \lim_{n\rightarrow \infty} \mathcal C_{X_{n,\epsilon}}(\tau)\\& = \int_{\mathbb R}e^{2i\pi \omega\tau}2\pi^2|\omega|\frac{T^2}{1+4\pi^2T^2\omega^2}e^{-4\pi^2\tau_\eta^2\omega^2}d\omega.\label{CorrXLimFour2}
  \end{align}
In particular, we have 
\begin{align}
\mathcal C_{X}(0) = \langle X^2 \rangle&= \int_{\mathbb R}2\pi^2|\omega|\frac{T^2}{1+4\pi^2T^2\omega^2}e^{-4\pi^2\tau_\eta^2\omega^2}d\omega \\
&\build{=}_{\tau_\eta\to 0}^{} \log\left(\frac{T}{\tau_\eta}\right)+O(1),\label{eq:VarIDXTauEtaDivLog},
\end{align}
where the $O(1)$ constant is equal to minus one-half the Euler-Mascheroni constant ($\approx -0.288$), and
\begin{align}\label{eq:ConvXtoX_1asTauEta}
\lim_{\tau_\eta\to 0}\mathcal C_{X}(\tau) = \mathcal C_{X_1}(\tau),
\end{align}
where $X_1$ is the single-layer fractional Ornstein-Uhlenbeck process depicted in Proposition \ref{Prop:StatX_FOU}.}

{Concerning the expression of this correlation function in the physical space, it can be written for numerical purposes as
\begin{equation}\label{eq:ExpressCXPhysical}
\mathcal C_{X}(\tau) = \frac{T}{4\tau_\eta^3}\int_{\mathbb R} e^{-\frac{|\tau-t|}{T}}\left[\tau_\eta-t\mathcal F\left(\frac{t}{2\tau_\eta}\right)\right]dt,
\end{equation}
where the so-called Dawson integral $\mathcal F(x)=e^{-x^2}\int_0^x e^{y^2}dy$ enters.}
\end{proposition}

  \textit{Proof.} 

The correlation function $\mathcal C_{X_{n,\epsilon}}$ (Eq. \ref{eq:CXNEpsilonFour}) corresponds to the successive linear operations made on a white noise $\widetilde{W}(dt)$: an OU process for a large time scale $T$, $n-2$ OU processes at the small time scale $\tau_\eta/\sqrt{n-1}$, and a fractional OU process of vanishing Hurst exponent at $\tau_\eta/\sqrt{n-1}$ (and defined in Proposition \ref{Prop:StatX_FOU}). Expressions \ref{CorrXLimFour} to \ref{eq:ConvXtoX_1asTauEta} follow from this spectral representation. The physical form of $\mathcal C_{X}$ (Eq. \ref{eq:ExpressCXPhysical}) is obtained  through inverse Fourier transformation of  Eq. \ref{CorrXLimFour2}.

  \textit{End of proof.}

  \begin{proposition}\label{propcorrelationsInfiniteMRW} {(Concerning the covariance structure of the infinitely differentiable causal MRW $u$ and the corresponding acceleration process)}

{Assume $\gamma^2<1$. The unique statistically stationary solution $u_{n,\epsilon}$ of the set of equations Eqs. \ref{eq:OU_MRW_infinite_N_1_Multi} to \ref{eq:OU_MRW_infinite_N_N_Multi} converges, as far as the average and variance are concerned, when both $\epsilon\to 0$ and $n\to \infty$ (the limiting procedure commutes) to a zero-average process that we note $u$. }

{Its correlation function reads
\begin{align}\label{eq:CorrUFRN}
\mathcal C_u(\tau) &= \int_{\mathbb R}G_T(h+\tau)\mathcal C_f(h)e^{\gamma^2\mathcal C_X(h)}dh\\
&=T e^{-\frac{|\tau|}{T}}\int_0^{|\tau|}\cosh\left(\frac{h}{T}\right)\mathcal C_f(h)e^{\gamma^2\mathcal C_X(h)}dh+T\cosh\left(\frac{\tau}{T}\right)\int_{|\tau|}^\infty e^{-\frac{h}{T}}\mathcal C_f(h)e^{\gamma^2\mathcal C_X(h)}dh,\label{eq:CorrelUInfinite}
\end{align}
where $\mathcal C_X$ corresponds to the correlation function of the infinitely differentiable Gaussian process $X$ depicted in Proposition \ref{propcorrelationsInfiniteX}, and $C_f$ the correlation function of the Gaussian force $f$ entering in the dynamics of $u_n$ (Eq. \ref{eq:OU_MRW_infinite_N_1_Multi}) once the limit $n\to \infty$ has been taken, and given by
\begin{equation}\label{eq:CorrfInfinity}
\mathcal C_f(\tau) = \frac{\sigma^2}{T\int_{0}^\infty e^{-\frac{h}{T}}e^{-h^2/(4\tau_\eta^2)}e^{\gamma^2\mathcal C_{X}(h)}dh}e^{-\frac{\tau^2}{4\tau_\eta^2}}.
\end{equation}
In the limit of infinite Reynolds numbers, i.e. as $\tau_\eta/T\to 0$, the correlation function $\mathcal C_u$ of $u$ coincides with the one of the single-layered MRW $u_1$, which was shown in {Section \ref{Sec:CausalMRW}}  to coincide itself with the one of the single-layered OU process $v_1$ (Eq. \ref{eq:OU}) of variance $\sigma^2$, and we have
\begin{equation}\label{eq:LimitBehaveCorrelU}
\lim_{\tau_\eta\to 0}\mathcal C_u(\tau) = \mathcal C_{u_1}(\tau)=\mathcal C_{v_1}(\tau)=\sigma^2e^{-\frac{|\tau|}{T}}.
\end{equation}
Rephrased in terms inherited from the phenomenology of turbulence, the asymptotic behaviour of the correlation function (Eq. \ref{eq:LimitBehaveCorrelU}) says that intermittent corrections observed at finite Reynolds numbers (Eq. \ref{eq:CorrUFRN}), and governed by the coefficient $\gamma$, disappear at infinite Reynolds numbers. In a similar spirit, these intermittent corrections only affect the dissipative range (i.e. $\tau$ of the order and smaller than $\tau_\eta$), and disappear in the inertial range $\tau_\eta\ll \tau\ll T$. }

{Going back to finite Reynolds number predictions, i.e. keeping $\tau_\eta$ finite and smaller than $T$, the expression of the Lagrangian integral time scale $T_L$ is of special interest, and we get
\begin{equation}
T_L=\int_0^\infty \frac{\mathcal C_u(\tau)}{\mathcal C_u(0)}d\tau = \frac{T^2}{\sigma^2}\int_{0}^\infty \mathcal C_f(h)e^{\gamma^2\mathcal C_X(h)}dh\build{\to}_{\tau_\eta\to 0}^{}T.
\end{equation}
The corresponding expression for the acceleration correlation function $\mathcal C_a$ is then obtained while taking (minus) the second derivatives of $\mathcal C_u$ (Eq. \ref{eq:CorrelUInfinite}), and reads
\begin{equation}\label{eq:CorrelAUInfinite}
\mathcal C_a(\tau) = \mathcal C_f(\tau) e^{\gamma^2\mathcal C_X(\tau)}-\frac{1}{T^2}\mathcal C_u(\tau).
\end{equation}
Incidentally, the acceleration variance, and its behaviour in the infinite Reynolds limit (i.e. while looking at the limit $\tau_\eta/T\to 0$), reads 
\begin{align}
\mathcal C_a(0)=\langle a^2\rangle &= \mathcal C_f(0)  e^{\gamma^2\mathcal C_X(0)}-\frac{\sigma^2}{T^2}\\
&\build{\sim}_{\tau_\eta/T\to 0}^{}\frac{\sigma^2}{\sqrt{\pi}T\tau_\eta},\label{eq:PredVarAcceIDMRW}
\end{align} 
consistent with standard dimensional predictions, with no further intermittent corrections. }
\end{proposition}

  \textit{Proof.} 

Start with  showing the form of the asymptotic correlation function $\mathcal C_f$  (Eq. \ref{eq:CorrfInfinity}) of the force term $f$, when the number of layers $n$ goes to infinity. Consider first this correlation at a finite $n$. We have, seeking for the stationary solution of Eq. \ref{eq:OU_MRW_infinite_N_2_Multi}  and computing its correlation function in the statistically steady regime,
\begin{align*}
\mathcal C_{f_{n-1}}(\tau) &= \beta_n \int_{\mathbb R}e^{2i\pi\omega \tau}\left[\frac{\frac{\tau_\eta^2}{n-1}}{1+\frac{4\pi^2\tau_\eta^2\omega^2}{n-1}}\right]^{n-1}d\omega.
\end{align*}
Remark that for all positive $x$ and integers $n$, by the binomial formula, $(1+x/n)^n$ is bounded from below by $1+x$, such that $(1+4\pi^2\tau_\eta^2\omega^2/(n-1))^{1-n}$ is bounded from above by $(1+4\pi^2\tau_\eta^2\omega^2)^{-1}$, which is an integrable function. This allows the use of dominated convergence to conclude on the convergence of $\mathcal C_{f_{n-1}}$ as $n\to \infty$, once we take for $\beta_n$ the expression in Eq. \ref{eq:ChoiceBetaNMulti}. Taking then the limit $n\to \infty$, the inverse Fourier transform of the obtained Gaussian function is computed to arrive at Eq. \ref{eq:CorrfInfinity}. 

Looking for the stationary solution of $u$ (Eq. \ref{eq:OU_MRW_infinite_N_1_Multi}), once the limit  $n\to \infty$ has been taken and keeping in mind that the log-correlated field $X$ is independent of the forcing term $f$, the velocity correlation function reads $\mathcal C_u(\tau) = (g_T\star g_T\star \mathcal C_f e^{\gamma^2\mathcal C_X})(\tau)$.  This corresponds to the expression provided in Eq. \ref{eq:CorrUFRN}. 

Whereas it is straightforward to show the convergence of the correlation function of the process as $\tau_\eta\to 0$ and then $\epsilon\to 0$, the convergence as $\epsilon\to 0$ and only then $\tau_\eta\to 0$, as it is stated in Eq. \ref{eq:LimitBehaveCorrelU}, deserves attention. In any case, both ordering of limits give the same convergence towards the one of the OU process (Eq. \ref{eq:LimitBehaveCorrelU}). The full demonstration of this is developed in Appendix \ref{App:SQMRWInfinity}, where the respective convergence of the second order structure function is studied.

Other assertions of Proposition \ref{propcorrelationsInfiniteMRW} follow from the expression of  $\mathcal C_u$.

  \textit{End of proof.} 

\begin{proposition}\label{propSFInfiniteMRW}{ (Concerning the scaling of the higher-order structure functions of the infinitely differentiable causal MRW $u$)}

{Without loss of generality, consider an infinite number of layers $n\to \infty$, and call $u_\epsilon$ the respective process. Define the velocity increment of the process $u_{\epsilon}$ as 
\begin{equation}
\delta_\tau u_{\epsilon}(t) = u_{\epsilon}(t+\tau)-u_{\epsilon}(t).
\end{equation}
Accordingly, define the respective asymptotic structure functions as
\begin{equation}
\mathcal S_{u,m} (\tau)=\lim_{\epsilon\rightarrow 0}  \left\langle \left(u_{\epsilon}(t+\tau)-u_{\epsilon}(t)\right)^m\right\rangle.
\end{equation}
As we have seen when presenting the correlation structure of $u$ in proposition \ref{propcorrelationsInfiniteMRW}, we have, for $\gamma^2<1$,
\begin{equation}\label{eq:PredTheoS2InfiniteMulti}
\mathcal S_{u,2} (\tau)=\lim_{\epsilon\rightarrow 0} \mathcal S_{u_\epsilon,2} (\tau) = 2\left[\sigma^2-\mathcal C_u(\tau)\right]\build{\longrightarrow}_{\tau_\eta\to 0}^{}2\sigma^2\left[1-e^{-\frac{|\tau|}{T}}\right].
\end{equation}
With respect to the convergence of the fourth-order structure function $ \mathcal S_{u_\epsilon,4} $, we have a more subtle behaviour related to the ordering of the limits. We can show that, taking first the limit $\tau_\eta\to 0$ and keeping $\epsilon$ finite, $ \mathcal S_{u_\epsilon,4} $ coincides with the fourth-order structure function of the single-layered MRW $u_1$ for which scaling properties are listed in {Section \ref{Sec:CausalMRW}}. More precisely, we can write for $4\gamma^2<1$
\begin{equation}\label{eq:PropS4MRWInfinity}
\lim_{\epsilon\rightarrow 0} \lim_{\tau_\eta\to 0} \mathcal S_{u_\epsilon,4} (\tau) =  \mathcal S_{u_1,4} (\tau),
\end{equation}
which exhibits an intermittent behaviour (see Eq. \ref{eq:PropS4MRW}, with $q=2\sigma^2/T$ such that $u$ and $u_1$ have same variance). In the reverse order of the limits, calculations get intricate, but under an approximation procedure, we obtain the following scaling behaviour
\begin{equation}\label{eq:PropS4MRWInfinityReverse}
 \lim_{\tau_\eta\to 0} \lim_{\epsilon\rightarrow 0} \mathcal S_{u_\epsilon,4} (\tau) = c_{\gamma,4}  \mathcal S_{u_1,4} (\tau),
\end{equation}
where $c_{\gamma,4}$ is a constant that depends only on the intermittency coefficient $\gamma$ which can be computed. We can notice that, in this approximation, the ordering of the limits has a consequence only on the value of the multiplicative constant entering in the power-laws (Eqs. \ref{eq:PropS4MRWInfinity} and  \ref{eq:PropS4MRWInfinityReverse}), whereas the power-law exponent is the same in both cases, and exhibits an intermittent correction.}

{In a similar way, whereas taking the limit $\tau_\eta\to 0$ and then $\epsilon\rightarrow 0$ has no difficulties, we can assert that 
\begin{equation}\label{eq:PredSmInfDiffProcU}
 \lim_{\tau_\eta\to 0} \lim_{\epsilon\rightarrow 0} \mathcal S_{u_\epsilon,2m} (\tau) = c_{\gamma,2m}  \mathcal S_{u_1,2m} (\tau),
\end{equation}
showing that $u$ exhibits a lognormal spectrum (take a look at \ref{eq:PropS2NMRW} with again $q=2\sigma^2/T$) when the Reynolds number becomes infinite.}
\end{proposition}

We gather all proofs in Appendix \ref{App:SQMRWInfinity}.

\section{Scaling properties of the structure functions of the causal multifractal random walk}
\label{App:SQMRW}

To set our notations, define various quantities that will enter in following calculations. The velocity increments read
\begin{align}
\delta_\tau u_{1,\epsilon}(t) &= u_{1,\epsilon}(t+\tau)-u_{1,\epsilon}(t)\\
&=\int_{\mathbb R}g_{\tau,T}(t-s)e^{\gamma X_{1,\epsilon}(s)-\gamma^2\langle X_{1,\epsilon}^2\rangle}W(ds),
\end{align}
where $g_{\tau,T}$ corresponds to the OU-kernel associated to velocity increments, that is
\begin{align}\label{eq:DefQCalcCausalMRW}
g_{\tau,T}(t) = \sqrt{q}\left[e^{-\frac{t+\tau}{T}}1_{t+\tau\ge 0}-e^{-\frac{t}{T}}1_{t\ge 0}\right].
\end{align}
We obtain
\begin{align}
\left\langle \left( \delta_\tau u_{1,\epsilon}\right)^2 \right\rangle&=\int_{\mathbb R^2}g_{\tau,T}(t-s_1)g_{\tau,T}(t-s_2)\left\langle e^{\gamma \left( X_{1,\epsilon}(s_1)+X_{1,\epsilon}(s_2) \right)-2\gamma^2\langle X_{1,\epsilon}^2\rangle}W(ds_1)W(ds_2)\right\rangle\\
&=\int_{\mathbb R^2}g_{\tau,T}(t-s_1)g_{\tau,T}(t-s_2)\left\langle e^{\gamma \left( X_{1,\epsilon}(s_1)+X_{1,\epsilon}(s_2) \right)-2\gamma^2\langle X_{1,\epsilon}^2\rangle}\right\rangle\left\langle W(ds_1)W(ds_2)\right\rangle\\
&=\int_{\mathbb R}g_{\tau,T}^2(t-s)\left\langle e^{2\gamma X_{1,\epsilon}(s)-2\gamma^2\langle X_{1,\epsilon}^2\rangle}\right\rangle ds\\
&=\int_{\mathbb R}g_{\tau,T}^2(s) ds,\label{eq:S2U1}
\end{align}
where we have used the independence of the fields $X_{1,\epsilon}$ and $W$, and the fact that 
$\langle e^x \rangle = e^{\frac{1}{2}\langle x^2 \rangle}$ for any zero-average Gaussian random variable $x$. It is then easy to see that the result (Eq. \ref{eq:S2U1}) would have been the same with the standard Ornstein-Uhlenbeck process $v_1$ (Eq. \ref{eq:OU}), which shows that the asymptotic process $u_1$ has no intermittent corrections up to second order. Performing the remaining integral that enters in Eq. \ref{eq:S2U1} leads to the result obtained in Eq. \ref{eq:PropS2U1}.

Concerning the fourth-order structure function, we have in a similar way
\begin{align}
\left\langle \left( \delta_\tau u_{1,\epsilon}\right)^4 \right\rangle&=3\int_{\mathbb R^2}g^2_{\tau,T}(t-s_1)g^2_{\tau,T}(t-s_2)\left\langle e^{2\gamma \left( X_{1,\epsilon}(s_1)+X_{1,\epsilon}(s_2) \right)-4\gamma^2\langle X_{1,\epsilon}^2\rangle}\right\rangle ds_1ds_2 \\
&=3\int_{\mathbb R^2}g^2_{\tau,T}(t-s_1)g^2_{\tau,T}(t-s_2) e^{4\gamma^2\mathcal C_{X_{1,\epsilon}}(s_1-s_2)} ds_1ds_2 \\
&=6\int_{0}^{\infty}\left( g^2_{\tau,T}\star g^2_{\tau,T}\right)(s) e^{4\gamma^2\mathcal C_{X_{1,\epsilon}}(s)} ds,
\end{align}
where we have used Isserlis' theorem to factorize the four-time correlator of $W$ in terms of products of its correlations, which gives rise to 3 symmetrical terms of equal contribution, made appropriate change of variables, and finally exploited the parity of the functions $\left( g^2_{\tau,T}\star g^2_{\tau,T}\right)$ and $\mathcal C_{X_{1,\epsilon}}$. Dominated convergence ensures that
\begin{align}
\mathcal S_{u_1,4}(\tau) &= \lim_{\epsilon \rightarrow 0} \left\langle \left( \delta_\tau u_{1,\epsilon}\right)^4 \right\rangle\\
&=6\int_{0}^{\infty}\left( g^2_{\tau,T}\star g^2_{\tau,T}\right)(s) e^{4\gamma^2\mathcal C_{X_{1}}(s)} ds.\label{eq:FirstExpressS4U1}
\end{align}
At this stage, remark that the integral provided in Eq. \ref{eq:FirstExpressS4U1} makes sense only if the singularity $\sim s^{-4\gamma^2}$ implied by $e^{4\gamma^2\mathcal C_{X_{1}}(s)}$ (as easily seen in Eq. \ref{eq:CXDivLog}) is integrable in the vicinity of the origin. This explains the bound on $\gamma$ required by the existence on the fourth order structure function, that is
\begin{equation}
4\gamma^2<1.
\end{equation}
Compute then the function $\left( g^2_{\tau,T}\star g^2_{\tau,T}\right)(s)$, namely, for $s\ge 0$ and $\tau\ge 0$,
\begin{align}
\left( g^2_{\tau,T}\star g^2_{\tau,T}\right)(s) = q^2 e^{-\frac{2s}{T}}\int_{\mathbb R}e^{-\frac{4x}{T}}\left[e^{-\frac{\tau}{T}}1_{x+\tau\ge 0}-1_{x\ge 0}\right]^2 \left[e^{-\frac{\tau}{T}}1_{x+\tau+s\ge 0}-1_{x+s\ge 0}\right]^2dx,
\end{align} 
which integrand is made up of simple exponentials over intricated domains, and get in an exact fashion (with the help of a symbolic calculation software),
\begin{align}
\left( g^2_{\tau,T}\star g^2_{\tau,T}\right)(s) = \frac{q^2T}{4} &\left[ \left(1-e^{-\frac{\tau}{T}}\right)^3\left( 2+e^{\frac{\tau}{T}}+e^{2\frac{\tau}{T}}\right) e^{-2\frac{s}{T}}\right. \label{eq:FirstTermG2StarG2}\\
&\left.+2\left( 2e^{-\frac{\tau}{T}}-1\right)\sinh\left( 2\frac{\tau-s}{T}\right)1_{\tau-s\ge 0}\right],\label{eq:SecTermG2StarG2}
\end{align} 
and inject it into the expression of $\mathcal S_{u_1,4}$ (Eq. \ref{eq:FirstExpressS4U1}). Observe that the decrease of $\mathcal S_{u_1,4}$ as $\tau\rightarrow 0$ is governed by the second term $\left( g^2_{\tau,T}\star g^2_{\tau,T}\right)$ (Eq. \ref{eq:SecTermG2StarG2}), since the first term (Eq. \ref{eq:FirstTermG2StarG2}) implies a decrease towards 0 as $\tau^3$. Thus, only considering the leading contribution entering in Eq. \ref{eq:SecTermG2StarG2}), using $\left( 2e^{-\frac{\tau}{T}}-1\right)\approx 1$, we have in good approximation as $\tau\rightarrow 0$
\begin{align}
\mathcal S_{u_1,4}(\tau) &\approx 3Tq^2\int_{0}^{\tau}\sinh\left(\frac{2 (\tau - s)}{T}\right)e^{4\gamma^2\mathcal C_{X_{1}}(s)} ds\\
&=3Tq^2\int_{0}^{1}\sinh\left(\frac{2 \tau(1 - s)}{T}\right)e^{4\gamma^2\mathcal C_{X_{1}}(\tau s)} \tau ds\\
&\build{\sim}_{\tau\rightarrow 0}^{}6q^2\tau^2\left(\frac{\tau}{T}\right)^{-4\gamma^2}e^{4\gamma^2c(0)}\int_{0}^{1}\left(1 - s\right)s^{-4\gamma^2}ds\\
&=\frac{3}{1-6\gamma^2+8\gamma^4}q^2\tau^2\left(\frac{\tau}{T}\right)^{-4\gamma^2}e^{4\gamma^2c(0)},\label{eq:DerivAppS4MRW}
\end{align}
where the constant $c(0)$ is explicitly known, and given in Eq. 
\ref{eq:EulerMarCst}. This entails Eq. \ref{eq:PropS4MRW}.

Let us now generalize former calculations up to any order. We get
\begin{align}
\left\langle \left( \delta_\tau u_{1,\epsilon}\right)^{2m} \right\rangle &=\frac{(2m)!}{2^mm!}\int_{\mathbb R^m}\prod_{k=1}^m g^2_{\tau,T}(t-s_k)\left\langle e^{2\gamma \sum_{k=1}^m X_{1,\epsilon}(s_k)-2m\gamma^2\langle X_{1,\epsilon}^2\rangle}\right\rangle \prod_{k=1}^m ds_k \\
&=\frac{(2m)!}{2^mm!}\int_{\mathbb R^m}\prod_{k=1}^m g^2_{\tau,T}(t-s_k) e^{4\gamma^2 \sum_{k<p=1}^m\mathcal C_{X_{1,\epsilon}}(s_k-s_p)} \prod_{k=1}^mds_k \\
&\build{=}_{\epsilon\rightarrow 0}^{}\frac{(2m)!}{2^mm!}\int_{\mathbb R^m}\prod_{k=1}^m g^2_{\tau,T}(t-s_k) e^{4\gamma^2 \sum_{k<p=1}^m\mathcal C_{X_{1}}(s_k-s_p)} \prod_{k=1}^mds_k.\label{eq:S2NMRWtemp}
\end{align}
Once again, the exponential entering in Eq. \ref{eq:S2NMRWtemp} gives both the condition of existence on $\gamma$, and intermittent corrections. The strongest singularity is encountered along the diagonal, that is when all dummy variables $s_k$ coincide. It is equivalent to say that it is necessary to take
\begin{equation}
2m(m-1)\gamma^2<1,
\end{equation}
to guarantee the existence of the integral given in Eq. \ref{eq:S2NMRWtemp}. Similarly, it implies an intermittent correction of order $(\tau/T)^{-2m(m-1)\gamma^2}$, as stated in Eq. \ref{eq:PropS2NMRW}, which concludes the proofs of {Section \ref{Sec:CausalMRW}}.

\section{Scaling properties of the structure functions of the infinitely differentiable causal Multifractal Random Walk}
\label{App:SQMRWInfinity}

To set our notations, define various quantities that will enter in following calculations. The velocity increments read 
\begin{align}
\delta_\tau u(t) &= u(t+\tau)-u(t)\\
&=\int_{\mathbb R}g_{\tau,T}(t-s)e^{\gamma X(s)-\frac{\gamma^2}{2}\langle X^2\rangle}f(s)ds,
\end{align}
where $g_{\tau,T}$ corresponds to the OU-kernel associated to velocity increments, that is
\begin{align}
g_{\tau,T}(t) = e^{-\frac{t+\tau}{T}}1_{t+\tau\ge 0}-e^{-\frac{t}{T}}1_{t\ge 0}.
\end{align}
We obtain
\begin{align}
\left\langle \left( \delta_\tau u\right)^2 \right\rangle&=\int_{\mathbb R^2}g_{\tau,T}(t-s_1)g_{\tau,T}(t-s_2)\mathcal C_f(s_1-s_2)\left\langle e^{\gamma \left( X(s_1)+X(s_2) \right)-\gamma^2\langle X^2\rangle}\right\rangle ds_1ds_2\\
&=\int_{\mathbb R^2}g_{\tau,T}(t-s_1)g_{\tau,T}(t-s_2)\mathcal C_f(s_1-s_2)e^{\gamma^2C_X(s_1-s_2)} ds_1ds_2\\
&=\int_{\mathbb R}\left(g_{\tau,T}\star g_{\tau,T}\right)(s) \mathcal C_f(s)e^{\gamma^2C_X(s)} ds\label{eq:S2UInfiniteBefore}\\ &=2\int_{\mathbb R^+}\left(g_{\tau,T}\star g_{\tau,T}\right)(s) \mathcal C_f(s)e^{\gamma^2C_X(s)} ds,\label{eq:S2UInfinite}
\end{align}
where we have used the independence of the fields $X$ and $f$, and the fact that 
$\langle e^x \rangle = e^{\frac{1}{2}\langle x^2 \rangle}$ for any zero-average Gaussian random variable $x$. This shows that, contrary to the MRW case $u_1$ (Eq. \ref{eq:PropS2U1}), the asymptotic process $u$ (once the limit $\epsilon\to 0$ has been taken) has an intermittent correction up to second order when $\tau_\eta/T$ is finite. We have, for $\tau\ge 0$ and $s\ge 0$,
\begin{equation}\label{eq:FirstExpressGstarG}
\left(g_{\tau,T}\star g_{\tau,T}\right)(s) = T\left(e^{-s/T} - e^{-\tau/T}\cosh(s/T) + 
   \sinh\left(\frac{s - \tau}{T}\right)1_{s-\tau\ge 0}\right),
\end{equation}
which shows that once injected in Eq. \ref{eq:S2UInfinite}, we recover in a consistent manner
\begin{equation}
\left\langle \left( \delta_\tau u\right)^2 \right\rangle = 2\left(\sigma^2-C_u(\tau)\right).
\end{equation}
To see the behaviour of the second-order structure function in the (non-commuting) limit $\tau_\eta\to 0$ (i.e. the infinite Reynolds number limit) and then $\tau\to 0$ (i.e. the limit at small scales), regroup terms in Eq. \ref{eq:FirstExpressGstarG} and obtain, using the definition of $\mathcal C_f$ (Eq. \ref{eq:CorrfInfinity}),
\begin{align}\label{eq:FirstExpressS2Multi}
\left\langle \left( \delta_\tau u\right)^2 \right\rangle=2\sigma^2\left[ 1-\cosh\left( \frac{\tau}{T}\right)\right]+2\sigma^2\frac{\int_0^{\tau}\sinh\left( \frac{\tau-s}{T}\right)e^{-\frac{s^2}{4\tau_\eta^2}}e^{\gamma^2\mathcal C_X(s)}ds}{\int_0^{\infty}e^{-\frac{s}{T}}e^{-\frac{s^2}{4\tau_\eta^2}}e^{\gamma^2\mathcal C_X(s)}ds}.
\end{align}
Rescale then the dummy variable entering the second term by $\tau_\eta$ and obtain
\begin{align}\label{eq:SecondExpressS2Multi}
\left\langle \left( \delta_\tau u\right)^2 \right\rangle=2\sigma^2\left[ 1-\cosh\left( \frac{\tau}{T}\right)\right]+2\sigma^2\frac{\int_0^{\tau/\tau_\eta}\sinh\left( \frac{\tau-s\tau_\eta}{T}\right)e^{-\frac{s^2}{4}}e^{\gamma^2\mathcal C_X(s\tau_\eta)}ds}{\int_0^{\infty}e^{-\frac{s\tau_\eta}{T}}e^{-\frac{s^2}{4}}e^{\gamma^2\mathcal C_X(s\tau_\eta)}ds},
\end{align}
such that we obtain the simple result
\begin{align}\label{eq:ThirdExpressS2Multi}
\lim_{\tau_\eta\to 0 }\left\langle \left( \delta_\tau u\right)^2 \right\rangle=2\sigma^2\left[ 1-e^{-\frac{\tau}{T}}\right],
\end{align}
showing that, up to second-order statistics, the infinitely differentiable causal multifractal walk $u$ coincides with the underlying OU process (Eq. \ref{eq:OU}) in the infinite Reynolds number limit $\tau_\eta\to 0$.

Concerning the fourth-order structure function, we have in a similar way
\begin{align}
&\left\langle \left( \delta_\tau u\right)^4 \right\rangle=3\int_{\mathbb R^4}\prod_{k=1}^4g_{\tau,T}(t-s_k)\left\langle e^{\gamma \sum_{k=1}^4X(s_k)-2\gamma^2\langle X^2\rangle}\right\rangle \mathcal C_f(s_1-s_2)  \mathcal C_f(s_3-s_4) \prod_{k=1}^4ds_k \\
&=3\int_{\mathbb R^4}\prod_{k=1}^4g_{\tau,T}(s_k)e^{\gamma^2 \sum_{k<p=1}^4\mathcal C_X(s_k-s_p)}\mathcal C_f(s_1-s_2)  \mathcal C_f(s_3-s_4) \prod_{k=1}^4ds_k \\
&=3\int_{\mathbb R^4}g_{\tau,T}(s)g_{\tau,T}(s-h_1)g_{\tau,T}(s-h_2)g_{\tau,T}(s-h_3)\\
&\times e^{\gamma^2 \left(\mathcal C_X(h_1)+\mathcal C_X(h_2)+\mathcal C_X(h_3)+\mathcal C_X(h_1-h_2)+\mathcal C_X(h_1-h_3)+\mathcal C_X(h_2-h_3)\right)}\\
&\times \mathcal C_f(h_1)  \mathcal C_f(h_3-h_2) ds\prod_{k=1}^3dh_k \\
&=3\int_{\mathbb R^3}G_{\tau,T}(h_1,h_2,h_3)\mathcal C_f(h_1)  \mathcal C_f(h_2-h_3) e^{\gamma^2 \left(\sum_{k=1}^3\mathcal C_X(h_k)+\sum_{k<l,1}^3\mathcal C_X(h_k-h_l)\right)}\prod_{k=1}^3dh_k,\label{eq:S4IDMRWInter}
\end{align}
where we have noted 
\begin{equation}\label{eq:DefGTauT}
G_{\tau,T}(h_1,h_2,h_3)=\int_{\mathbb R}g_{\tau,T}(s)g_{\tau,T}(s+h_1)g_{\tau,T}(s+h_2)g_{\tau,T}(s+h_3)ds.
\end{equation}
The exact expression of the function $G_{\tau,T}$ (Eq. \ref{eq:DefGTauT}) could be obtained using a symbolic calculation software, although it is intricate. Instead, we will do an approximative calculation, based on an ansatz for the correlation function $\mathcal C_X$ entering in the expression of the moment of velocity increments (Eq. \ref{eq:S4IDMRWInter}), get then an equivalent at infinite Reynolds number (i.e. $\tau_\eta\to 0$), from which we deduce the scaling behaviour as $\tau$ goes to zero.

As we have seen, the correlation function $\mathcal C_X(\tau)$ of $X$ (Eq. \ref{CorrXLimFour2}) has several obvious limiting behaviours. First, it goes to zero at large arguments $\tau\gg T$. Secondly, as $\tau_\eta\to 0$, its value at the origin blows up logarithmically with $\tau_\eta$ (Eq. \ref{eq:VarIDXTauEtaDivLog}), and in the same limit, point-wise, for strictly positive arguments $\tau>0$, it behaves logarithmically with $\tau$ as $\tau\to 0$.  A simple ansatz for $\mathcal C_X(\tau)$ consistent with these limiting behaviours could be written in an approximative and simple way as
\begin{equation}\label{eq:AnsatzCX}
\mathcal C_X(\tau)\approx \frac{1}{2}\log \frac{T^2}{\tau_\eta^2+\tau^2}1_{|\tau|\le T} + d_{\tau_\eta}(\tau),
\end{equation}
where $d_{\tau_\eta}(\tau)$ is a bounded function of $\tau$ and $\tau_\eta$, that goes to zero at large arguments. Furthermore, we know that $d_{\tau_\eta}(0)\to d(0)$ coincides with minus one-half the Euler-Mascheroni constant (i.e. $\approx -0.288$) as $\tau_\eta\to 0$ (Eq. \ref{eq:VarIDXTauEtaDivLog}). Henceforth, calculations will not be performed in a rigorous way since the ansatz (Eq. \ref{eq:AnsatzCX}) in only an approximative, although realistic, form of $\mathcal C_X$. 

Find now the point-wise behaviour of the correlation function $C_f$ of $f$ (Eq. \ref{eq:CorrfInfinity}). We have, looking for an equivalent of the multiplicative factor entering in Eq. \ref{eq:CorrfInfinity}, using the ansatz proposed in Eq. \ref{eq:AnsatzCX},
\begin{equation}\label{eq:EquivCfAnsatz}
\frac{T(T/\tau_\eta)^{\gamma^2}e^{\gamma^2d(0)}g(\gamma)}{\sigma^2}\mathcal C_f(\tau) \build{\sim}_{\tau_\eta\to 0}^{} \frac{1}{\sqrt{4\pi\tau_\eta^2}}e^{-\frac{\tau^2}{4\tau_\eta^2}},
\end{equation}
where
\begin{equation}
g(\gamma)=\frac{1}{\sqrt{4\pi}}\int_{0}^\infty e^{-h^2/4}\frac{1}{(1+h^2)^{\gamma^2}}dh.
\end{equation}
From the equivalent derived in Eq. \ref{eq:EquivCfAnsatz}, we can see that $\mathcal C_f$, properly weighted, will participate to the fourth-order moment of increments (Eq. \ref{eq:S4IDMRWInter}) similarly to a distributional Dirac function, and will greatly simplify its expression. Check the realism of the ansatz (Eq. \ref{eq:EquivCfAnsatz}) on the second-order structure function (Eq. \ref{eq:S2UInfiniteBefore}) and obtain $\left\langle \left( \delta_\tau u\right)^2 \right\rangle \sim \frac{\sigma^2}{Tg(\gamma)}\left(g_{\tau,T}\star g_{\tau,T}\right)(0)=\frac{\sigma^2}{g(\gamma)}(1-e^{-\tau/T})$ as $\tau_\eta\to 0$. We can see that the approach based on the ansatz (Eq. \ref{eq:AnsatzCX}) introduces an error compared to the exact result given in Eq. \ref{eq:ThirdExpressS2Multi}: instead of the exact factor $2$ entering in Eq. \ref{eq:ThirdExpressS2Multi}, we find the factor $1/g(\gamma)\approx 2.1388$ once is used the empirical intermittency coefficient given in Eq. \ref{eq:ChoiceGamma}, corresponding thus to an overestimation of order $1/(2g(\gamma))\approx 7\%$ of the multiplicative constant, the remaining power-law dependence on $\tau$ being correct. 

Having justified the good performance of this approximative procedure, inject then Eq. \ref{eq:EquivCfAnsatz} into Eq. \ref{eq:S4IDMRWInter}, use the limiting behaviour of $\mathcal C_X$ as $\tau_\eta\to 0$ (Eq. \ref{eq:ConvXtoX_1asTauEta}), and get in a heuristic fashion the following expression
\begin{align}
\left\langle \left( \delta_\tau u\right)^4 \right\rangle \build{\sim}_{\tau_\eta\to 0}^{} 6\frac{\sigma^4}{g^2(\gamma)T^2}\int_{0}^{\infty}G_{\tau,T}(0,h,h)e^{4\gamma^2 \mathcal C_{X_1}(h)}dh.
\end{align}
Noticing that $G_{\tau,T}(0,h,h)=\left(g^2_{\tau,T}\star g^2_{\tau,T}\right)(h)$, we recover the fourth-order structure function of the MRW process (Eq. \ref{eq:FirstExpressS4U1}) using $q=2\sigma^2/T$ in Eq. \ref{eq:DefQCalcCausalMRW} (to make sure that we are comparing two processes of same variance $\sigma^2$) up to a multiplicative factor such that 
\begin{equation}
\left\langle \left( \delta_\tau u\right)^4 \right\rangle \build{\sim}_{\tau_\eta\to 0}^{}\frac{1}{4g^2(\gamma)}\left\langle \left( \delta_\tau u_1\right)^4 \right\rangle.
\end{equation}
The numerical value of this factor is $\frac{1}{4g^2(\gamma)}\approx 1.1436$ working with the empirical value for $\gamma$ (Eq. \ref{eq:ChoiceGamma}), saying that $\left\langle \left( \delta_\tau u\right)^4 \right\rangle$ is very similar to $\left\langle \left( \delta_\tau u_1\right)^4 \right\rangle$ at large Reynolds number, in particular its (intermittent) scaling behaviour with $\tau$ (see Eq. \ref{eq:DerivAppS4MRW}).

Let us end this appendix with computing, under the very same approximation based on Eq. \ref{eq:AnsatzCX}, higher-order structure functions. We have
\begin{align}
&\left\langle \left( \delta_\tau u\right)^{2m} \right\rangle\\
&=\frac{(2m)!}{2^m m!}\int_{\mathbb R^{2m}}\prod_{k=1}^{2m}g_{\tau,T}(t-s_k)\left\langle e^{\gamma \sum_{k=1}^{2m}X(s_k)-m\gamma^2\langle X^2\rangle}\right\rangle\prod_{k=1}^{m} \mathcal C_f(s_{2k-1}-s_{2k}) \prod_{k=1}^{2m}ds_k \\
&=\frac{(2m)!}{2^m m!}\int_{\mathbb R^{2m}}\prod_{k=1}^{2m}g_{\tau,T}(t-s_k)e^{\gamma^2 \sum_{k<l,1}^{2m}\mathcal C_X(s_k-s_l)}\prod_{k=1}^{m} \mathcal C_f(s_{2k-1}-s_{2k}) \prod_{k=1}^{2m}ds_k \\
&\build{\sim}_{\tau_\eta\to 0}^{}\frac{(2m)!}{2^m m!}\left( \frac{\sigma^2}{g(\gamma)T}\right)^m\int_{\mathbb R^{m}}\prod_{k=1}^{m}g^2_{\tau,T}(t-s_k)e^{\gamma^2 \sum_{k<l,1}^{m}\mathcal C_{X_1}(s_k-s_l)}\prod_{k=1}^{m}ds_k,
\end{align}
showing that
\begin{align}
\left\langle \left( \delta_\tau u\right)^{2m} \right\rangle \build{\sim}_{\tau_\eta\to 0}^{}\frac{1}{2^mg^m(\gamma)}\left\langle \left( \delta_\tau u_1\right)^{2m} \right\rangle,
\end{align}
which entails Eq. \ref{eq:PredSmInfDiffProcU}.

\bibliographystyle{jfm}

\end{document}